%% file: main.tex
\documentclass[fleqn,usenatbib]{mnras}

\usepackage{newtxtext,newtxmath}

\usepackage[T1]{fontenc}
\usepackage{ae,aecompl}

\usepackage{graphicx}	

\usepackage{siunitx}
\usepackage[normalem]{ulem}

\newcommand{\app}[1]{Appendix~\ref{sec:#1}}
\newcommand{\eq}[1]{Equation~\ref{eq:#1}}
\newcommand{\fig}[1]{Figure~\ref{fig:#1}}
\renewcommand{\sec}[1]{Section~\ref{sec:#1}}
\newcommand{\tab}[1]{Table~\ref{tab:#1}}

\newcommand{\simba}{\mbox{\sc{Simba}}}
\newcommand{\eagle}{\mbox{\sc{Eagle}}}
\newcommand{\galform}{\mbox{\sc{Galform}}}
\newcommand{\powderday}{\mbox{\sc{Powderday}}}
\newcommand{\eightfifty}{\SI{850}{\micro\meter}}
\newcommand{\mumetre}{\SI{}{\micro\meter}}

\title[SMGs in cosmological hydrodynamic simulations]{Reproducing sub-millimetre galaxy number counts with cosmological hydrodynamic simulations}

\author[C. C. Lovell et al.]{Christopher C. Lovell,$^{1}$\thanks{E-mail: c.lovell@herts.ac.uk (CCL)}
James E. Geach,$^{1}$
Romeel Dav\'{e},$^{2,3,4}$
\newauthor
Desika Narayanan$^{5,6,7}$ \&
Qi Li$^{5}$
\\
$^{1}$Centre for Astrophysics Research, School of Physics, Astronomy \& Mathematics, University of Hertfordshire, Hatfield AL10 9AB\\
$^{2}$Institute for Astronomy, Royal Observatory, University of Edinburgh, Edinburgh EH9 3HJ\\
$^{3}$University of the Western Cape, Bellville, Cape Town 7535, South Africa\\
$^{4}$South African Astronomical Observatories, Observatory, Cape Town 7925, South Africa\\
$^{5}$Department of Astronomy, University of Florida, 211 Bryant Space Sciences Center, Gainesville, FL, USA\\
$^{6}$University of Florida Informatics Institute, 432 Newell Drive, CISE Bldg E251, Gainesville, FL, USA\\
$^{7}$Cosmic Dawn Center, Niels Bohr Institute, University of Copenhagen and DTU-Space, Technical University of Denmark
}

\date{Accepted XXX. Received YYY; in original form ZZZ}

\pubyear{2020}

\begin{document}
\label{firstpage}
\pagerange{\pageref{firstpage}--\pageref{lastpage}}
\maketitle

\begin{abstract}
Matching the number counts of high-$z$ sub-millimetre-selected galaxies (SMGs) has been a long standing problem for galaxy formation models.
In this paper, we use 3D dust radiative transfer to model the sub-mm emission from galaxies in the \simba\ cosmological hydrodynamic simulations, and compare predictions to the latest single-dish observational constraints on the abundance of \eightfifty-selected sources.
We find good agreement with the shape of the integrated \eightfifty\ luminosity function, and the normalisation is within 0.25 dex at $> 3 \; \mathrm{mJy}$, unprecedented for a fully cosmological hydrodynamic simulation, along with good agreement in the redshift distribution of bright SMGs.
The agreement is driven primarily by \simba's good match to infrared measures of the star formation rate (SFR) function between $z = 2-4$ at high SFRs.
Also important is the self-consistent on-the-fly dust model in \simba, which predicts, on average, higher dust masses (by up to a factor of 2.5) compared to using a fixed dust-to-metals ratio of 0.3.
We construct a lightcone to investigate the effect of far-field blending, and find that 52\% of sources are blends of multiple components, which makes a small contribution to the normalisation of the bright-end of the number counts.
We provide new fits to the \eightfifty\ luminosity as a function of SFR and dust mass.
Our results demonstrate that solutions to the discrepancy between sub-mm counts in simulations and observations, such as a top-heavy IMF, are unnecessary, and that sub-millimetre-bright phases are a natural consequence of massive galaxy evolution.
\end{abstract}



\input{intro}
\input{methods}
\input{selection}
\input{results}

\input{conc}

\section*{Acknowledgements}

The authors wish to thank the referee for comprehensive comments that greatly improved this manuscript.
We also wish to thank Gian Luigi Granato and Claudia Lagos for providing their number counts, and James Trayford, Maarten Baes, Gergo Popping, Ian Smail, Christopher Hayward and Rob Ivison for helpful comments and suggestions.
C.C.L. and J.E.G. acknowledge financial support from the Royal Society by way of grants RGF\textbackslash{}EA\textbackslash{}181016 and URF\textbackslash{}R\textbackslash{}180014.
\simba\ was run at the DiRAC@Durham facility managed by the Institute for Computational Cosmology on behalf of the STFC DiRAC HPC Facility (www.dirac.ac.uk). The equipment was funded by BEIS capital funding via STFC capital grants ST/P002293/1, ST/R002371/1 and ST/S002502/1, Durham University and STFC operations grant ST/R000832/1. DiRAC is part of the National e-Infrastructure.
Partial support for D.N. and Q.L. were provided from the US National Science Foundation via NSF AST-1715206 and AST-1909153.

We used the following open source software packages in the analysis, unless already mentioned: \textsc{Astropy} \citep{robitaille_astropy:_2013}, \textsc{Scipy} \citep{2020SciPy-NMeth} and Matplotlib \citep{Hunter:2007}.

\section*{Data Availability}
The data underlying this article will be shared on reasonable request to the corresponding author.



\bibliographystyle{mnras}
\bibliography{smg_paper,custom,pd_refs}



\appendix
\input{appendix}


\bsp	
\label{lastpage}
\end{document}

%% file: intro.tex
\section{Introduction}

Sub-millimeter (sub-mm) galaxies \citep[SMGs;][]{smail_deep_1997,hughes_high-redshift_1998,blain_submillimeter_2002} are a rare cosmological population of galaxies with significant emission in the 250--1000 \mumetre\ wavelength range.
This emission comes from the re-processing of ultraviolet (UV) emission by dust grains within the galaxy, which is reemitted in the far-infrared and subsequently redshited to the sub-mm \citep{hildebrand_determination_1983}.
Due to the negative $K$-correction, SMGs have the observationally unique property that for a given luminosity, their measured flux density in the sub-mm remains constant over a large range in redshift.
This makes them an ideal source population to study galaxy evolution over the first few billion years of the Universe's history  \citep[for a review, see][]{casey_dusty_2014}.

A number of surveys over the past 30 years have discovered and characterised large numbers of SMGs.
The first samples were revealed with the Sub-millimetre Common User Bolometer Array (SCUBA) installed on the James Clerk Maxwell Telescope \citep[JCMT;][]{smail_deep_1997,hughes_high-redshift_1998}.
These were subsequently followed up with a number of additional SCUBA surveys in different extragalactic survey fields \citep{chapman_redshift_2005,coppin_scuba_2006} as well as with other instruments such as the Large APEX BOlometer CAmera \citep[LABOCA;][]{siringo_large_2009,weis_large_2009}.
However, such surveys were typically pencil-beams, detecting small samples of objects and susceptible to cosmic variance.
SCUBA's successor, SCUBA-2 \citep{holland_scuba-2_2013}, increased the number of bolometers by two orders of magnitude, increasing mapping speeds by an order of magnitude and making much larger sub-mm surveys possible.
The SCUBA-2 Cosmology Legacy Survey \citep[S2CLS;][]{geach_scuba-2_2017} was the largest of the first JCMT Legacy Surveys, mapping $\sim 5 \;\mathrm{deg^{2}}$ over a number of well studied extragalactic fields close to the \eightfifty{} confusion limit.

Recently, interferometers such as the Atacama Large Millimetre/sub-millimetre Array (ALMA) have afforded unprecedented angular resolution, allowing for detailed studies of resolved properties of SMGs \citep[for a recent review, see][]{hodge_high-redshift_2020}.
These studies have shown that at least some sources observed with single-dish instruments are `blends' of multiple components, both associated and unassociated \citep[e.g.][]{wang_sma_2011,smolcic_millimeter_2012,hodge_alma_2013,danielson_alma_2017,stach_alma_2018,wardlow_alma_2018,hayward_observational_2018}.
However, blank field surveys with ALMA have so far covered much smaller areas than those accessible by single-dish observatories.
Follow up of individual bright sources from single-dish surveys have been performed \citep[e.g. ALESS;][]{hodge_alma_2013,karim_alma_2013} but such surveys suffer from incompleteness at the faint end.

Studies with both single-dish and interferometric instruments are beginning to form a consistent picture of SMGs properties.
The simplest way to characterise the populations from single-dish surveys that does not rely on obtaining redshifts or matching with counterparts in other bands is to measure the number counts, i.e.\ the projected number density as a function of flux density.
For the SMG population the counts are now well-constrained and not dominated by cosmic variance effects \citep{geach_scuba-2_2017}.
Matching with counterparts observed at other wavelengths allows redshifts and other intrinsic properties to be determined \citep[e.g.][]{dudzeviciute_alma_2020}.
SMGs with flux densities $> 1\; \mathrm{mJy}$ are relatively rare ($\sim 10^{-5} \; \mathrm{cMpc^{-3}}$ at $z \sim 2$), peak at cosmic noon \citep[$z \sim 2-3$;][]{chapman_redshift_2005,simpson_alma_2014,dudzeviciute_alma_2020}, and have large stellar masses \citep{swinbank_rest-frame_2004,michalowski_stellar_2012,da_cunha_alma_2015}, halo masses \citep{hickox_laboca_2012,chen_faint_2016,an_multi-wavelength_2019,lim_scuba-2_2020}, gas reservoirs \citep{riechers_massive_2010,engel_most_2010,carilli_imaging_2010,bothwell_survey_2013} and central black hole masses \citep{alexander_weighing_2008,wang_alma_2013}.
However, many of the details of this picture are still uncertain, and often the subject of selection and incompleteness effects.

The high sub-mm fluxes in SMGs have been attributed to both a high star formation rate (SFR), leading to substantial UV emission, and a large dust reservoir attenuating that emission.
Using simple local calibrations between the SFR and the thermal IR emission
\citep{kennicutt_jr_star_2012,wilkins_recalibrating_2019},
or multi-band spectral energy density (SED) fitting to stellar population synthesis (SPS) models, the inferred SFRs of SMGs are of the order of hundreds, sometimes thousands of solar masses per year \citep[e.g.][]{rowan-robinson_extreme_2018}.
What causes these extremely high SFRs is subject to debate.
Local Ultra Luminous Infra-Red Galaxies \citep[ULIRGs; $L_{\mathrm{bol}} \geqslant 10^{11} \, L_{\odot}$;][]{sanders_luminous_1996}, which exhibit similar observational properties to SMGs, are predominantly the result of gas-rich major mergers.
It has been proposed that similar merger events at high-$z$ could be the cause of SMG populations \citep[e.g.][]{narayanan09a,narayanan_formation_2010,narayanan10b}.
However, the frequency of such events alone is too low to explain the observed number densities \citep{hayward_submillimetre_2013}.
Alternatively, sustained gas accretion, and starbursts triggered by instabilities in disks and bars (where present), have also been proposed as candidate processes for triggering significant rest-frame FIR emission \citep{fardal01a,dave_nature_2010,narayanan_formation_2015}.

Cosmological simulations of galaxy evolution provide a unique tool for studying these questions.
When combined with appropriate radiative transfer models, the sub-mm emission from galaxies can be predicted.
Comparisons can then be made to observed number counts as an additional modelling constraint, as well as allowing one to investigate the physical properties of SMGs and the origin of their bright sub-mm emission.
Unfortunately, it has been notoriously difficult for many modern cosmological models to match the observed number counts of SMGs, or to generate the large SFRs seen in observed sources, without invoking novel modelling assumptions.

A number of semi-analytic models (SAMs) have attempted to reproduce sub-mm number counts \citep[\textit{e.g.}][]{granato_infrared_2000,fontanot_reproducing_2007,somerville_galaxy_2012}.
One such model is the \galform\ (SAM), which has been tuned to successfully reproduce the number counts of \eightfifty\ and $\mathrm{1.1 \,mm}$ selected galaxies\footnote{as well as the rest-frame UV luminosity function of Lyman-break galaxies at $z = 3$ and the $z = 0$ $K$-band luminosity function}.
However, in order to achieve this good agreement \galform\ invokes a top-heavy Initial Mass Function (IMF).
Early versions of the model used a flat IMF above $1 \, \mathrm{M_{\odot}}$, in sub-$L_{*}$ mergers \citep{baugh_can_2005,swinbank_properties_2008}.
This is required to produce sufficiently bright sub-mm emission during frequent low-mass merger events.
Later versions of the model used a more moderately top-heavy IMF in starbursts, triggered by disk instabilities rather than mergers, and found similarly good agreement with the number counts \citep{cowley_simulated_2015,park_clustering_2016,lacey_unified_2016,cowley_evolution_2019}.
However, such IMF variability is still controversial, particularly extreme forms and any dependence on merger state \citep{bastian_universal_2010,hopkins_variations_2013,krumholz_big_2014}, and is inconsistent with the constraints on the IMF in massive star-forming galaxies which is significantly less extreme \citep[e.g.][]{tacconi_submillimeter_2008}, though there is tentative evidence of a bottom-light/top-heavy IMF in both local star-forming region analogues \citep{motte_unexpectedly_2018,schneider_excess_2018} and some gravitationally lensed high-redshift starbursts \citep{zhang_stellar_2018}.
\cite{safarzadeh_is_2017} showed that a variable IMF is degenerate with a number of other modelling processes in SAMs, such as the form of stellar feedback.
They highlight that taking in to account dust mass allows for a good fit to the number counts without resorting to a variable IMF.
Most recently, the \textsc{Shark} SAM \citep{lagos_shark:_2018} is able to broadly reproduce the \eightfifty\ counts (whilst slightly overestimating the bright end counts compared to S2CLS; \citealt{geach_scuba-2_2017}) using a fixed \cite{chabrier_galactic_2003} IMF \citep{lagos_far-ultraviolet_2019}.
They attribute the good agreement to their use of physically motivated attenuation curves obtained from a self-consistent galaxy evolution model \citep[\eagle;][]{trayford_fade_2020}.

This said, SAMs require relatively simplified assumptions regarding the star-dust geometry in galaxies.
Because the observed sub-mm flux density depends in large part on the extent of the dust (i.e.\ in order to produce a sufficiently cold peak in the thermal dust SED such that the galaxy would be detectable in the sub-mm), hydrodynamic simulations of galaxy formation provide an attractive alternative for modelling dusty galaxies at high-$z$.
However, hydrodynamic simulations, which self-consistently model physical processes above the sub-grid scale \citep{somerville_physical_2015}, have typically struggled to reproduce sub-mm number counts, commonly underpredicting by factors of up to 1 dex or more.
The disparity with observational constraints has been variously attributed to the choice of a fixed IMF, the lack of `bursty' star formation on short time scales, and the well known offset in the normalisation of the star-forming sequence at $z \sim 2$ seen in such simulations, at the epoch of peak SMG activity \citep{madau_cosmic_2014}.
The smaller volumes necessary for such simulations, due to the increased computational complexity, have also been highlighted as a potential source for the offset.
\cite{dave_nature_2010} found that galaxies rapidly forming stars through secular gas accretion processes, rather than mergers, can explain the number densities of SMGs, quantifying the suggestion in \cite{dekel_cold_2009} that SMGs can be fed via steady cold accretion rather than mergers.
However, the abundance-matched SMGs in \cite{dave_nature_2010} have SFRs $\sim 2-4 \times$ lower than observed SMG's SFRs inferred using local calibrations.
\cite{shimizu_submillimetre_2012} model the sub-mm emission using a spherically symmetric dust screen model, finding reasonably good agreement with observed number counts, and use a lightcone to measure the angular correlation function of sub-mm sources.

While the \citet{dave_nature_2010} and \cite{shimizu_submillimetre_2012} cosmological hydrodynamic simulations represented major steps forward in modelling sub-mm galaxies in bona fide cosmological hydrodynamic simulations, they did not explicitly couple their models with dust radiative transfer (RT) in order to translate the simulations to observer-space.
As a result, direct comparisons with sub-mm surveys are fraught with uncertainty.
Recently, \citet{mcalpine_nature_2019} advanced this effort via self-consistent predictions for the sub-mm emission using sophisticated 3D dust RT.
They used the \eagle\ simulations \citep{schaye_eagle_2015,crain_eagle_2015} combined with the \textsc{Skirt} RT code \citep{camps_data_2018} and found good agreement between \textsc{Eagle} and the observed SMG redshift distribution.
However, they form very few high flux density ($> 3 \; \mathrm{mJy}$) sources, and the luminosity function at IR-wavelengths has been shown to be in tension with observational constraints \citep{wang_multi-wavelength_2019,cowley_evolution_2019}.

In this paper we use RT to model the sub-mm emission from galaxies in the \simba\ simulation \citep{dave_simba:_2019}, a state-of-the-art cosmological hydrodynamical simulation.
\simba\ reproduces key galaxy demographics from early epochs until today in a sufficiently large volume to produce substantial numbers of SMGs, making it an ideal platform to investigate the SMG population within a cosmological context.
A novel element of \simba\ is its self-consistent dust model, which accounts for the growth and destruction of dust from various physical processes \citep{li_dust--gas_2019}.
We use this feature of \simba\ together with the \textsc{Powderday} 3D dust RT code \citep{narayanan_powderday_2020} to produce self-consistent predictions for the \eightfifty\ sub-mm emission.
We focus on the number density of sub-mm sources, using a lightcone to account for blending in a large single-dish beam and to quantify cosmic variance in pencil-beam surveys, and then compare to recent observational constraints.

This paper is laid out as follows.
In \sec{method} we describe the \simba\ simulations in detail, our SED modelling framework, our galaxy selection criteria, and our method for constructing lightcones.
In \sec{all_counts} we present our results for the \eightfifty\ number counts, including an assessment of the contribution of blends, an analysis of the redshift distribution of sources and comparisons with the latest observational and modelling constraints.
In \sec{drivers} we explore the drivers of sub-mm emission in \simba, focusing on the distribution of star formation rates and dust masses.
Finally, we summarise our conclusions in \sec{conc}.
Throughout we assume a \cite{planck_collaboration_planck_2016} concordant cosmology, with parameters $\mathrm{\Omega_{m}}=0.3$,  $\mathrm{\Omega_{\Lambda}}=0.7$, $\mathrm{\Omega_{b}}=0.048$, $\mathrm{H_{0}} = 68 \mathrm{km \, s^{-1} \, Mpc^{-1}}$, $\sigma_{8} = 0.82$, and $\mathrm{n_{s}} = 0.97$.

%% file: methods.tex
\section{Simulations \& Methods}
\label{sec:method}

\begin{figure*}
\includegraphics[width=\textwidth]{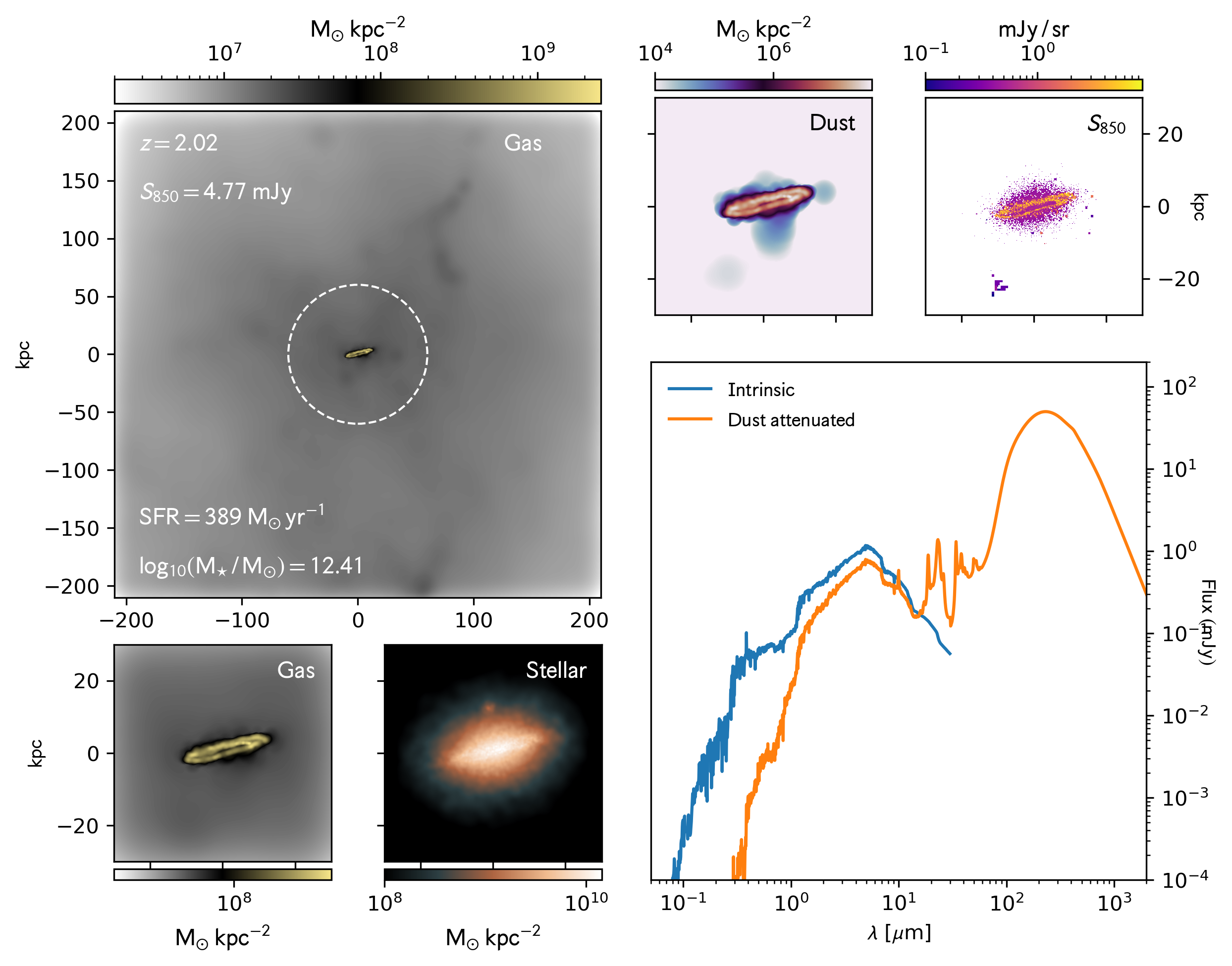}
    \caption{A \simba\ SMG at $z = 2$, with $S_{850} = 4.77 \; \mathrm{mJy}$.
    \textit{Top left}: surface density of gas. The $D = 120\,\mathrm{pkpc}$ aperture through which the spectrum is measured is shown by the dashed white circle.
    \textit{Bottom left}: zoom on the surface density of gas (left) and stars (right).
    \textit{Top right}: zoom on the surface density of dust (left) and the resolved $S_{850}$ emission (right).
    \textit{Bottom right}: the intrinsic (blue) and dust reprocessed integrated SED (orange) over the $120\,\mathrm{pkpc}$ aperture.
    }
    \label{fig:render}
\end{figure*}

\subsection{The \simba\ Simulations}
\label{sec:simba}

The \simba\ simulations are a series of state-of-the-art cosmological hydrodynamical simulations of galaxy formation \citep{dave_simba:_2019}.
They are the successor to the \textsc{Mufasa} simulations \citep{dave_mufasa:_2016,dave_mufasa:_2017} with improvements to the sub-grid prescriptions for both star formation and AGN feedback.
Both \textsc{Mufasa} and \simba\ are built on \textsc{Gizmo} \citep{hopkins_new_2015}, a gravity plus hydrodynamics code based on \textsc{Gadget-3} \citep{springel_simulations_2005}, and use its Meshless Finite Mass (MFM) method.

Non-equilibrium radiative cooling from H, He and metals is handled by \textsc{Grackle} \citep{smith_grackle_2017}, with the \cite{rahmati_evolution_2013} self-shielding prescription applied to a spatially uniform ionizing background \citep{haardt_radiative_2012}.
Star formation is based on the H$_2$ Schmidt-Kennicutt relation \citep{kennicutt_global_1998}, calculated using the \cite{krumholz_comparison_2011} sub-grid models with minor modifications \citep[see][]{dave_mufasa:_2016}.
Stellar wind-driven feedback is modelled as a decoupled kinetic outflow with a 30\% hot component, where the mass loading factor scales as measured in \cite{angles-alcazar_cosmic_2017} from the {\sc fire} simulations, and gas elements are locally enriched in the instantaneous enrichment approximation.

Black holes are seeded dynamically within Friends-of-Friends (FOF) halos where the stellar mass $M_{\star} \gtrsim 10^{\,9.5} \; \mathrm{M_\odot}$.
These black holes are then grown via two modes: a torque driven cold-accretion mode based on \cite{angles-alcazar_gravitational_2017}, and Bondi accretion from the hot halo \citep{bondi_mechanism_1944}.
The resulting energetic feedback is modelled kinetically depending on the Eddington ratio $f_{\rm Edd}$, where high accretion rates  ($f_{\rm Edd}>0.2$) represent multiphase winds and low accretion rates ($f_{\rm Edd}<0.02$) result in collimated jets, with a transition region in between.
Radiative feedback from X-ray emission is also included guided by the model introduced in \cite{choi_radiative_2012}, where a spherically-symmetric kinetic push is added to star-forming gas and heat is added to non-star-forming gas.

\simba\ also includes a unique self-consistent on-the-fly dust framework that models the production, growth and destruction of grains~\citep{dave_simba:_2019,li_dust--gas_2019}.
Dust grains are assumed to have a single size, 0.1 \mumetre, and are passively advected along with gas elements.
Metals ejected from SNe and AGB stars condense into grains following the \cite{dwek_evolution_1998} prescription.
The condensation efficiencies for each process are updated based on the theoretical models of \cite{ferrarotti_composition_2006} and \cite{bianchi_dust_2007}, respectively, the latter to match the low metallicity regime of the dust-to-gas mass ratio \citep[DTG;][]{remy-ruyer_gas--dust_2014}.
The amount of dust can increase through two-body processes by accreting gas-phase metals \citep{dwek_evolution_1998,hirashita_dust_2000,asano_dust_2013}.
Grains can be destroyed by high velocity ions in hot, dense environments via `thermal sputtering,' as well as in SNe shocks following the \cite{mckinnon_dust_2016} prescription.
Hot-phase winds, star formation and any gas subject to X-ray or jet feedback from AGN also completely destroy dust in a given gas element.
This prescription results in dust-to-metal ratios in good agreement with observations in \simba, and dust mass functions broadly in agreement with data albeit somewhat low at $z\sim 2$~\citep{li_dust--gas_2019}, although coming much closer than previous models~\citep[e.g.][]{mckinnon17a}.  Thus it appears that \simba\ may mildly underestimate the dust content of dusty SFGs during Cosmic Noon, which is relevant for this work.

\simba\ was tuned primarily to match the evolution of the overall stellar mass function and the stellar mass--black hole mass relation \citep{dave_simba:_2019}.
The model reproduces a number of key observables at both low and high redshift that do not rely on this tuning, and are bona fide predictions of the model, including SFR functions, the cosmic SFR density, passive galaxy number densities \citep{rodriguez_montero_mergers_2019}, galaxy sizes and star formation rate profiles \citep{appleby_impact_2020}, central supermassive black hole properties \citep{thomas_black_2019}, damped Lyman-$\alpha$ abundances \citep{hassan_testing_2020}, star formation histories \citep{mamon_frequency_2020}, the reionisation-epoch UV luminosity function \citep{wu_photometric_2020}, and the low-redshift Ly$\alpha$ absorption \citep{christiansen_jet_2019}.
Importantly for this study, \simba\ reproduces the bright-end CO luminosity function at $z = 2$ \citep{dave_galaxy_2020}, which has been difficult to match in other recent models \citep[see][]{riechers_coldz_2019,popping_alma_2019}.\footnote{though these comparisons are sensitive to the choice of $\alpha_{\,\mathrm{CO}}$ conversion factor and/or conversion between higher J-order CO transitions to CO(1-0) \citep{decarli_alma_2019}.}

This fiducial physics model was run on a number of volumes with different resolutions.
The largest has a side length of 147\,Mpc with 1024$^3$ dark matter particles and 1024$^3$ gas elements in the volume, and an adaptive gravitational softening length covering 64 neighbours with a minimum value of $0.5 \,h^{-1}\,\mathrm{kpc}$.
We use this simulation in the present study, because we wish to study rare massive SMGs.
While MFM is effectively an unstructured mesh hydro scheme, its gas elements are mass-conserving so can be regarded as particles.
The gas element mass is $1.2\times 10^7 M_\odot$ and the dark matter particle mass is $6.3\times 10^7 M_\odot$, which for the present study means that our SMGs are resolved with thousands of gas elements at minimum.

Our tests indicate that this is sufficient to reliably predict the far-infrared spectrum with RT, which we describe next.

\subsection{Sub-millimetre Emission Modelling}

\subsubsection{Dust continuum radiative transfer}
\label{sec:method_rt}

We estimate the sub-mm fluxes through dust continuum RT using \powderday\citep{narayanan_powderday_2020}\footnote{Maintained at \href{https://github.com/dnarayanan/powderday}{github.com/dnarayanan/powderday}}.
\powderday\ provides a convenient Python framework for modelling the dust-attenuated SEDs of galaxies in cosmological simulations, with support for parallelism through multithreading and MPI.
The code is modular and includes the Flexible Stellar Population Synthesis model for source populations \citep[FSPS,][]{conroy_propagation_2009,conroy_propagation_2010}\footnote{Using Python-FSPS \citep{dan_foreman-mackey_python-fsps:_2014} to interface with the Fortran FSPS code.}, \textsc{Hyperion} for Monte Carlo RT \citep{robitaille_hyperion_2011}, and the \textsc{Yt} toolkit \citep{turk_yt_2010} for interfacing with cosmological simulation data, including \textsc{Gizmo}.
Below we describe the main components of \powderday, and any modifications made for this project.
A full description of \powderday\ is provided in \cite{narayanan_powderday_2020}.

Each star particle is treated as a Simple Stellar Population (SSP), with a fixed age and metallicity.
These properties are provided directly to FSPS (without relying on grid interpolation), which generates an SED assuming an IMF combined with theoretical isochrones.
We use the default MILES empirical spectral library \citep{sanchez-blazquez_medium-resolution_2006} combined with the BPASS isochrones \citep{eldridge_binary_2017,stanway_re-evaluating_2018}, which take into account binary evolution pathways in the determination of the emission.
For consistency with \simba\ we use a \cite{chabrier_galactic_2003} IMF; we modified FSPS to include BPASS models assuming a Chabrier IMF\footnote{Grids provided at \href{https://github.com/christopherlovell/fsps}{github.com/christopherlovell/fsps}}.
In \app{bpass} we investigate the dependence of our results on the choice of SSP model for sources; it is quite mild, typically resulting in $\sim 5\%$ variation in the $850\mumetre$ flux.
We do not include a contribution from AGN activity to the intrinsic flux, since AGN are generally found to be bolometrically sub-dominant in SMGs~\citep{alexander_x-ray_2005,coppin_mid-infrared_2010}.
We also do not explicitly model subgrid absorption and emission, since this would introduce a significant number of extra free parameters in to our modelling pipeline, however we plan to evaluate the impact of such processes in future work.
The intrinsic emission for an example galaxy at $z = 2$ is shown as the blue line in \fig{render}.

Once the radiation is emitted from sources it propagates through the dusty ISM, which acts to scatter, absorb and re-emit the incident radiation.
\textsc{Hyperion} solves this dust RT problem using a Monte Carlo approach. Note that \powderday\ includes heating from the CMB, which can be non-negligible in galaxies at high redshift \citep[$z \geqslant 4$; see][]{privon_interpretation_2018}.
Photon packets are released with random direction and frequency, and propagate until they escape the grid or reach some limiting optical depth $\tau$.
The dust mass is represented on an octree grid, where each cell has a fixed dust mass and temperature.
We use the \cite{draine_interstellar_2003} dust models to determine the wavelength dependence of the absorption, scattering and emission cross-sections, with $R_V = 3.1$\footnote{see \href{docs.hyperion-rt.org/en/stable/dust/d03.html}{http://docs.hyperion-rt.org/en/stable/dust/d03.html} for details}.
An iterative procedure is used to calculate the equilibrium dust temperature.
The output SEDs are then calculated through ray tracing.
The post-processed SED for an example galaxy is shown in orange in \fig{render}; the far-UV is attenuated and re-emitted at IR wavelengths.

There are a number of free parameters in \textsc{Hyperion} that can be tuned to the size and resolution of the simulation being processed.
We set the number of photons used for calculating initial temperatures and specific energies, ray tracing source and dust emission, and calculating output SEDs to $n_{\mathrm{phot}} = 1 \times 10^{6}$.
The octree grid is refined until each cell contains fewer than $n_{\mathrm{ref}} = 16$ gas elements.
To test the convergence we ran a number of galaxies with increased photon number ($n_{\mathrm{phot}} = 1 \times 10^{7}$) and a finer octree grid ($n_{\mathrm{ref}} = 8$) and found that, for galaxies with $S_{850} \geqslant 1 \; \mathrm{mJy}$ for the original parameters, the mean fractional difference in the flux densities was $\sim 18\%$, or $< 0.1 \; \mathrm{dex}$, sufficient for this work.

\begin{figure}
\includegraphics[width=\columnwidth]{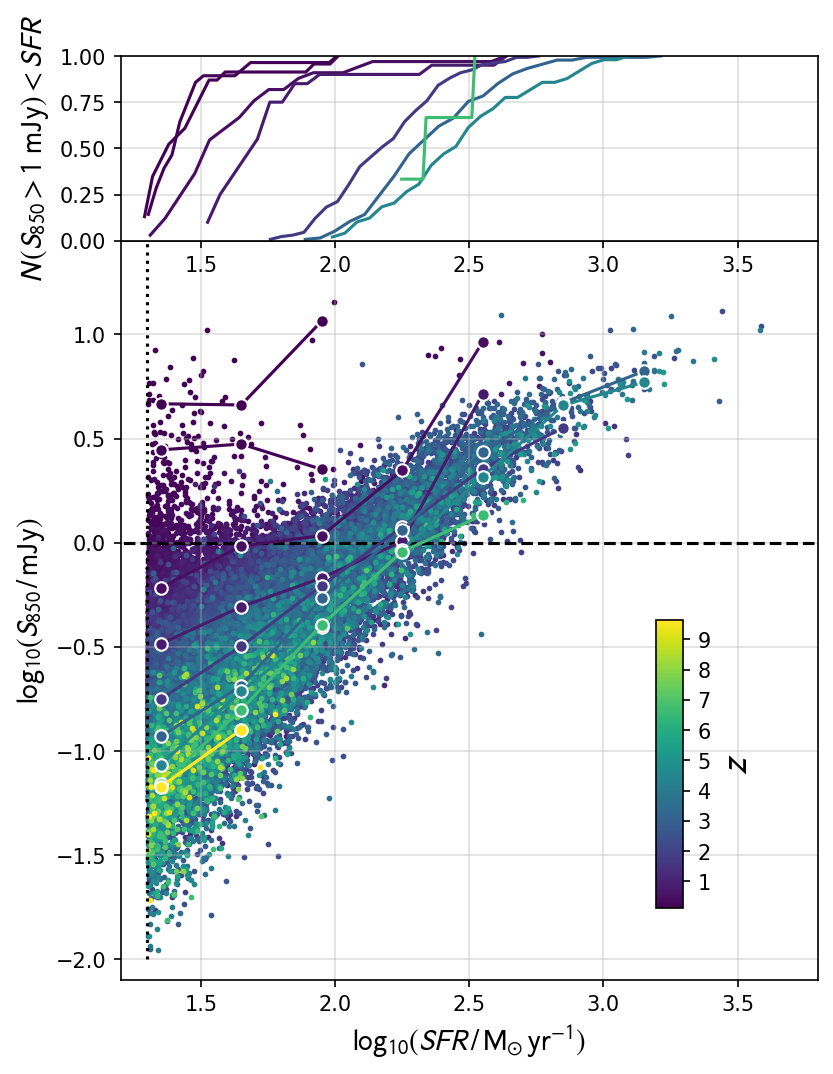}
    \caption{\textit{Bottom panel:} Instantaneous star formation rate against \eightfifty\ flux density for all selected galaxies in each snapshot, coloured by redshift.
    The dashed horizontal line marks $S_{850} = 1 \mathrm{mJy}$.
    Binned medians are shown by the large points, at the following redshifts: $z_{\mathrm{bin}} = [0.12,0.2,0.5,0.8,1.7,3.2,4.5,6.7]$.
    There is a correlation between \eightfifty\ flux density and SFR at all redshifts, but this is strongest at cosmic noon ($z \sim 2$).
    \textit{Top panel:} cumulative fraction of galaxies with $S_{850} > 1 \mathrm{mJy}$ greater than the given SFR, at $z \in z_{\mathrm{bin}}$.
    At lower redshifts, a small number of low-SFR galaxies have high (> 1 mJy) fluxes, but at $z > 0.5$ the snapshots are complete above this flux density limit.
    }
    \label{fig:sfr_s850_completeness}
\end{figure}

\subsubsection{Aperture modelling}

Sub-mm observations of the high redshift universe can either be performed using single dish observations with instruments such as SCUBA-2 on the JCMT providing large area coverage, or through interferometric studies with facilities such as ALMA for improved resolution and sensitivity.
When comparing to models it is important to take account of these different observational approaches, and to mimic the actual detection of sub-mm emission in the appropriate way.
Since we are most concerned with the global demographics of SMGs such as number counts, we mock the \textit{single-dish} approach in this work.
Specifically, we focus on counts measured by the SCUBA-2 camera on the JCMT at \eightfifty\ \citep{geach_scuba-2_2017}\footnote{We use the SCUBA-2 filter profiles provided at \url{https://www.eaobservatory.org/jcmt/instrumentation/continuum/scuba-2/filters/}} with an angular resolution of 14.8$''$ (FWHM).
This corresponds to a physical resolution of $\sim 120 \; \mathrm{pkpc}$ at $z \sim 2$ (see \app{size_evolution} for details).
Therefore, we adopt a fixed aperture diameter of $D = 120 \; \mathrm{pkpc}$ at all redshifts, within which we measure the emergent sub-mm emission.
This does not follow the true evolution of the SCUBA-2 beam size with redshift, but allows us to fairly compare the emission properties between galaxies at different redshifts.
Note that the aperture is typically much larger than individual galaxies, and often includes the contribution from satellites or near-neighbours; we will investigate the effects of beam confusion in \sec{blending}.
This aperture scale is shown for an example galaxy in \fig{render}.

%% file: selection.tex
\subsection{Galaxy Selection}
\label{sec:selection}

\begin{figure*}
\includegraphics[width=\textwidth]{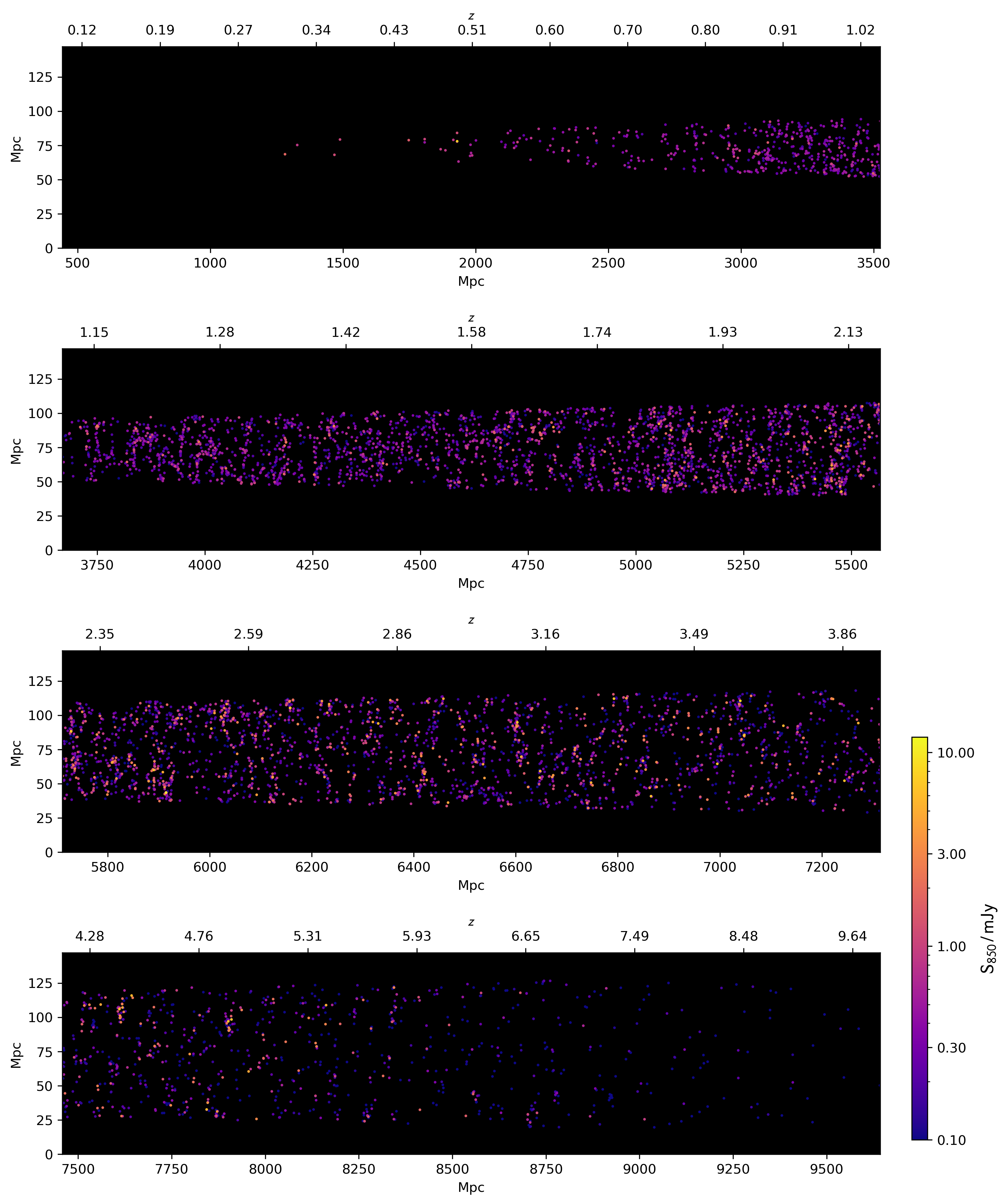}
    \caption{
    \simba\ reconstructed lightcone over 0.5 deg$^{2}$, between $0.1 < z < 10$.
    Each point shows a galaxy coloured by \eightfifty\ flux density.
    The radial distance on the $x$-axis is the comoving distance.
    The transverse distance $d$ on the $y$-axis is the comoving distance, and is $d = [42.6,67.0,80.4,91.2,103.72,116.3]$ at $z = [1,2,3,4,6,9]$, respectively.
    }
    \label{fig:lightcone}
\end{figure*}

We apply {\sc Powderday} to every other snapshot between $z = 0.1-10$, in order to allow for the construction of lightcones (see \sec{methods:lightcone} below).
From these snapshots, we select galaxies on which to run the RT via a conservative SFR cut.
It has been seen in other studies that there is a strong correlation between a galaxy's SFR and its \eightfifty\ flux \citep[\textit{e.g.}][]{hayward_submillimetre_2013}.
To avoid the computational expense of performing RT on tens of thousands of galaxies with undetectable sub-mm fluxes, we perform a cut by instantaneous star formation rate,
\begin{equation}
  \mathrm{SFR_{inst}} > 20 \; \mathrm{M_{\odot} \, yr^{-1}} \;\;,
\end{equation}
which roughly corresponds to $S_{850} = 0.25 \; \mathrm{mJy}$, well below the observational limit of our primary comparison dataset ($S_{850} \gtrsim 1 \;\mathrm{mJy}$).
This gives 1670 galaxies at $z = 2$ within our $100 \, h^{-1} \, \mathrm{Mpc}$ volume.
To avoid accounting for the same emission twice we ignore galaxies that lie within 60\,pkpc of another galaxy in the selection, and use an aperture centred on the most highly star forming object of the two.
At $z = 2$ approximately 5\% of the selection is accounted for within other apertures.

\fig{sfr_s850_completeness} shows the correlation between $\mathrm{SFR_{inst}}$ and $S_{850}$ for all galaxies in our selection in all snapshots.
There is a clear positive correlation except at the lowest redshifts.
We therefore conclude that our sample is complete down to $\la 1\; \mathrm{mJy}$, except for a few galaxies at low redshifts ($z\la 0.5$) with low SFRs that have significant $S_{850}$ emission owing to their proximity, but these galaxies contribute negligibly to the overall number counts (see \sec{int_counts}).

\subsection{Lightcone Construction}
\label{sec:methods:lightcone}

The \simba\ simulations output times were chosen in such a way that every other consecutive snapshot lines up in redshift space, \textit{i.e.} the comoving distance between every other snapshot is the same as the side length of the simulation box.
This makes creating lightcones relatively simple.
We first assume some sky area, $A = \ell^{2}$.
At each snapshot we then find the comoving distance covered by $\ell$.
Due to the small comoving volume of the fiducial \simba\ run the same structures can appear multiple times if a sufficiently large sky area is chosen.
To mitigate this effect, we randomly choose a line-of-sight alignment axis, and randomly translate the volume along the plane of the sky direction.
We use an area $A = 0.5 \; \mathrm{deg^{2}}$ comparable to single S2CLS fields \citep{geach_scuba-2_2017}.
Once the selection has been made for each snapshot, the lightcone is created by stitching each consecutive snapshot along the chosen $z$-direction.

\begin{figure}
\includegraphics[width=\columnwidth]{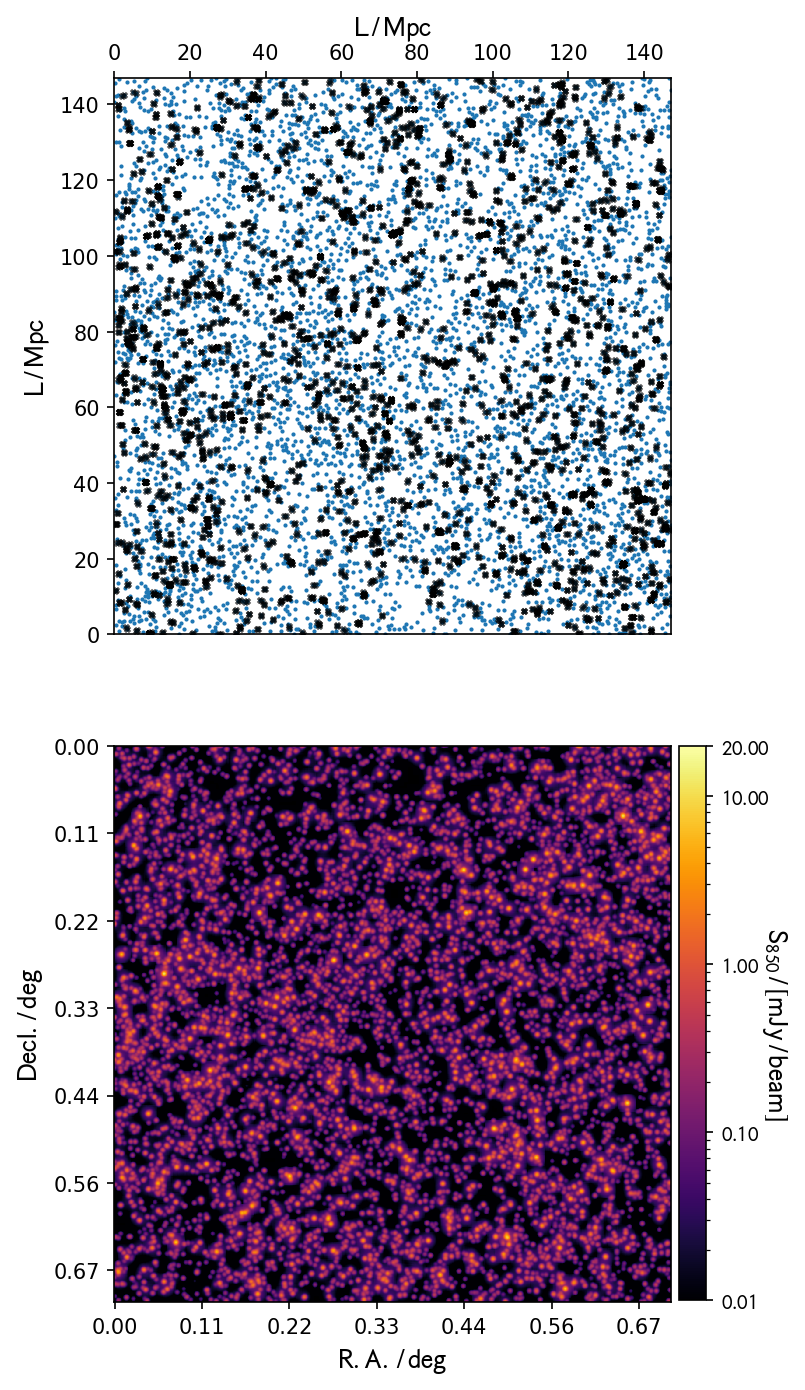}
    \caption{\textit{Top}: map of a single lightcone realisation using all selected objects ($\mathrm{SFR_{inst}} > 20 \; \mathrm{M_{\odot} \, yr^{-1}}$) on the sky plane (blue points).
    Sources blended along the line of sight are shown by black crosses.
    \textit{Bottom}: The \eightfifty\ map convolved with the SCUBA-2 beam \citep{dempsey_scuba-2_2013}, coloured by flux density.
    }
    \label{fig:map}
\end{figure}

\fig{lightcone} shows the distribution of galaxies in a single lightcone realisation.
The number density increases with redshift to cosmic noon ($z \sim 2$), and then decreases gradually toward $z = 10$, broadly as observed.
The total volume of the lightcone between $0 < z < 10$ is $4.6 \times 10^{7} \; \mathrm{Mpc^{3}}$, which is $\sim 14.5 \, \times$ larger than the simulation box size.
\fig{map} shows the projected map from this lightcone realisation.
The `observed' map is produced by convolving the projected \simba\ \eightfifty\ lightcone with the SCUBA-2 point spread function \citep{dempsey_scuba-2_2013}.
Note that it does not include instrumental noise, however this could be trivially added to mimic real SCUBA-2 observations if needed.
We explore the effect of source blending, both associated and unassociated, in \sec{blending}.
The effect of cosmic variance can also be investigated by taking multiple realisations of the lightcone; we investigate this in \sec{int_counts}.

%% file: results.tex
\section{Sub-millimetre Number Counts}
\label{sec:all_counts}

\input{counts}

\input{associated_blends}
\input{redshift_distribution}

\input{counts_models}

\section{Drivers of sub-millimetre emission in Simba}
\label{sec:drivers}

What is the explanation for the reasonably close match between the single-dish observational constraints on the integrated sub-mm number counts and those predicted by \simba, particularly at the bright end?
We investigate this by looking at the two primary physical sources for sub-mm emission: ongoing star formation generating UV emission, and a large dust reservoir to attenuate and re-radiate that emission.
We begin by examining the combination of these properties, and evaluating the strength of any correlations.

\input{alltogether}
\input{sfrf}
\input{dust}

%% file: counts.tex
\subsection{Integrated Number Counts}
\label{sec:int_counts}

\begin{figure*}
\includegraphics[width=\textwidth]{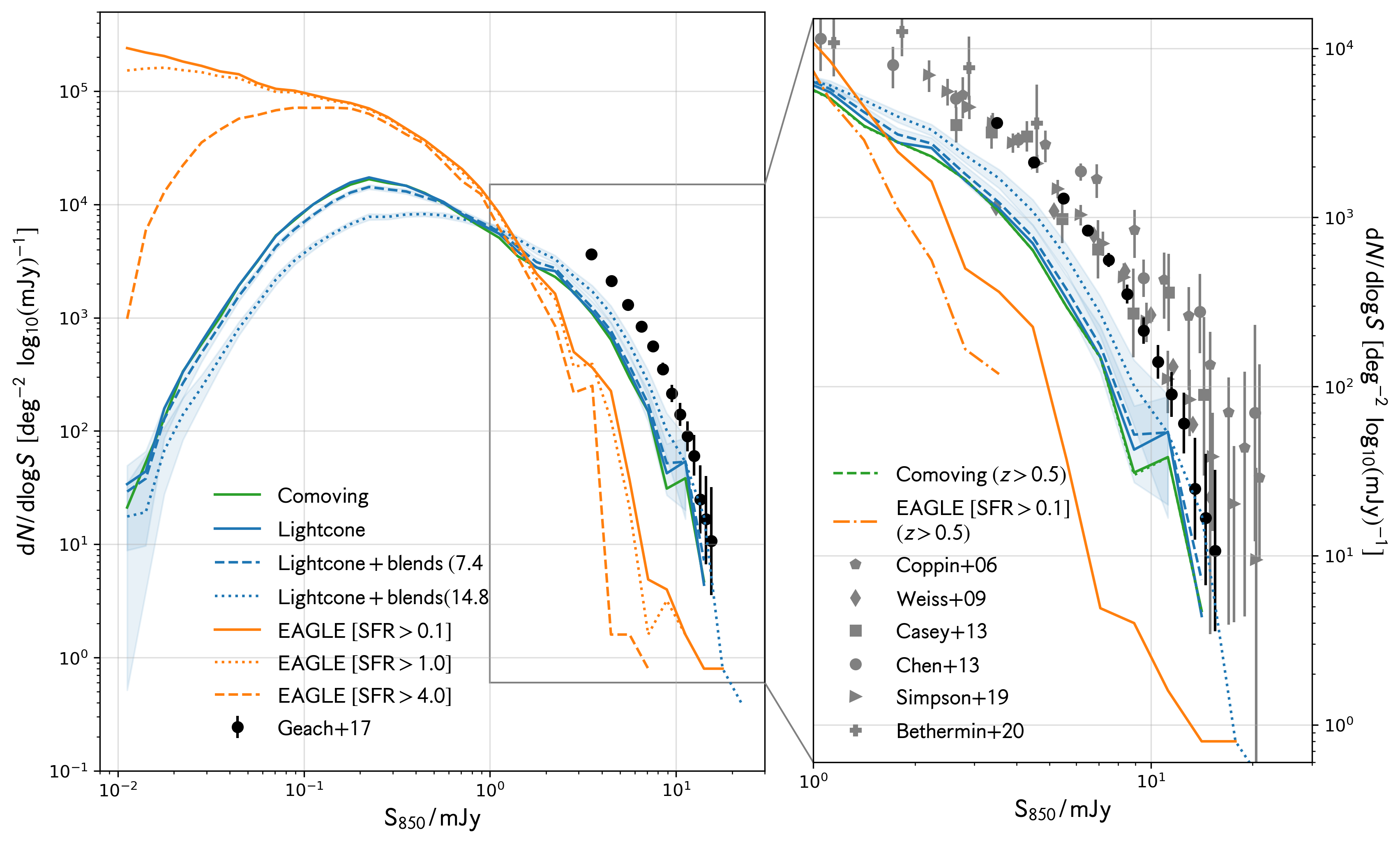}
    \caption{\eightfifty\ differential number counts in \simba.
    Results from the comoving method (solid green) and the lightcones (mean of 50 realisations, solid blue) are shown.
    Including the effects of blends leads to an increase in the normalisation at $> 1 \; \mathrm{mJy}$ (dashed and dotted blue, for 14.8" and 7.4" beam sizes, respectively).
    The shaded blue region shows the quadrature combination of the poisson errors and the inter-lightcone realisation scatter on the blended counts.
    We show observational constraints from S2CLS \protect\citep[black;][]{geach_scuba-2_2017} as well as a number of previous studies in the inset panel
    \protect\citep[grey;][]{coppin_scuba_2006,weis_large_2009,casey_characterization_2013,chen_alma_2015,simpson_east_2019,bethermin_alpine-alma_2020}.
    We also present results from the \eagle\ simulations \protect\citep[orange,][]{mcalpine_nature_2019}, generated using the comoving method, for different SFR cuts; these are converged for $\text{SFR} > 0.1 \; \mathrm{M_{\odot} \, yr^{-1}}$.
    The inset panel show the effect of excluding galaxies with $z \leqslant 0.5$ from the \simba\ (green dashed) and \eagle\ (orange dashed dotted) counts.
    }
    \label{fig:diff_counts}
\end{figure*}

We begin by comparing \simba\ SMG predictions to the observed integrated number counts. Recent SMG surveys tightly constrain the number counts for $S_{850} \gtrsim 3 \; \mathrm{mJy}$ \citep{coppin_scuba_2006,scott_combined_2006,weis_large_2009,austermann_aztec_2010,scott_source_2012,geach_scuba-2_2017,simpson_east_2019}, and this has traditionally been a major challenge for models to reproduce.  We examine this in two ways: using the individual snapshots assembled based on a weighting function which we call the ``comoving'' method, and using the lightcone method described in \sec{methods:lightcone}.

For the comoving method, we first define the volume-normalised number density at that redshift, \smash{$\mathrm{d}N(z) \,/\, \mathrm{d}S \, \mathrm{d}V \; [\mathrm{mJy^{-1} \, Mpc^{-3}}]$}.
We then scale this by the volume defined by the midpoint redshifts between the nearest neighbouring snapshots,
\begin{align*}
  z_{i,\mathrm{low}} &= (z_{i} - z_{i-1}) / 2 \\
  z_{i,\mathrm{upp}} &= (z_{i+1} - z_{i}) / 2 \;\;.
\end{align*}
These can be used to find the volume by integrating the differential comoving volume \citep[defined in][]{hogg_distance_2000} between these limits,
\begin{align*}
  V_{\mathrm{C}}^{z} &= \int_{z_{i,\mathrm{upp}}}^{z_{i,\mathrm{low}}} \frac{\mathrm{d}V_{\mathrm{C}}}{\mathrm{d}z} \,\mathrm{d}z
\end{align*}
The total number counts are then given by summing the contribution from each snapshot,
\begin{equation}
  \frac{dN}{dS} = \sum_{z=z_{0}}^{z_{max}} \, \frac{\mathrm{d}N(z)}{\mathrm{d}S \, \mathrm{d}V} \, V_{\mathrm{C}}^{z} \;\;.
\end{equation}

The advantage of using the comoving approach is that the whole volume is used, which maximises the dynamic range of the number counts by including the most extreme galaxies at all redshifts.
The lightcone approach, however, is more useful to account for observational effects such as blending along the line of sight.
Blending of associated (near-field) and unassociated (far-field) sources can increase the apparent fluxes of individual detections in single dish maps \citep[see][for a discussion]{hodge_high-redshift_2020}; we examine this in more detail in \sec{blending}.

We compare our results primarily to the latest constraints from the S2CLS \eightfifty\ counts \citep{geach_scuba-2_2017}.
This large survey covered 5 deg$^{2}$ over the UKIDSS-UDS, COSMOS, Akari-NEP, Extended Groth Strip, Lockman Hole North, SSA22 and GOODS-North fields to a depth of $\sim$1\,mJy.

\fig{diff_counts} shows the differential number counts of \eightfifty\ sources using our three approaches:  comoving (solid green), lightcone (solid blue), and lightcone including blends for 7.4" (dashed blue) and 14.8" (dotted blue) apertures.
In the left panel, we compare to \cite{geach_scuba-2_2017} observations, while in the blow-up plot on the right which focuses on the observationally probed regime, we additionally compare to a number of other single-dish surveys \citep{coppin_scuba_2006,weis_large_2009,casey_characterization_2013,chen_resolving_2013,simpson_east_2019} as well as the interferometric constraints from \cite{bethermin_alpine-alma_2020}.
Notable among these are the results from \cite{chen_resolving_2013}, which utilise cluster lensing fields to extend to lower flux densities than accessed in \cite{geach_scuba-2_2017}.
The turnover at very low fluxes arises from incompleteness below $1 \; \mathrm{mJy}$ (see \sec{selection}) owing to our $\mathrm{SFR} > \, 20 \; \mathrm{M_{\odot}  \; yr^{-1}}$ sample selection; we are not concerned with this regime at present, since it lies below the depth of current single-dish SMG surveys, though we note that such a turnover has been constrained in the semi-empirical models of \cite{popping_alma_2020}.

The blue shaded region shows the uncertainty in the \simba\ prediction, calculated from two sources.
The first is from Poisson errors on the raw counts.
The second is from the spread in counts over 50 different lightcone realisations, encoding the effect of cosmic variance on the counts.
The shaded region shows the quadrature combination of these from the blended lightcone counts (described in detail in the next section).
We find that field-to-field variance is approximately equal to Poisson variance at all flux densities, similar to that found for \textsc{Galform} in \cite{cowley_simulated_2015} (for $ > 5 \; \mathrm{mJy}$).  \fig{diff_counts} shows that the lightcone and comoving approaches (green and blue lines) are in excellent agreement with each other over the flux density range probed ($\sim 0.01 - 15 \; \mathrm{mJy}$).
This is unsurprising since they come from the same underlying simulation data, but it is a useful check.

\simba\ matches the shape of the latest observed \eightfifty\ number counts from \citet{geach_scuba-2_2017}, and the normalisation is within 0.25 dex at $> 3 \; \mathrm{mJy}$.
The agreement at the bright end ($>12 \; \mathrm{mJy}$), where cosmological hydrodynamic simulations have traditionally struggled, is particularly good.
Table~1 details the predicted differential and cumulative number counts from \simba; we note that the cumulative number counts provide a less robust comparison to data since we do not model the impact of lensing which strongly increases the number counts at the most extreme luminosities.
\simba's level of agreement is unprecedented from cosmological hydrodynamic simulations \citep[for a review, see][]{casey_dusty_2014}.
For comparison, we also show the results from the \eagle\ simulation~\citep{mcalpine_nature_2019}, which illustrates that \eagle\ does not come as close to matching the \eightfifty\ number counts~\citep[see also][]{wang_multi-wavelength_2019,cowley_evolution_2019}.
In particular, \eagle\ does not produce any bright ($>4 \; \mathrm{mJy}$) sources at $z > 0.5$.
We discuss the comparison to \eagle\ and other models in more detail in \S\ref{sec:modelcomp}.

The unprecedented close agreement between \simba\ and observations of SMG number counts is the primary result in this paper.
We note that \simba\ was not tuned specifically to match SMGs, or the SFRs in massive high-$z$ galaxies; this model was tuned primarily to match the evolution of the overall stellar mass function and the stellar mass--black hole mass relation~\citep{dave_simba:_2019}.
Our result thus demonstrates that a hierarchical structure formation model, analysed using dust RT and accounting for observational effects, is capable of broadly matching SMG number counts without the need for any ad hoc physics modifications such as IMF variations.

\begin{table}
  \caption{\simba\ differential number counts $\mathrm{d}\,N \,/\, \mathrm{d \,log}\,S$ and cumulative number counts $N(>S)$, from the comoving and lightcone methods (including blends within a 14.8" aperture).}
	\centering
	\label{tab:counts}
	\begin{tabular}{ccccc}
		\hline
    & & Comoving & Lightcone & Comoving \\
    & &  &  + blends (14.8") & \\
    \hline
		$S_{\mathrm{850}}$ & $\mathrm{log_{10}}(S_{\mathrm{850}})$ & \multicolumn{2}{c}{$\mathrm{d}\,N \,/\, \mathrm{d \,log}\,S$} & $N(>S)$ \\
    $\mathrm{mJy}$ & $\mathrm{log_{10}(mJy)}$ & \multicolumn{2}{c}{$(\mathrm{deg^{-2}}\;\; \mathrm{log_{10}(mJy)^{-1}})$} & $\mathrm{deg^{-2}}$ \\
		\hline
    1.12  & 0.05 & 5070.93 & 6028.80 & 1502.45 \\
    1.41  & 0.15 & 3483.30 & 4935.20 & 1074.74 \\
    1.78  & 0.25 & 2791.47 & 3951.60 & 761.00 \\
    2.24  & 0.35 & 2293.71 & 3294.80 & 506.74 \\
    2.82  & 0.45 & 1678.16 & 2331.60 & 308.15 \\
    3.55  & 0.55 & 1083.91 & 1704.40 & 170.04 \\
    4.47  & 0.65 & 639.59 & 1087.20 & 83.87 \\
    5.62  & 0.75 & 295.89 & 581.20 & 37.09 \\
    7.08  & 0.85 & 148.95 & 270.00 & 14.85 \\
    8.91  & 0.95 & 31.04 & 100.80 & 5.85 \\
    11.22 & 1.05 & 38.29 & 53.20 & 2.39 \\
    14.13 & 1.15 & 4.72 & 17.60 & 0.24 \\
    17.78 & 1.25 & 0.0 & 0.80 & 0.0 \\
    22.39 & 1.35 & 0.0 & 0.40 & 0.0 \\
		\hline
	\end{tabular}
\end{table}

%% file: associated_blends.tex
\subsection{Unassociated and Associated Blends}
\label{sec:blending}

Owing to the relatively large beam of single-dish instruments, it has been suggested that blending may play an important role in setting the SMG number count distribution, particularly at the bright end \citep[e.g.][]{hayward_spatially_2013,hayward_observational_2018,cowley_simulated_2015,hodge_high-redshift_2020}.
We investigate the effect of two types of blends, physically \textit{associated} blends of near-field objects (within the same large scale structure), and \textit{unassociated} blends of far-field objects that align along the line-of-sight.

The lightcone method can be used to directly evaluate the impact of unassociated blending.
To do so, we combine all sources with an on-sky separation less than $R$ arcseconds.
We simply sum the contributions within this aperture, rather than a more sophisticated method using a matched-filtered PSF \citep[as performed in][]{cowley_simulated_2015}.
We test two aperture sizes, $R = [7.4,14.8]''$, equal to the SCUBA-2 beam HWHM and FWHM, which bound the true contribution.

We find that, for all sources in a given lightcone ($8275^{+124}_{-135}$), 35\% (11\%) contribute to the flux of another source for the $14.8''$ ($7.4''$) aperture, where our uncertainties are the $16^{\mathrm{th}} - 84^{\mathrm{th}}$ percentile range on the 50 lightcone realisations.
This leaves the number of sources post-blending at $5589_{-82}^{+79}$ ($7295_{-120}^{109}$), and the fraction of those remaining sources that are blends of more than one galaxy is 52\% (11\%).
This is for \textit{all} sources in our lightcone; restricting to $S_{850} > 1 \; \mathrm{mJy}$ gives a blended fraction of 69\% (28\%), and for $S_{850} > 3 \; \mathrm{mJy}$ this rises to 74\% (30\%).
This is somewhat higher than the fraction measured in \cite{hayward_spatially_2013} ($\gtrsim 50\%$ for $S_{850} > 1 \; \mathrm{mJy}$).

\begin{figure}
\includegraphics[width=\columnwidth]{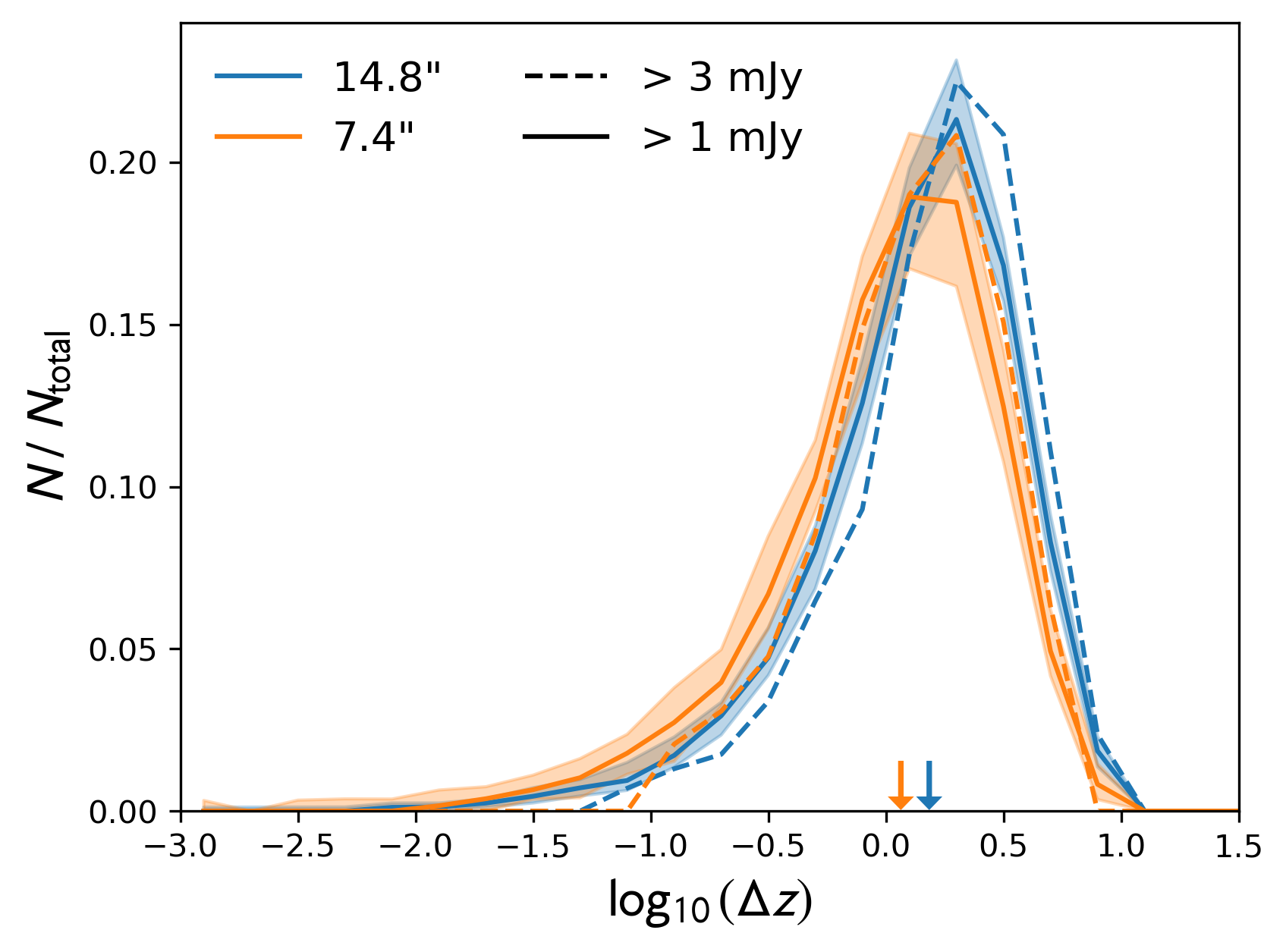}
    \caption{Normalised distribution of $\Delta \, z$ for $R = 14.8''$ and $7''$ aperture sizes, in blue and orange respectively.
    Each aperture size is shown for a different lower flux density limit, $\geqslant 1 \; \mathrm{mJy}$ and $\geqslant 3 \; \mathrm{mJy}$ (solid and dashed, respectively).
    The arrows on the $x$-axis show the median of the distribution for each aperture size for the $\geqslant 1 \; \mathrm{mJy}$ selection.
    }
    \label{fig:deltaz}
\end{figure}

We can also study the redshift separation of our blended sources, measured as the sum of the redshift separations of each source in quadrature, with respect to the primary source,
\begin{equation}
  \Delta \, z = \left( \sum^{N}_{\,i \,>\, 1} (z_{i} - z_{1})^{2} \right)^{1/2} \;\;,
\end{equation}
where $N$ is the total number of components contributing to the blended source, and $z_{i}$ is the redshift of component $i$.
\fig{deltaz} shows the normalised distribution of $\Delta z$.
There is a single strong peak in the distribution around unity, tailing off at lower and higher separations.
For the $14.8''$ ($7.4''$) aperture, the median $\Delta z = 1.53$ ($1.16$) for $\geqslant 1 \; \mathrm{mJy}$ sources.
This increases for higher flux densities, $\Delta z = 1.86$ ($1.34$) for  $\geqslant 3 \; \mathrm{mJy}$ sources.
The general shape of the $\Delta z$ distribution is in good agreement with that found in previous studies (minus a low separation peak, discussed below), as well as the trends with aperture size and lower flux density limit, however there are quantitative differences.
\cite{hayward_spatially_2013} measure $\Delta z = 0.99$ for $\geqslant 3 \; \mathrm{mJy}$ sources in a $15''$ aperture, almost a factor of 2 smaller than that seen in \simba, however they do see an increase for their brightest sources ($\Delta z = 1.46$ for $\geqslant 7 \; \mathrm{mJy}$ sources).
Similarly, \cite{cowley_simulated_2015} find $\Delta z \sim 1$ for $\geqslant 4 \; \mathrm{mJy}$ sources in a $15''$ aperture.
The higher median $\Delta z$ seen in \simba\ may be due to the higher redshift distribution of sources (see \sec{redshift_distribution}).

\fig{diff_counts} depicts the impact of unassociated blends in the \simba\ lightcone on the integrated number counts, via comparing the blue solid line without blending and the blue dashed and dotted lines for blending with 7.4" and 14.8" apertures, respectively.
In general, blending tends to increase the normalisation above $1 \; \mathrm{mJy}$ by a small factor, compensated by a decrease in the normalisation at the faint end.
The larger aperture has a more significant effect on the normalisation, increasing it by $\sim 0.15 \; \mathrm{dex}$ at $7 \; \mathrm{mJy}$, which leads to excellent agreement with the \cite{geach_scuba-2_2017} results in this bright flux density regime.
The dynamic range is also significantly extended, with the brightest source for the 14.8" aperture lightcone having $S_{850} = 22.4 \; \mathrm{mJy}$, compared to $15.2 \; \mathrm{mJy}$ for the 7.4" aperture.
In summary, unassociated blending provides a small but significant contribution to the bright end of the number counts.

\begin{figure}
\includegraphics[width=\columnwidth]{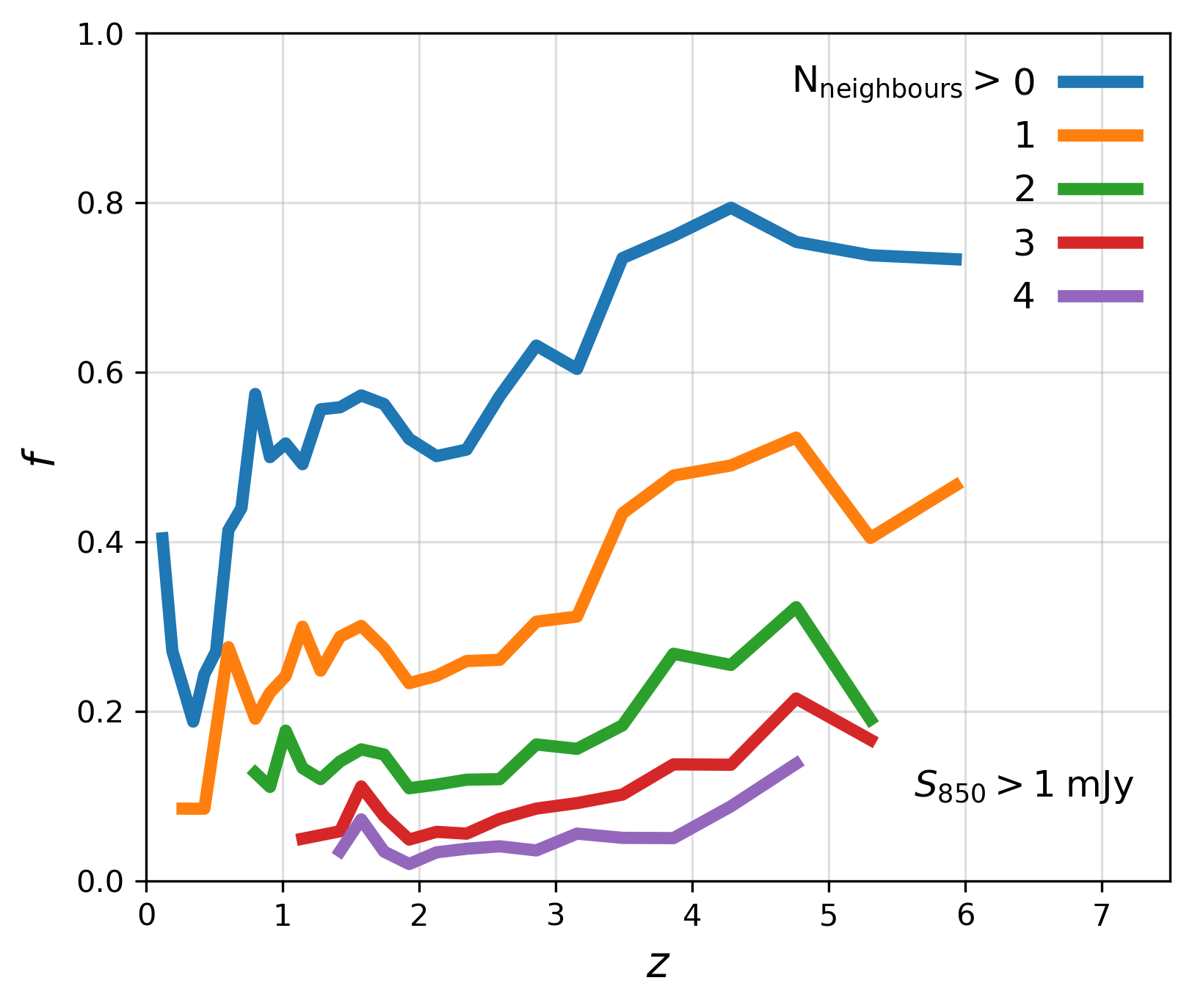}
    \caption{Fraction of sources with different numbers of neighbours for the $S_{850} > 1 \; \mathrm{mJy}$ population.
    We only show redshifts where there are at least 5 sources.
    }
    \label{fig:neighbours}
\end{figure}

Of course, there may be significant near-field blending from multiple galaxies interacting in the same halo or clumpy sub-structure within a single galaxy, which could boost the flux of `individual' sources \citep{bussmann_hermes_2015,simpson_scuba-2_2015,stach_alma_2018}.
Evidence of this has been seen with ALMA \citep[see][]{hodge_high-redshift_2020}.
We cannot directly investigate this since we compute the SMG flux within the entire SCUBA-2 beam; this also explains why we do not see a low redshift peak in the $\Delta z$ distribution in \fig{deltaz}.
However, we can examine the environment of SMGs in order to determine whether the brightest objects are likely to have neighbours that can contribute significant sub-mm flux.\footnote{Strictly, galaxies with larger separations may still be `associated', since they may reside within the same large scale structure, but for our purposes we class all galaxies within 120 kpc as `associated'.}

\fig{neighbours} shows the fraction of our selected sources above some flux density limit with neighbours, where a `neighbour' is defined as any galaxy with a stellar mass $M_{\star} > 5.8 \times 10^{8}M_\odot$ that lies with 60\,pkpc of the source.
Greater than 50\% of sources with $S_{850} > 1 \; \mathrm{mJy}$ have at least one neighbour at all redshifts, dropping at $z < 1$.
For the brighter, $S_{850} > 3.6 \; \mathrm{mJy}$ population the fraction is even higher, at least 60\% at all redshifts where there are sufficient sources.
Evidence of greater multiplicity of high flux density sources has been seen in observations \citep{bussmann_hermes_2015}.

While we do not compute RT fluxes in smaller galaxies owing to these systems being too poorly resolved for RT, we can roughly estimate the impact of blending by examining the fraction of the SFR in a halo contributed by the central galaxy.
SFR does not translate directly into $S_{850}$, but there is some correlation (see \sec{alltogether}), and since smaller galaxies are likely to be lower metallicity and thus likely contain less dust, one expects that their contribution to the blended $S_{850}$ flux will be overestimated by just considering their contribution to the SFR.
Thus we can place an upper limit on the impact of associated blends.

For galaxies with $S_{850} > 1\; \mathrm{mJy}$ at $z=2$, we find that the central galaxy contributes 95\% of the total SFR, on average.
At higher redshifts, and for higher $S_{850}$ cuts, the corresponding numbers are even smaller.
This suggests that associated blends will only contribute at most $\sim5$\% to the $S_{850}$ flux in SMGs.

In short, whilst SMGs are rare, unassociated blends are still common, and have a small but significant effect on the number counts.
Associated blends cannot be directly estimated here, but using the SFR as a proxy shows that the central galaxy in the beam contributes more than 95\% of the $S_{850}$ flux on average.
We will perform a more detailed comparison with high-resolution interferometric observations in future work, utilising high-resolution zoom simulations of individual \simba\ galaxies.

%% file: redshift_distribution.tex
\subsection{Redshift Distribution of SMGs}
\label{sec:redshift_distribution}

An orthogonal constraint to number counts on galaxy formation models is the redshift distribution of SMGs.  This tests whether the models' SMGs are appearing at the right cosmic epochs.
We investigate this by examining in \simba\ the redshift distribution of SMGs above a flux limit chosen to match current observational constraints.

\begin{figure}
\includegraphics[width=\columnwidth]{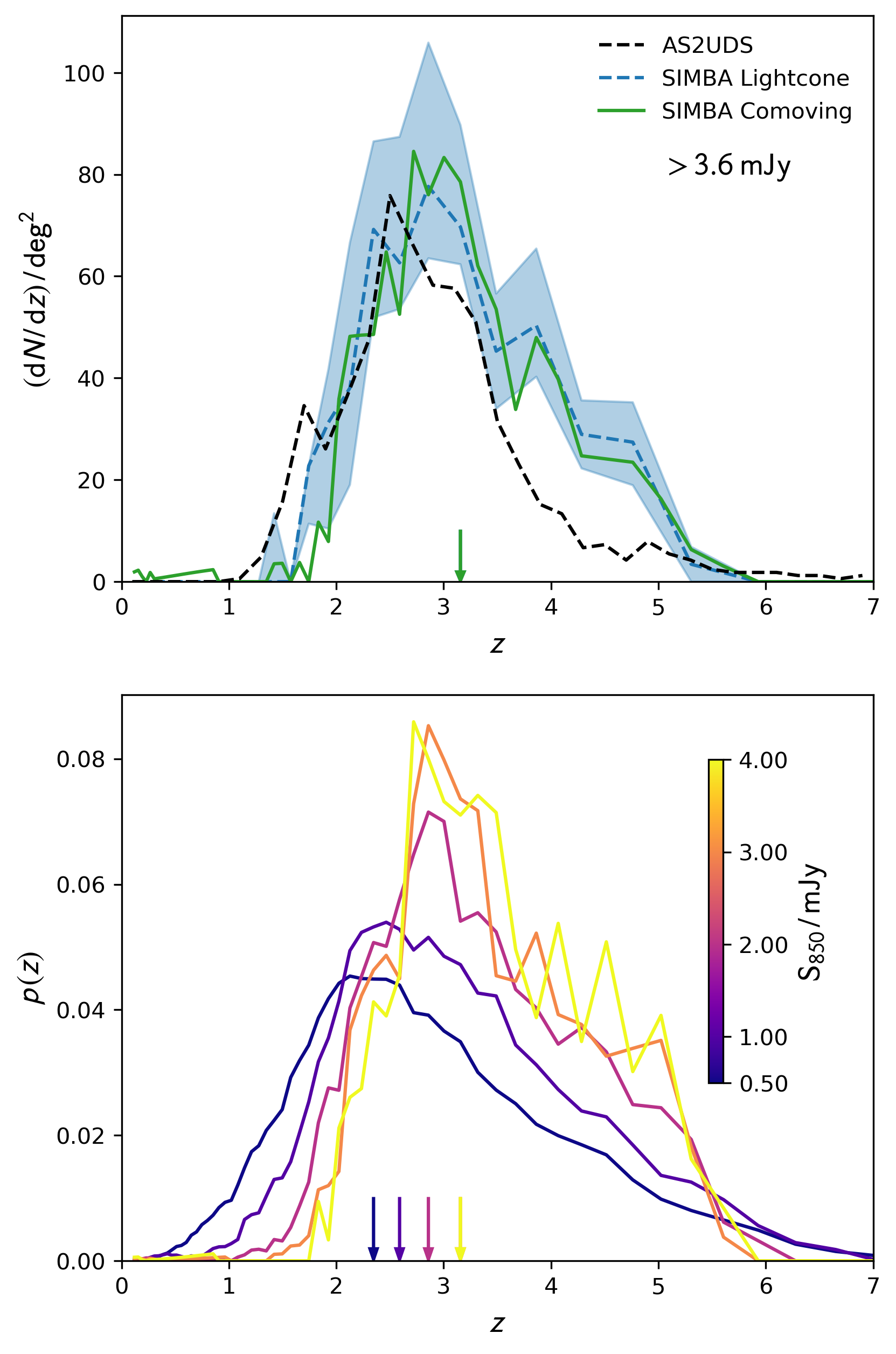}
    \caption{\textit{Top:} differential number count evolution with redshift per square degree.
    Comoving counts are shown in green, and the median lightcone counts as the dahed blue line, with the $16^{\mathrm{th}}-84^{\mathrm{th}}$ percentile range shown by the shaded region.
    We shows observational constraints from AS2UDS \citep{dudzeviciute_alma_2020}, corrected for incompleteness \citep{geach_scuba-2_2017}, by the black dashed line.
    The medians for both approaches is shown by the arrow on the $x$-axis.
    \textit{Bottom:} the normalised redshift distribution from the lightcone method for different flux density cuts.
    Medians are again shown by arrows.
    }
    \label{fig:number_density_evolution}
\end{figure}

The top panel of \fig{number_density_evolution} shows the differential number counts per square degree for the $S_{850} > 3.6 \; \mathrm{mJy}$ SMG population.
We show the distribution for the full comoving snapshots, as well as the median and $16^{\mathrm{th}}$--$84^{\mathrm{th}}$ spread for the 50 lightcone realisations.
We compare to observations from
AS2UDS \citep{dudzeviciute_alma_2020}, an ALMA follow up survey of S2CLS sources from the 0.96 deg$^2$ UKIDSS Ultra-Deep Survey field \citep{stach_alma_2019}.
We correct for incompleteness using a conservative upper estimate from \cite{geach_scuba-2_2017}.

In \simba, the median redshift for these SMGs, with $16-84\%$ range, is $z = 3.16_{-0.69}^{+1.12}$, for both the lightcone and comoving methods.
The 1$\sigma$ spread from different lightcone realisations is shown to illustrate the impact of field-to-field variance on the distribution; the comoving method predictions lie generally within the variance of the lightcone method.

Overall, \simba's redshift distribution peaks at $z\sim 3$, which is somewhat higher than observed.
\cite{dudzeviciute_alma_2020} measure a median redshift of $z = 2.79_{-0.07}^{+0.07}$, lower than that obtained from both our lightcone and comoving methods.
There is a clear excess of sources in \simba\ at $3.5 \ga z \ga 5$.
A number of other studies measure similar median redshifts for similar flux density cuts, particularly where estimates are made for the redshifts of optical/IR undetected sources \citep{hodge_high-redshift_2020}.
This suggests that \simba\ overproduces SMGs at higher redshifts.

Interestingly, the existence of SMGs at high redshifts has sometimes been presented as a challenge to hierarchical galaxy formation models, since high-$z$ SMGs are forming stars so rapidly at early times.  \simba\ not only meets this challenge, but notably overshoots it. As we will see later, \simba\ routinely predicts galaxies with $\mathrm{SFR}\, \ga 1000 \; \mathrm{M_{\odot} \,yr^{-1}}$ as high as $z\gtrsim 4$, with high dust contents.

\begin{figure}
\includegraphics[width=\columnwidth]{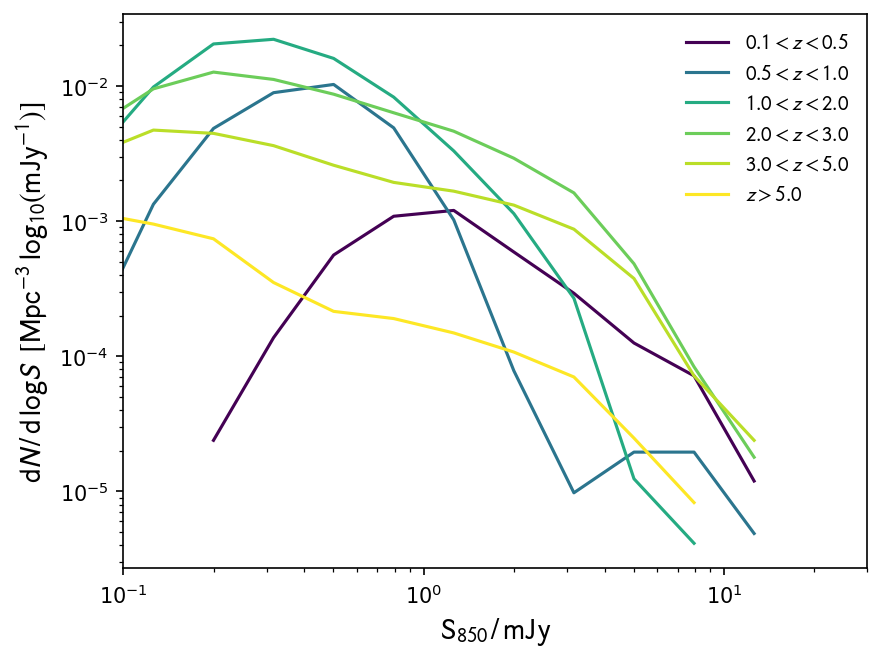}
    \caption{Comoving differential number counts in bins of redshift.
    }
    \label{fig:redshift_diff_counts}
\end{figure}

There is some observational evidence for positive evolution in the median redshift with increasing flux density cut \citep{chapman_redshift_2005,wardlow_laboca_2011,simpson_alma_2014,da_cunha_alma_2015,simpson_scuba-2_2017}, a form of SMG downsizing.
To test whether we see similar flux density-dependent evolution, the bottom panel of \fig{redshift_diff_counts} shows the normalised redshift distribution from the lightcone method for different flux density limits, ranging from $S_{850} > 4 \; \mathrm{mJy}$ to ranging from $S_{850} > 0.5 \; \mathrm{mJy}$ (yellow to blue).

In general, \simba's redshift distribution becomes shallower and broader when including lower flux density sources.  The median redshift decreases (from $z = 3.15$ for $S_{850} > 4 \; \mathrm{mJy}$, to $z = 2.34$ for $S_{850} > 0.5 \; \mathrm{mJy}$).
The percentage of galaxies at $z > 3$ for $S_{850} > [0.5,1,2,3,4,5] \; \mathrm{mJy}$ is [12,20,30,39,44,62]\%, respectively.
Even at $z > 6$, when the universe was just a billion years old, \simba\ predicts 8 sources with $S_{850} > 1 \; \mathrm{mJy}$ within the whole comoving volume, which is broadly in agreement with AS2UDS.

The variation in the median redshift with flux density cut qualitatively agrees with that seen in observations, and with empirical models such as that of \cite{bethermin_evolution_2015,casey_brightest_2018}.
However, such variation is not seen in the \cite{lagos_far-ultraviolet_2019} \textsc{SHARK} semi-analytic model \citep[see][for a review]{hodge_high-redshift_2020}.

A complementary view of the redshift distribution of SMGs is provided by the comoving differential number counts in different redshift intervals.
This is shown in \fig{redshift_diff_counts}, from $z>5$ down to $z=0.1$.
In order to boost statistics, we combine all snapshots within the listed redshift interval, and construct a volume-normalised number count distribution from this.

As expected from the integrated redshift distribution in Figure~\ref{fig:number_density_evolution}, the differential number counts show a rapid rise at early epoch, and then drop past $z\sim 2$.  \fig{redshift_diff_counts} additionally shows that the shape of the number count distribution changes significantly.
At $z > 5$ the luminosity function is power law-like, with no faint end turnover above 0.1 mJy.
However, at lower redshifts, the distribution appears more Schechter-like, with a more prominent knee.
The faint-end turnover owing to our selection limit also becomes evident; we remind the reader that these differential counts are only expected to be complete above $\sim 1 \; \mathrm{mJy}$.
The redshift variation in the shape of the number count distribution represents a prediction from \simba\ that can be tested with future observations.

Overall, \simba\ broadly reproduces the observed redshift distribution of SMGs, albeit with a significant excess at $z\sim 4-5$.
Moreover, \simba\ also produces SMG downsizing in qualitative accord with observations, with fainter SMGs peaking in number density at a lower redshift.
\simba\ produces detectable ($\sim 1 \; \mathrm{mJy}$) SMGs as early as $\sim 6$, and predicts that the shape of the number count distribution evolves with redshift.

%% file: counts_models.tex
\subsection{Model Comparisons}
\label{sec:modelcomp}

\begin{figure}
\includegraphics[width=\columnwidth]{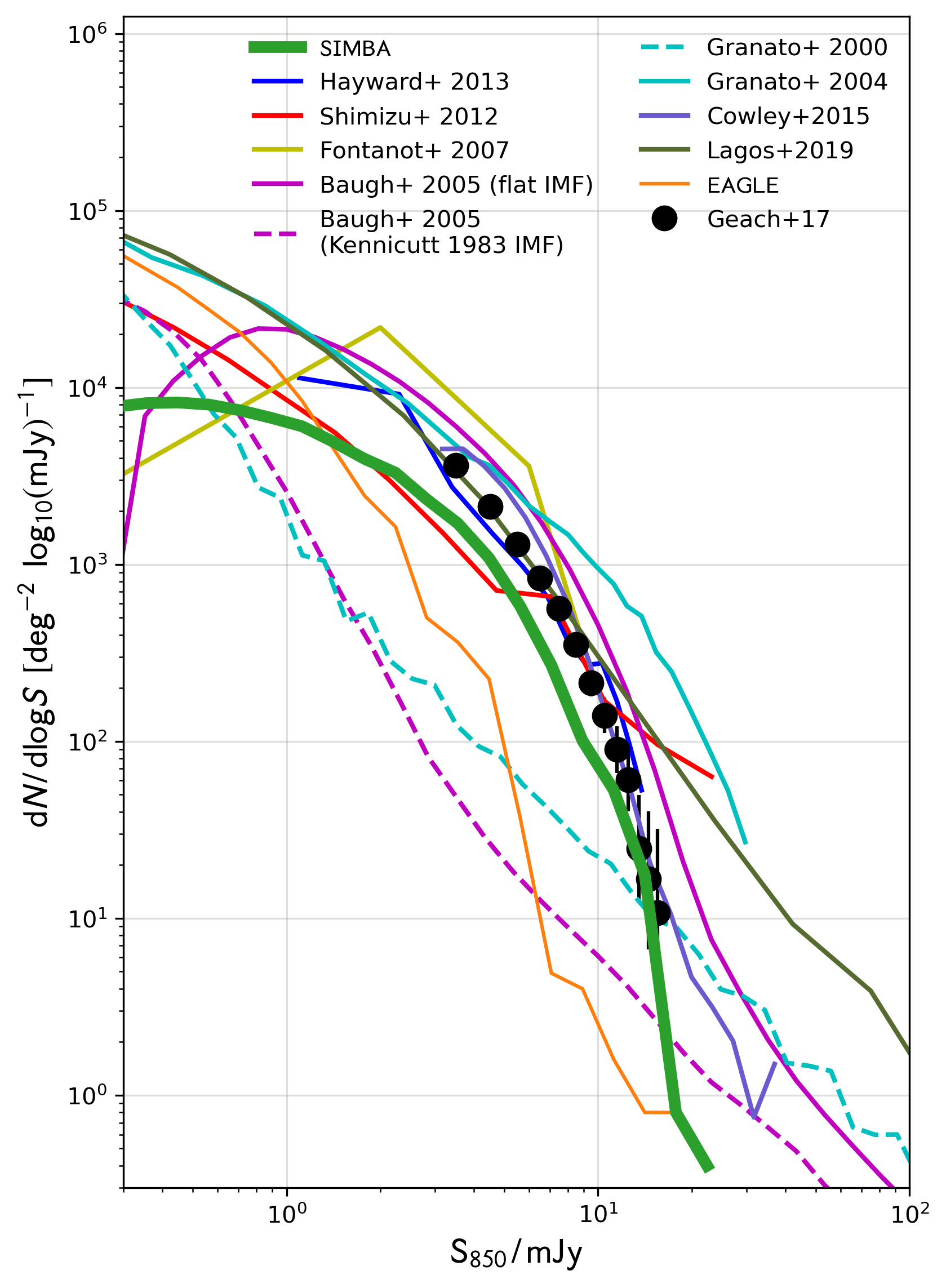}
    \caption{Differential number counts comparison with other models in the literature from \protect\cite{casey_dusty_2014}.
    We also plot the SHARK semi-analytic model results \protect\citep{lagos_far-ultraviolet_2019}, and the updated GALFORM results from \protect\cite{lacey_unified_2016}, including the effect of the SCUBA-2 beam \protect\citep{cowley_simulated_2015} (labelled Cowley+15).
    The \simba\ counts are represented by the comoving method (green line).
    Observational SCUBA-2 CLS counts \protect\citep[grey,][]{geach_scuba-2_2017} are shown in grey.
    The \eagle\ simulation \protect\citep[orange,][]{mcalpine_nature_2019} counts are identical to those in \fig{diff_counts}.
    }
    \label{fig:diff_counts_models}
\end{figure}

To contextualise our results within the current landscape of hierarchical models for SMGs, we now compare \simba's \eightfifty\ counts with various other semi-analytic and hydrodynamic model predictions from the literature over the past twenty years.
While hierarchically-based models have generally not matched the number counts ``out of the box'', they have over the years developed various modifications that have resulted in better agreement.
It is thus interesting to highlight such models, particularly when in \sec{drivers} we discuss the physical reasons why \simba\ appears to be broadly successful at matching the \eightfifty\ number counts and redshift distribution without ad hoc modifications.

\eagle\ is a recent cosmological hydrodynamic simulation showing good agreement with a number of key galaxy distribution functions \citep{schaye_eagle_2015,crain_eagle_2015}.
The Ref-100 fiducial run, with box volume $(100\;\mathrm{Mpc})^{3}$, contains 1504$^3$ dark matter particles and 1504$^3$ gas elements.
UV to sub-mm photometry for all galaxies in 20 snapshots covering the redshift range $0 \geqslant z \geqslant 20$, have been produced using version 8 of the SKIRT dust-radiative transfer code \citep{camps_data_2018}.\footnote{available at \href{http://icc.dur.ac.uk/Eagle/database.php}{http://icc.dur.ac.uk/Eagle/database.php}}
These show good agreement with low redshift optical colours \citep{trayford_optical_2017} and FIR dust-scaling relations \citep{camps_far-infrared_2016}.
\cite{mcalpine_nature_2019} also investigated the sub-mm source population, finding reasonable agreement with the observed redshift distribution as measured by \citet{simpson_alma_2014}.

We have calculated the \eagle\ \eightfifty\ luminosity function as follows.
Using the publicly available \eightfifty\ fluxes for each galaxy, we sum the fluxes of galaxies that lie within 60\,pkpc of each other to mimic our $D = 120\,\mathrm{pkpc}$ aperture.
We then combine all snapshots between $0.1 \geqslant z \geqslant 20$ using the comoving technique, described above, to give the number density per unit solid angle.
To be conservative, we use a lower SFR limit than that used for \simba{} to allow us to pick up objects with lower SFR within the $D = 120\,\mathrm{pkpc}$ aperture of another galaxy that may contribute to its total flux.
To test the convergence with SFR limit we show three different SFR limits: $\mathrm{SFR} > [0.1, 1.0, 4.0] \; \mathrm{M_{\odot} yr^{-1}}$.

\fig{diff_counts} shows the \eagle\ predictions as the orange line for each of these selections.  The normalisation is significantly lower than in \simba\ (and even lower compared to the observational constraints), by around 0.5 dex at $3 \, \mathrm{mJy}$ and up to 1 dex at $10\,\mathrm{mJy}$.
There are also no bright sources ($>4 \; \mathrm{mJy}$) in \eagle\ at $z > 0.5$.
Our number counts derived for \eagle\ are in agreement with those presented by \cite{wang_multi-wavelength_2019,cowley_evolution_2019}.

The counts are reasonably converged for $\mathrm{SFR} > 1 \; \mathrm{M_{\odot} \, yr^{-1}}$, but demonstrate that there is a significant contribution at observable SMG fluxes from $1 < \mathrm{SFR} < 4 \; \mathrm{M_{\odot} \, yr^{-1}}$ galaxies.
In contrast, in \simba\ we find minimal contribution from SFR$<20 \; \mathrm{M_{\odot} \, yr^{-1}}$ galaxies (see \sec{blending}).

It has been suggested that part of the offset in \eightfifty\ counts between \eagle\ and the observations is due to the small simulation volume \citep{wang_multi-wavelength_2019}.
Smaller periodic volumes naturally do not contain massive clusters or their protocluster progenitors, which have been proposed as regions of preferential SMG activity, are also less likely to sample galaxies in the act of starbursting.
Our results tentatively suggest that this cannot account for the offset entirely; our \simba\ volume is only $\sim$3$\times$ larger than that of \eagle, and still does not contain a large number of clusters -- there is only a single $10^{15} M_\odot$ system at $z=0$ in the \simba\ volume.
Moreover, the deficit in \eagle\ counts extends to low fluxes, whose galaxies would be quite well represented in a 100~Mpc box.
We show in \app{convergence} that in \simba\ we do not see any greater deficit at the faint end in a higher resolution 50~Mpc box at $z = 3.7$.

It has also been suggested that the offset in the \eagle\ counts is a result of not tuning to the statistical properties of dusty star-forming populations \citep{mcalpine_nature_2019}.
Equally, \simba\ has not been directly tuned to such properties.
We will demonstrate in \sec{drivers} that the increased star formation and self-consistent dust model lead indirectly to \simba's better agreement.
In \simba, the increased star formation likely occurs because early galaxies have very high mass loading factors that elevate substantial gas into the halo, which then coalesces into massive systems at $z\sim 2-3$, fueling particularly vigorous star formation during Cosmic Noon.

The same effect was noted in both \protect\citet{finlator06a} and \citet{narayanan15a}, using fairly different feedback schemes.
\simba\ includes AGN quenching feedback, primarily due to AGN jets that rely on low black hole accretion rates.
At $z\sim 2-3$, some massive galaxies satisfy this and fall off the main sequence, while others do not and end up vigorously forming stars, appearing at the top end of the main sequence.
We note that \simba\ agrees well with the number density of galaxies that lie $\ga 1$~dex below the main sequence at these epochs~\citep{rodriguez_montero_mergers_2019}, though it fails to sufficiently quench those galaxies since it does not match the counts lying $\ga 2$~dex below the main sequence (Merloni et al, submitted; Finkelstein et al, submitted).
So it appears that \simba's AGN feedback is approximately striking the correct balance between quenching sufficient galaxies at $z\sim 2$, while not quenching too many massive galaxies which would eliminate the SMG population entirely.

Finally, it has been suggested that \eagle\ may underestimate the FUV attenuation \citep{baes_cosmic_2019}.
This may be a result of the constant dust-to-metals ratio governing the diffuse dust mass, the modelling of dust in HII regions \citep{trcka_reproducing_2020}, or the global star-dust geometry \citep[e.g.][]{narayanan18a,salim_dust_2020}.
We address the impact of the self-consistent dust model in \simba\ in \sec{dust}.

\fig{diff_counts_models} shows a comparison of \simba\ to a wider suite of models using various techniques~\citep{granato_infrared_2000,granato04a,baugh05a,fontanot_reproducing_2007,shimizu_submillimetre_2012,hayward_submillimetre_2013,lagos_far-ultraviolet_2019}, alongside the \cite{geach_scuba-2_2017} observational constraints as grey diamonds.
The \simba\ and (the most optimistic) \eagle\ results are reproduced from \fig{diff_counts} in green and orange, respectively.

\begin{figure*}
\includegraphics[width=\textwidth]{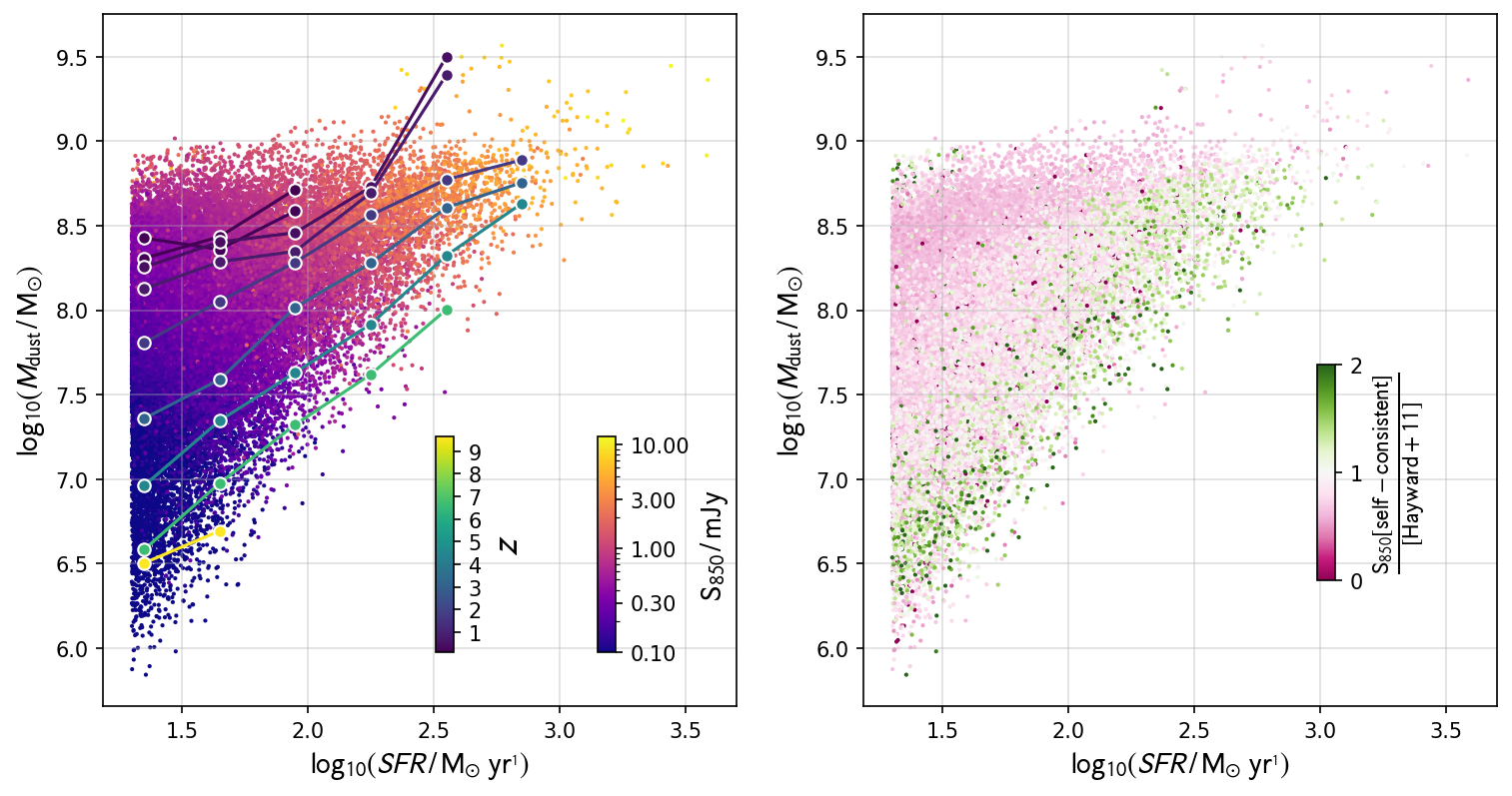}
    \caption{
    Dust mass against SFR for each galaxy at all redshifts.
    \textit{Left panel:} each galaxy is coloured by its \eightfifty\ flux density.
    Larger connected points show the median relations (at redshifts $z_{\mathrm{bin}} = [0.12,0.2,0.5,0.8,1.7,3.2,4.5,6.7]$) coloured by their redshift.
    \textit{Right panel:} Each point is coloured by the ratio of its \eightfifty\ luminosity in \simba\ and that predicted by the parametric form of \protect\cite{hayward_what_2011}.
    }
    \label{fig:dust_sfr}
\end{figure*}

A pioneering attempt to predict SMG number counts in a hierarchical framework was made using an early version of the \galform SAM \citep{granato_infrared_2000}, presented in \cite{baugh_can_2005}, but fell dramatically short (dashed cyan line).
An independent SAM was presented in \citet{granato04a} (solid cyan line), which overshoots the number counts at the bright end, owing to updated cooling and star formation modules combined with RT using GRASIL~\citep{silva98a}.

\citet{baugh05a} produced an update to the \galform model, and presented results when assuming a canonical IMF (dashed purple line), then went on to demonstrate that assuming a flat IMF above one solar mass within merging galaxies could mitigate this issue and produce sufficient SMGs (solid purple line).
While impressive in its agreement, such an IMF is somewhat controversial \citep{bastian_universal_2010,hopkins_variations_2013,krumholz_big_2014,tacconi_submillimeter_2008,motte_unexpectedly_2018,schneider_excess_2018,zhang_stellar_2018}.
\cite{lacey_unified_2016} presented an update to the \citep{baugh05a} model, in particular using a much less top-heavy IMF in mergers (slope $x = 1$).
\cite{cowley_simulated_2015} presented the \eightfifty\ number counts subject to blending with a beam size identical to the JCMT, and we show these predictions in \fig{diff_counts_models} (dark purple line).
The agreement with the \cite{geach_scuba-2_2017} results is exceptional over the flux density range probed, though this is still reliant on a top-heavy IMF in mergers.
At higher flux densities \cite{cowley_simulated_2015} predict an upturn in the number counts, which we do not see in our results including blending.

\cite{fontanot_reproducing_2007} (solid yellow line) attempted to reproduce the observed counts in the \textsc{Morgana} SAM, without implementing a variable IMF.
They found good agreement with the sub-mm LF, attributing this to their cooling model.
However, their model overestimated number counts of local massive galaxies.
This corroborates the suggestion of \citet{dekel09a} that assuming highly efficient conversion of gas into stars, it is possible to achieve the SFRs required for SMGs at $z\sim 2$; but such near-unity conversion efficiencies are well above the $\sim 5-10\%$ conversion efficiencies inferred for today's massive ellipticals that are putatively SMG descendants~\citep{behroozi_average_2013,moster_emerge_2018}.

\cite{lagos_far-ultraviolet_2019} (solid dark green line) presented results for the SHARK SAM \citep{lagos_shark:_2018}, using attenuation curves computed from \eagle\ using the SKIRT RT code \citep{trayford_fade_2020} and parametrised in terms of dust column density.
They also use a fixed \cite{chabrier_galactic_2003} IMF, and this gives reasonably good agreement at the faint end, whilst overestimating the number of bright sources by $>\, 1\; \mathrm{dex}$.

\cite{hayward_submillimetre_2013} (solid blue line) ran idealised (i.e. non-cosmological) hydrodynamic simulations of disc galaxies and mergers, and then weighted their contributions with a hierarchical model to estimate the sub-mm number counts.
They get good agreement with observations, albeit with perhaps optimistic assumptions about the contributions of mergers to the SMG population.
For instance, they attribute 30-50\% of $S_{850} > 1 \; \mathrm{mJy}$ sources to associated blends, which is much higher than our more direct modelling suggests (\sec{blending}).

The \cite{shimizu_submillimetre_2012} results are particularly interesting, in the sense that they are the first cosmological hydrodynamic simulations that do a reasonable job of matching \eightfifty\ number counts (solid red line).
They used a $100 \,h^{-1} \, \mathrm{Mpc}$ {\sc Gadget-3} simulation, and implemented a simplified dust model of a spherical dust shell around each galaxy, out to 9\% of the virial radius, where this value was tuned to match the UV luminosity function at $z=2.5$.
While their model did not include AGN quenching feedback so likely did not produce a viable $z=0$ galaxy population (although this was not tested directly), they were able to get within striking distance of observed SMG counts, albeit with too shallow a slope that strongly over-predicted the brightest systems and under-predicted by $\sim 0.3$ dex the number of $S_{850} \sim 3 \;\mathrm{mJy}$ sources.

In summary, hierarchical models have -- to date -- had some difficulty in reproducing SMG counts.
Agreement is possible in SAMs by tuning parameters accordingly, albeit sometimes with questionable physical motivation.
Both the \cite{shimizu_submillimetre_2012} simulations and \eagle\ use cosmological hydrodynamics models to produce large populations of sub-mm galaxies, but still show significant discrepancies compared to the observed \eightfifty\ counts.
This highlights that \simba's agreement with SMG number counts is not trivial.
It is thus interesting to examine why \simba\ performs so well in this regard: what are the physical drivers of the \eightfifty\ emission in \simba?

%% file: alltogether.tex
\subsection{The Star Formation Rate--Dust Mass Plane}
\label{sec:alltogether}

The left panel of \fig{dust_sfr} shows the SFR--dust mass relation in \simba.
There is a clear dependence of \eightfifty\ emission along both the SFR and dust-mass dimensions.
Dust masses tend to increase with redshift for our $\mathrm{SFR} > 20 \; \mathrm{M_{\odot} \; yr^{-1}}$ selection, and it is the galaxies with lower SFRs that show the largest relative increase.

The dependence of \eightfifty\ emission on SFR and dust mass has been parametrised as a power-law relation using  idealised simulations with simplified geometries by \protect\cite{hayward_what_2011} with the following form,
\begin{equation}\label{eq:hayward11}
  S_{\mathrm{850}} \,/\, \mathrm{mJy} = \mathrm{a} \; \left( \frac{\rm SFR}{100 \; \mathrm{M_\odot \; yr^{-1}}} \right)^{\mathrm{b}} \, \left( \frac{M_{\mathrm{dust}}}{10^{8} \; \mathrm{M_\odot}} \right)^{\mathrm{c}}
\end{equation}
where $\mathrm{a}$, $\mathrm{b}$ and $\mathrm{c}$ are free parameters.
\cite{hayward_what_2011} found the following best fits, $\mathrm{a} = 0.65$, $\mathrm{b} = 0.42$ and $\mathrm{c} = 0.58$.
The right panel of \fig{dust_sfr} shows the ratio of the \eightfifty\ flux predicted from the full RT and that from the \cite{hayward_submillimetre_2013} parametric model (using dust masses from the self-consistent model (see \sec{simba}) and instantaneous SFRs directly from \simba).
There are clear gradients along the SFR and dust-mass directions.
There is a population of galaxies at fixed SFR with low dust masses for which the \cite{hayward_what_2011} model under-predicts the \eightfifty\ emission compared to \simba\ by up to a factor of two.
At higher dust masses, however, \cite{hayward_what_2011} over-predicts the emission by approximately the same factor.
Similarly, at a fixed dust mass of $10^{8.5} \; \mathrm{M_{\odot}}$ the most star-forming galaxies under-predict the emission by a factor of two compared to the Hayward model, whereas the lowest star-forming galaxies (in this sample) over-predict the emission by a factor of two.
The difference between \simba{} and \citet{hayward_what_2011} can likely be attributed to the significantly more complex star-dust geometries in \simba, combined with a relatively sophisticated dust model \citep{li_dust--gas_2019}.

We use \simba\ to generate new fits to \eq{hayward11}, and find the following best-fit parameters: $\mathrm{a} = 0.58 \pm 0.0023$, $\mathrm{b} = 0.51 \pm 0.0022$ and $\mathrm{c} = 0.49 \pm 0.0031$, with $1 \sigma$ uncertainties $< 0.01$ for each parameter.
While broadly similar, our fit suggests a stronger dependence of the sub-mm emission on SFR than in \cite{hayward_what_2011}, and a weaker dependence on dust mass.
Whilst the relation is reasonably tight, with a median fractional residual of $19.5 \%$ for galaxies where $S_{850} \geqslant 1 \; \mathrm{mJy}$, we caution that when computing quantities such as number count distributions, it is important to account for the scatter in the distribution, which can particularly impact the bright end.
However, our results suggest that a reasonably tight relation does exist, and can be used to cheaply predict the \eightfifty\ emission in other models.

Our best fit relation demonstrates that dust mass and SFR have an almost equally strong role in governing the strength of sub-mm emission.
Hence to understand the origin of \simba's high \eightfifty\ fluxes compared to many other models, we must investigate what is unique about the star formation rates and dust masses predicted for high-redshift galaxies in \simba.

%% file: sfrf.tex
\subsection{Contribution to the Star Formation Rate Function}
\label{sec:sfrf}

\begin{figure*}
\includegraphics[width=\textwidth]{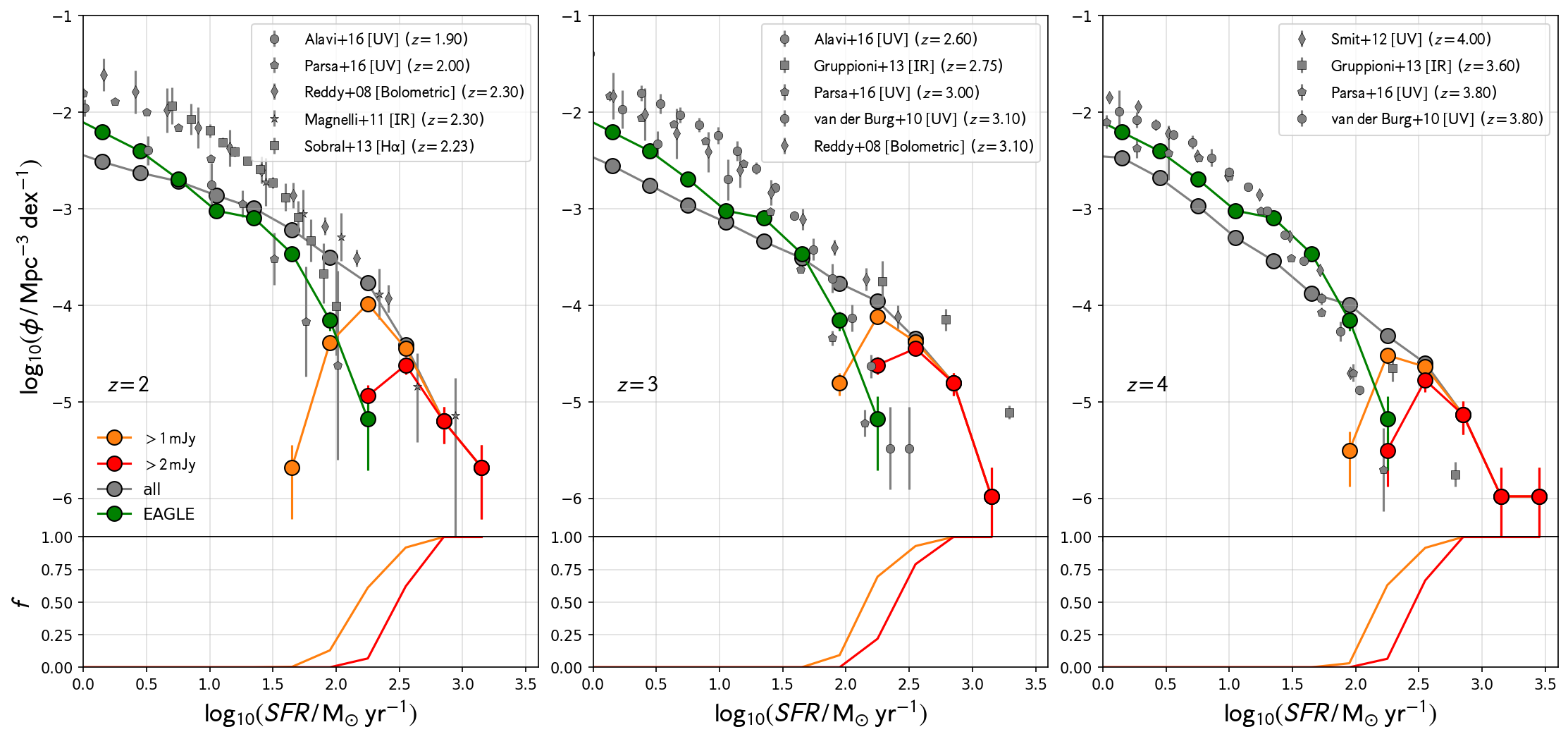}
    \caption{
    Star formation rate function at $z = [2,3,4]$ (left to right panels) for the whole population (grey), and for the sub-mm population $> 1 \, \mathrm{mJy}$ (orange) and $> 2 \, \mathrm{mJy}$ (red).
    \eagle\ is shown in green.
    Bottom panels show the fraction of all galaxies that satisfy the two sub-mm flux density thresholds at a given SFR.
    Observational constraints from the \protect\cite{katsianis_evolution_2017-1} compilation in the UV \protect\citep{van_der_burg_uv_2010,smit_star_2012,alavi_ultra-faint_2014,parsa_galaxy_2016}, H$\alpha$ \protect\citep{sobral_large_2013} and IR tracers \protect\citep{reddy_multiwavelength_2008,magnelli_evolution_2011,gruppioni_herschel_2013} are also shown, with the measurement redshift in the legend.
    }
    \label{fig:sfrf}
\end{figure*}

We begin by examining \simba's star formation rates, quantified by the Star Formation Rate Function (SFRF).
\fig{sfrf} shows the SFRF in \simba\ at $z = [2,3,4]$.
The sub-mm contribution for two flux density cuts, $> 1 \; \mathrm{mJy}$ (orange) and $> 2 \; \mathrm{mJy}$ (red) is shown, as well as the SFRF for the full population (grey).  For comparison, the \eagle\ SFRF is shown in green.

SMGs are strongly biased to the most star-forming systems, as we have already seen in \fig{sfr_s850_completeness}, accounting for \textit{all} galaxies where $\mathrm{SFR} > 10^{\,3} \; \mathrm{M_{\odot} \; yr^{-1}}$.
The sub-mm SFRF turns over at lower $\mathrm{SFRs}$ ($\sim 10^{\,2} \; \mathrm{M_{\odot} \; yr^{-1}}$), and galaxies with $\mathrm{SFR} < 30 \; \mathrm{M_{\odot}} \, \mathrm{yr^{-1}}$ do not produce currently observable sub-mm emission at these redshifts.
This justifies our use of a $\mathrm{SFR} > 20 \; \mathrm{M_{\odot}} \, \mathrm{yr^{-1}}$ selection for examining SMGs, which conservatively ensures a complete sample at $S_{850}>1$~mJy during the main SMG epoch.

\fig{sfrf} also shows a number of observational constraints to the SFRF.  We used the \cite{katsianis_evolution_2017-1} compilation of constraints from UV \citep{van_der_burg_uv_2010,smit_star_2012,alavi_ultra-faint_2014,parsa_galaxy_2016}, H$\alpha$ \citep{sobral_large_2013} and IR selected samples \citep{reddy_multiwavelength_2008,magnelli_evolution_2011,gruppioni_herschel_2013}.
The authors use SFR indicators at these wavelengths from \cite{kennicutt_star_1998} obtained from SPS models, and dust-correct the UV measurements using the \cite{smit_star_2012} and \protect\cite{hao_dust-corrected_2011} prescriptions.
They assume a \cite{salpeter_luminosity_1955} IMF, which we convert to \cite{chabrier_galactic_2003} by multiplying by a factor of 0.63 \citep{madau_cosmic_2014}.
This compilation gives a comprehensive census of star forming galaxies, tracing both dust-poor and low-mass systems, as well as massive, highly star-forming, dust-obscured systems.

IR-selected SFR measurements tend to extend the SFRF to higher SFRs by up to an order of magnitude compared to those from UV-selected samples, since rapidly star-forming galaxies at this epoch tend to be quite dust-obscured.
\simba\ is in good agreement with these IR-selected constraints at $z \sim 2$ \citep{magnelli_alma_2020} and $ z \sim 4$ \citep{gruppioni_herschel_2013}.
At $z \sim 3 $ the \cite{gruppioni_herschel_2013} constraints have a higher normalisation, but these are in tension with those from \cite{reddy_multiwavelength_2008}, highlighting the inter-study scatter at the high-SFR end.
UV-selected samples, where they do extend to high-SFRs, significantly underestimate the normalisation compared to IR-selected constraints.

While \simba\ has success in matching the high-SFR end ($\mathrm{SFR} > 20 \, \mathrm{M_{\odot} \; yr^{-1}}$; of importance for this paper), it generally falls well short of producing enough low-SFR galaxies, falling short in number density by up to $\sim 0.7$~dex at $\mathrm{SFR} \la 10 \, \mathrm{M_{\odot} \; yr^{-1}}$.
In part this is an issue of resolution.
If we examine a $25 \,h^{-1} \, \mathrm{Mpc}$ \simba\ box with identical physics, we find a better match to the SFRF for $\mathrm{SFR} \la 10 \; \mathrm{M_{\odot} \; yr^{-1}}$ (see \app{convergence}).
This is due to both an intrinsic non-convergence in the model, as well as the scatter in the SFR--$M_*$ relation.
To clarify the latter, note that the large-volume \simba\ simulation has a galaxy stellar mass completeness limit of $5.8\times 10^8\; M_\odot$, which at $z\sim 2$ corresponds broadly to an SFR limit of $\mathrm{SFR} \sim 1 \; \mathrm{M_{\odot} \; yr^{-1}}$.
However, the substantial scatter in the SFR--$M_{\star}$ relation~\citep{dave_simba:_2019} means that we will begin losing galaxies to our $M_{\star}$ cut at significantly higher SFR.
However, this non-convergence appears to be more prominent at $z=2$ than at higher redshifts, suggesting that this cannot fully explain the discrepancies at all epochs.

Another potential source of the discrepancy is the well-known offset in the SFR--$M_{\star}$ relation between all types of hierarchical models and observations at $z\sim 2$, in which models tend to under-predict SFRs by factors $\sim\times 2-3$.
If this is due to systematics in inferring SFRs from SED data~\citep[e.g.][]{leja19a}, then this would shift the observational data points to the left by up to 0.5~dex.
Again, this would help, but would not fully mitigate the discrepancy.
Thus we conclude that \simba\ likely falls somewhat short at reproducing enough low-SFR galaxies at Cosmic Noon, although perhaps not as egregiously as \fig{sfrf} naively suggests.
These low-SFR galaxies may contribute to the faint-end ($\sim 3 \mathrm{mJy}$) of the number counts, which could improve the agreement with observations, however they will have minimal effect at brighter flux densities.

\fig{sfrf} also shows the SFRF in the \eagle\ model, in green.
\eagle\ does not produce galaxies with extremely high ($\ga 300 \; \mathrm{M_{\odot} \; yr^{-1}}$) SFRs, tending to follow the UV-selected constraints at the high-SFR end.
This has been variously attributed to the lack of `bursty' star formation in the \eagle\ model \citep{furlong_evolution_2015}, or to the strength of the AGN feedback \citep{katsianis_evolution_2017}.
Whatever the cause, we speculate that the lack of highly star-forming galaxies is the primary reason for the corresponding dearth of bright \eightfifty\ sources in \eagle, as has recently been suggested by \cite{baes_infrared_2020}.
Indeed, the discrepancy between \eagle's SFRF and IR observations at SFR$> 100 \; \mathrm{M_{\odot} \; yr^{-1}}$ is broadly similar to the discrepancy seen in their \eightfifty\ number counts at $S_{850}>1$~mJy.

We note that simulation volume effects do not play a role in the \simba\ SFRF prediction.
We have checked the SFRF against a $50 \,h^{-1} \, \mathrm{Mpc}$ box size \simba\ run with the same resolution and input physics but one-eighth the volume (and approximately one-third that of \eagle), and the SFRF is indistinguishable up to the point that the small-volume run runs out of galaxies (SFR$\sim 400 \; \mathrm{M_{\odot} \; yr^{-1}}$).
This is even true in the $25  \,h^{-1} \, \mathrm{Mpc}$ \simba\ box with $8\times$ higher mass resolution.
Hence the SFRF is quite well converged versus volume effects (see \app{convergence} for details).
We correspondingly infer that the lack of high-SFR galaxies in \eagle\ does not owe to its smaller volume relative to \simba's.

Overall, \simba\ does a good job at reproducing the SFRF at the high-SFR end, generally tracking well the far-IR derived SFRF constraints at $z\sim 2-4$.  This is a major driver of its success in reproducing the \eightfifty\ number counts.  However, the far-IR emission is also strongly dependent on the amount of dust in the galaxy.  Thus next we examine the role that \simba's dust model plays in setting the \eightfifty\ counts.

%% file: dust.tex
\subsection{Dust-to-Metal and Dust-to-Gas Ratios}
\label{sec:dust}

\begin{figure}
\includegraphics[width=\columnwidth]{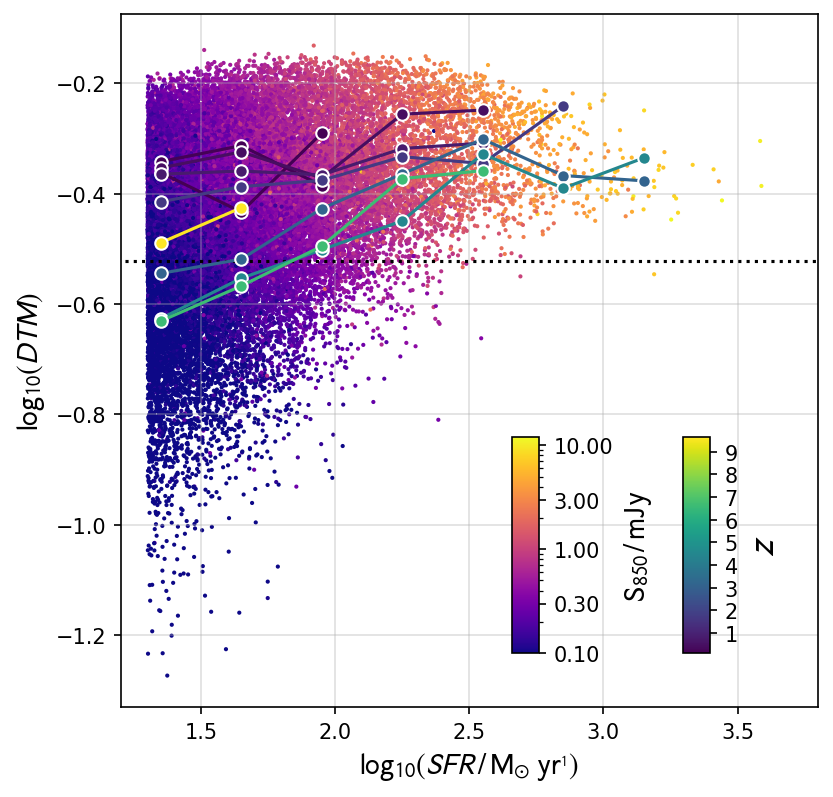}
    \caption{
    Dust-to-metals ratio ($f_{\,\mathrm{DTM}}$) against SFR for each galaxy, coloured by $S_{850}$ luminosity.
    Larger connected points show the median relations at the following redshifts: $z_{\mathrm{bin}} = [0.12,0.2,0.5,0.8,1.7,3.2,4.5,6.7]$.
    The black dotted line shows a fixed $\mathrm{DTM} = 0.3$.
    }
    \label{fig:dtm_sfr}
\end{figure}

We have already described the self-consistent dust model in \simba\ (see \sec{simba}).
This allows for both the creation and destruction of dust, meaning that the dust content of a galaxy does not directly scale with either the gas or metallicity evolution, but can evolve independently.
The dust-to-metal ($f_{\,\mathrm{DTM}}$) and dust-to-gas ($f_{\,\mathrm{DTG}}$) ratios are therefore direct predictions of the model, and can influence the sub-mm emission.

$f_{\,\mathrm{DTM}}$ describes the fraction of all ISM metals locked in dust grains, which for the self-consistent model is given by
\begin{equation} \label{eq:1}
  f_{\,\mathrm{DTM}} = \frac{M^{\mathrm{self-consistent}}_{\,\mathrm{dust}}}{M^{\mathrm{self-consistent}}_{\,\mathrm{dust}} + Z_{\,\mathrm{gas}} \, M_{\,\mathrm{gas}}} \;\;.
\end{equation}
where $M^{\mathrm{self-consistent}}_{\,\mathrm{dust}}$ is the total dust mass in the self-consistent model, $M_{\,\mathrm{gas}}$ is the total gas mass, and $Z_{\,\mathrm{gas}}$ is the gas-phase mass-weighted metallicity.
\fig{dtm_sfr} shows $f_{\,\mathrm{DTM}}$ versus SFR for all galaxies in our comoving selection at a range of redshifts.
Rather than all galaxies having identical values for $f_{\,\mathrm{DTM}}$, there is a large range in $f_{\,\mathrm{DTM}}$ at fixed SFR, and the median relation evolves with redshift.
Whilst $S_{850}$ is primary correlated with $\mathrm{SFR}$, there is also an apparent secondary correlation with $f_{\,\mathrm{DTM}}$.

Simulations that do not model the dust self-consistently must infer the dust mass from other galaxy properties, typically the metal content of the gas.
$f_{\,\mathrm{DTM}}$ is then the fraction of those gas-phase metals assumed to be in the form of dust.
This can complicate comparisons between simulations.
In the absence of a dedicated dust model, many simulations arbitrarily reduce the enrichment of the ISM in order to match the mass-metallicity relation \citep[MZR; \textit{e.g.} \mbox{\sc{MUFASA}} ][]{dave_mufasa:_2016}.
Applying a fixed $f_{\,\mathrm{DTM}}$ to the metal enriched gas in such models will give artificially lower dust masses.
The \eagle\ simulation does not arbitrarily reduce enrichment, and this is one potential cause of the high normalisation of the MZR in this model at $z = 0$ \citep[see][]{somerville_physical_2015}.
It also means that all ISM metals are in the gas, so $f_{\,\mathrm{DTM}}$ can directly be applied.

A fixed value of $f_{\,\mathrm{DTM}} = 0.3$ was assumed in the \eagle\ sub-mm predictions \citep{camps_data_2018,mcalpine_nature_2019}.
\fig{dtm_sfr} shows this value as a horizontal dotted line.
A large fraction of galaxies in \simba{} have a higher $f_{\,\mathrm{DTM}}$, particularly at $z < 5$.
This may explain in some part the general offset in infrared luminosity functions seen in the \eagle\ model at $z > 1$ \citep{baes_infrared_2020}.

\begin{figure}
\includegraphics[width=\columnwidth]{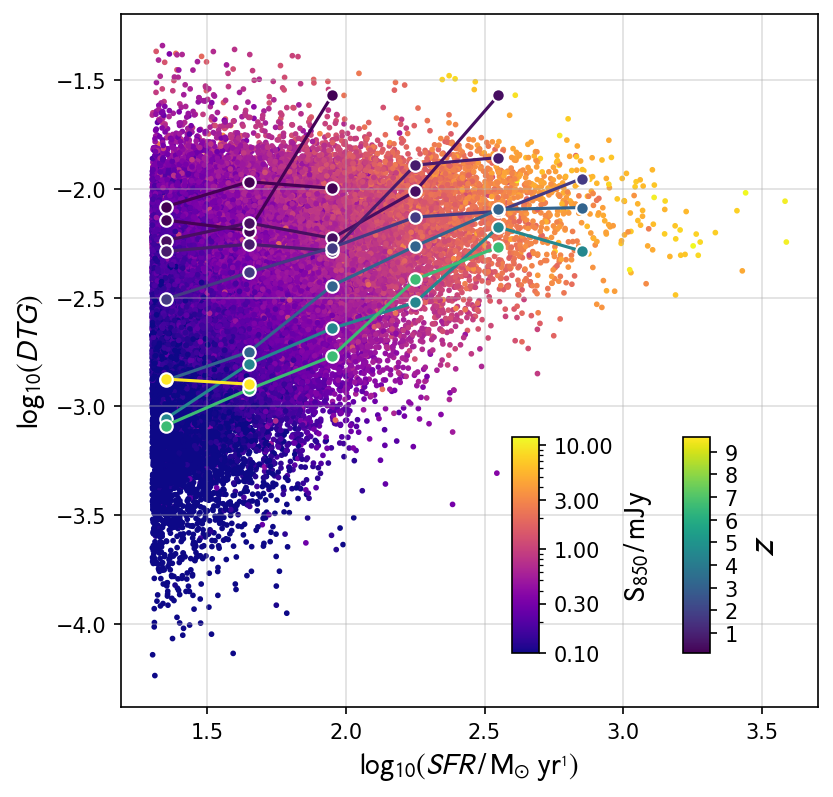}
    \caption{
    Dust-to-gas ratio (DTG) against SFR for each galaxy, coloured by $S_{850}$ luminosity.
    Larger connected points show the median relations at the following redshifts: $z_{\mathrm{bin}} = [0.12,0.2,0.5,0.8,1.7,3.2,4.5,6.7]$.
    }
    \label{fig:dtg_sfr}
\end{figure}

$f_{\,\mathrm{DTG}}$ relates the dust mass to the total gas mass of the galaxy.
\fig{dtg_sfr} shows ($f_{\,\mathrm{DTG}}$) versus SFR for all sub-mm galaxies in the comoving selection.
There is a much larger dynamic range in $f_{\,\mathrm{DTG}}$ than $f_{\,\mathrm{DTM}}$, and this appears to be due to stronger positive redshift evolution in the former, particularly for $\mathrm{SFR} \, < \, 100 \; \mathrm{M_{\odot} \; yr^{-1}}$.
This suggests that, whilst the fraction of metals locked in dust remains relatively constant with redshift, the consumption of gas in galaxies through star formation boosts $f_{\,\mathrm{DTG}}$ considerably.

\begin{figure}
\includegraphics[width=\columnwidth]{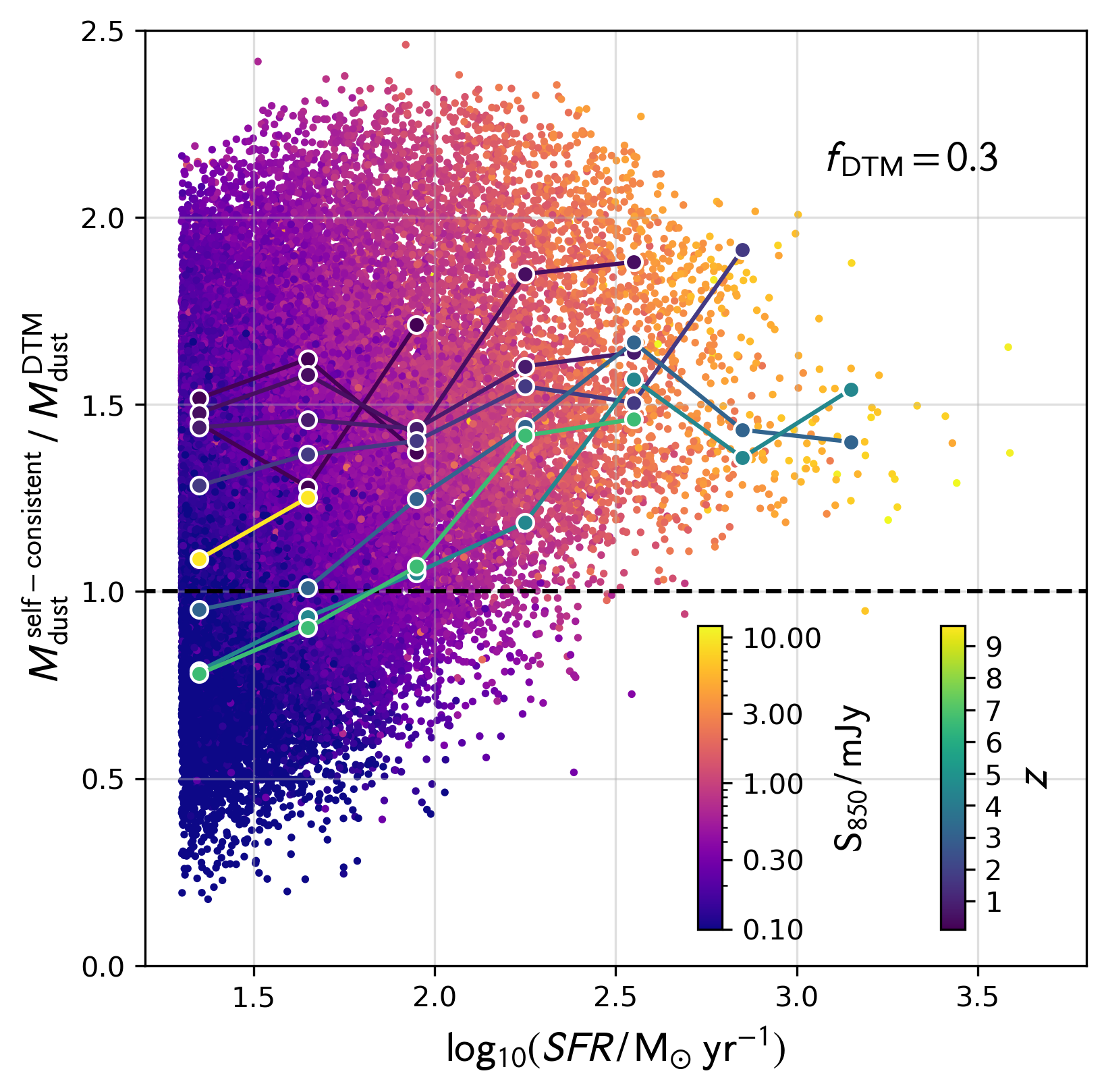}
    \caption{
    Ratio of the dust mass predicted by the self-consistent dust model, and that implied by using a fixed $f_{\,\mathrm{DTM}}$ ratio of 0.3, as a function of SFR.
    Each point shows a single galaxy coloured by $S_{850}$ luminosity.
    Larger connected points show the median relations at the following redshifts: $z_{\mathrm{bin}} = [0.12,0.2,0.5,0.8,1.7,3.2,4.5,6.7]$.
    }
    \label{fig:dtm_dust_mass}
\end{figure}

\begin{figure}
\includegraphics[width=\columnwidth]{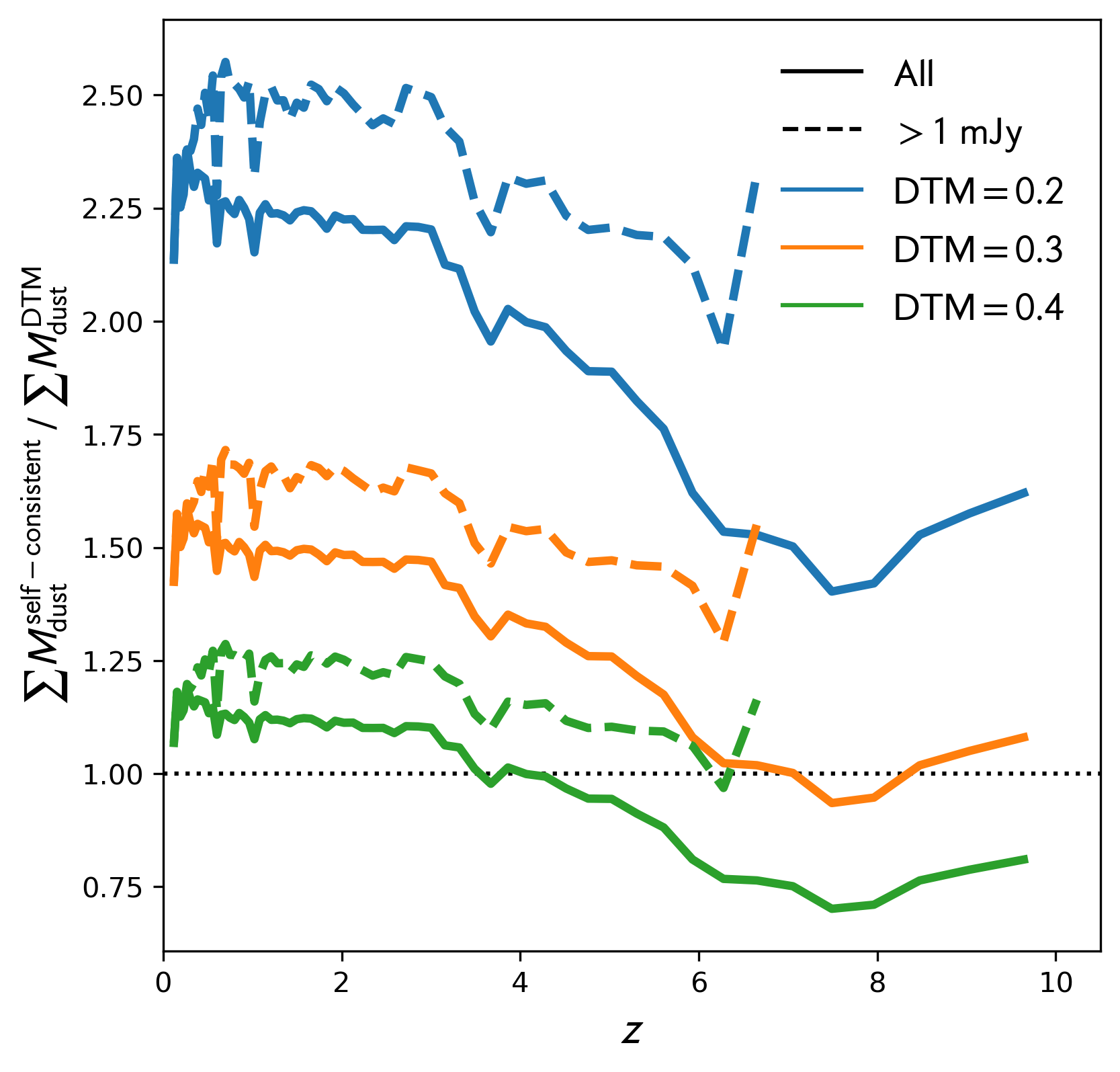}
    \caption{
    The ratio of \textit{total} dust mass in the self-consistent dust model to that implied by a model with fixed DTM, and its evolution with redshift.
    We show this ratio for a range of DTM values.
    We show all galaxies in the comoving volume (solid) regardless of $\mathrm{SFR}$, as well as a subset of sub-mm galaxies where $S_{850} \, > \, 1 \; \mathrm{mJy}$ (dashed).
    }
    \label{fig:total_dust_mass}
\end{figure}

Whilst \fig{dtm_sfr} shows the significant spread in $f_{\,\mathrm{DTM}}$, it does not tell us how much dust there is in comparison to using a fixed $f_{\,\mathrm{DTM}}$.
In order to best compare with the $f_{\,\mathrm{DTM}}$ used in \eagle\ we include the dust mass from the self-consistent model,
\begin{equation} \label{eq:2}
  M_{\,\mathrm{dust}}^{\,\mathrm{DTM}} = f_{\,\mathrm{DTM}}  \times (M^{\mathrm{self-consistent}}_{\mathrm{dust}} + \, M_{\mathrm{gas}} \, Z_{\,\mathrm{gas}}) \;\;.
\end{equation}
where $M_{\,\mathrm{dust}}^{\,\mathrm{DTM}}$ is the dust mass implied with a fixed $f_{\,\mathrm{DTM}}$.
\fig{dtm_dust_mass} shows the ratio of the dust mass from the self-consistent model, $M_{\,\mathrm{dust}}^{\,\mathrm{self-consistent}}$, and that implied by using a fixed $f_{\,\mathrm{DTM}} = 0.3$ as a function of SFR.
As implied by \fig{dtm_sfr}, a large number of galaxies in \simba{} have higher dust masses than would be obtained using a fixed DTM ratio, by factors of up to $2.5$.

To see how this affects the total mass of dust in all galaxies, in \fig{total_dust_mass} we plot the sum of all dust in the self-consistent model and in that implied by using a fixed $f_{\,\mathrm{DTM}}$.
When looking at all galaxies in the comoving volume, regardless of $\mathrm{SFR}$, we see that the self-consistent model gives higher dust masses at lower redshift, and this is proportional to the value of $f_{\,\mathrm{DTM}}$.
At $z = 0.1$, $f_{\,\mathrm{DTM}} = 0.3$ leads to 50\% less total dust compared to the self-consistent model.
We also consider just the SMGs with $S_{850} \, > \, 1 \; \mathrm{mJy}$, and find that these galaxies have even higher dust masses in the self-consistent model compared to using a fixed $f_{\,\mathrm{DTM}}$.
This reflects the higher normalisation of the $f_{\,\mathrm{DTM}}$ ratio in the high-SFR regime.

\begin{figure}
\includegraphics[width=\columnwidth]{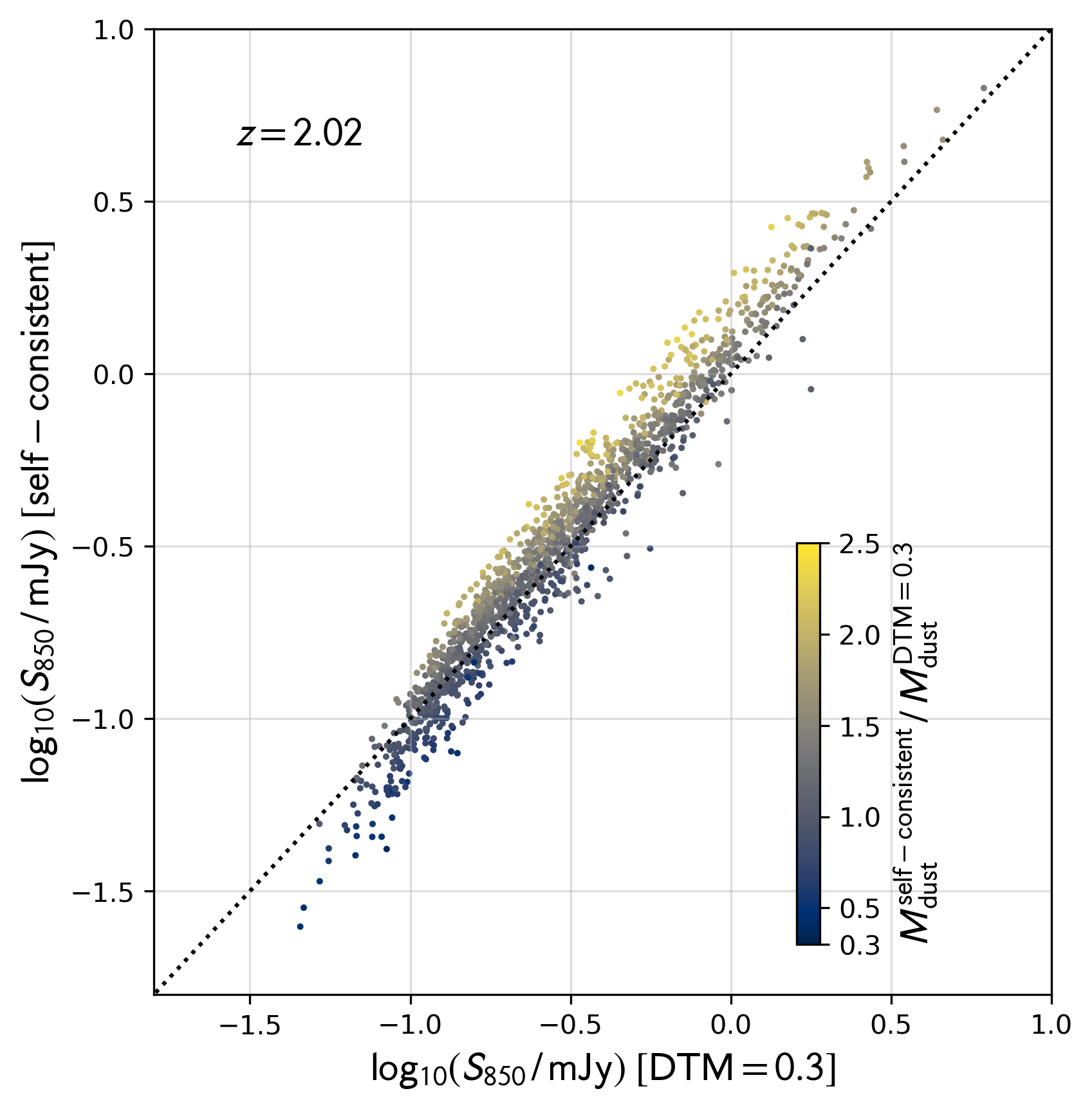}
    \caption{
    $S_{850}$ for the self-consistent dust model against $S_{850}$ using a fixed $f_{\,\mathrm{DTM}} = 0.3$, at $z = 2.02$.
    Each point shows a galaxy coloured by the ratio of its dust mass in the self-consistent model against that implied using a fixed $f_{\,\mathrm{DTM}} = 0.3$.
    The dotted line delimits where the flux densities are equal in both models.
    }
    \label{fig:dtm_comparison}
\end{figure}

To test how this higher dust mass in the self-consistent model translates into predicted \eightfifty\ emission, we re-ran the RT for all galaxies in a single snapshot ($z = 2.02$).
We modified \powderday\ to take account of the metals locked up in dust in the self-consistent model when calculating DTM, rather than just the metals in the gas.
We assumed a fixed $f_{\,\mathrm{DTM}} = 0.3$ to compare to \eagle.
\fig{dtm_comparison} shows the \eightfifty\ emission obtained in both the self-consistent and fixed $f_{\,\mathrm{DTM}}$ models.
There is some spread in the relation, and this is directly proportional to the ratio of the dust mass in the two models.
Where the self-consistent model predicts a higher dust mass, there is higher \eightfifty{} emission, by up to $+0.3$ dex.
This is slightly higher than that expected from the sub-linear scaling with dust mass measured in \eq{hayward11}, which may be attributable to the non-uniform dust distribution possible in the self-consistent model, as well as differences with redshift.
Assuming that the difference in predicted $S_{850}$ seen at $z = 2.02$ due to the self-consistent model translates to other redshifts, this could account for a reduction in the number density of the brightest sources via a systematic shift to lower flux densities of $\sim \; 0.3 \; \mathrm{dex}$.

%% file: conc.tex
\section{Conclusions}
\label{sec:conc}

We have modelled the sub-mm emission from galaxies in the \simba\ cosmological hydrodynamic simulation by using dust continuum radiative transfer with {\sc Powderday} in post-processing.
Our main findings are as follows:

\begin{itemize}
    \item We find good agreement with the shape of single-dish observational constraints on the integrated \eightfifty\ number counts, and the normalisation is within $-0.25 \; \mathrm{dex}$ at $S_{850} > 3 \; \mathrm{mJy}$. At the bright end ($S_{850} > 10 \; \mathrm{mJy}$) the agreement is excellent, within the observational errors.
    \item The number of $S_{850} > 3.6 \; \mathrm{mJy}$ sources peaks at $z = 3.16^{+1.12}_{-0.69}$ and drops off rapidly towards higher and lower redshifts, with brighter SMGs peaking at earlier epochs.
    These predictions broadly agree with observations, but \simba\ notably overpredicts sources at $3.5 < z < 5$.
	  \item Using a lightcone, we find that the multiplicity fraction is high; 52\% of sources are blends of unassociated components, which marginally increase the normalisation of the number counts for single-dish data.
    Associated blends are common, but unlikely to add significantly to the \eightfifty\ flux of individual sources.
	  \item The strength of the sub-mm emission is correlated with the level of star formation.
    The SFR function at $z\sim 2-4$ in \simba\ extends to very high SFRs, $ > 10^{3} \; \mathrm{M_{\odot} \; yr^{-1}}$, in good agreement with IR-inferred observational constraints, and it is these galaxies that dominate the bright end of the sub-mm luminosity function.
    \item \simba\ implements a self-consistent dust model, allowing for varying and evolving dust-to-metal (DTM) ratios.
    Compared to a fixed DTM ratio of 0.3, \simba\ predicts higher dust masses in the majority of galaxies.
    This increased dust mass leads to higher \eightfifty\ emission.
    \item The combination of higher SFRs and dust masses explains the good agreement with observed number counts.
    We provide fits for the \eightfifty\ emission as a function of these intrinsic parameters.
\end{itemize}

Given the unprecedented agreement with observational number count constraints for a cosmological hydrodynamic simulation, and good agreement with the redshift distribution, \simba\ represents an ideal test bed for exploring the nature of SMGs across cosmic time.
In future work we will explore the intrinsic properties of sub-mm sources, their relation to the wider high redshift galaxy population, and their fate at lower redshifts.
However, \simba\ remains limited by poor resolution, owing to its large random volume required to produce significant numbers of rapidly star-forming galaxies.
Hence we will also select individual galaxies and perform `zoom' simulations to explore the resolved line and continuum emission properties of SMGs, providing a direct comparison with the latest and up-coming ALMA observations of the dusty star-forming galaxy population.

%% file: appendix.tex
\section{Output details}
\label{app:output}
\tab{regions} details the snapshots from the $100 \; h^{-1} \; \mathrm{Mpc}$ volume used in this work, and the number of galaxies selected at each snapshot in the whole comoving volume as well as in the 50 lightcone realisations.

\begin{table*}
	\centering
	\caption{\simba\ snapshots on which the RT was run.
  We list the number of galaxies satisfying the selection criteria (see \sec{selection}) in the whole snapshot, as well as the median and 16-84$^{\mathrm{th}}$ percentiles of the number in the 50 lightcone realisations.
	}
	\label{tab:regions}
	\begin{tabular}[t]{cccc} 
		\hline
		Snapshot & $z$ & $N_{\mathrm{galaxy,\,comoving}}$ & $N_{\mathrm{galaxy,\,lightcone}}$ \\
		 & & $\mathrm{(100 \; cMpc)^{-3}}$ & $\mathrm{(0.707 \; deg)^{-2}}$ \\
		\hline
		$020$ & $9.64$ & $9$ & $7^{7}_{5}$ \\
		$022$ & $9.03$ & $18$ & $12^{14}_{10}$ \\
		$024$ & $8.48$ & $31$ & $17^{20}_{14}$ \\
		$026$ & $7.96$ & $41$ & $24^{25}_{21}$ \\
		$028$ & $7.49$ & $48$ & $26^{30}_{21}$ \\
		$030$ & $7.05$ & $71$ & $36^{42}_{32}$ \\
		$032$ & $6.65$ & $100$ & $45^{51}_{39}$ \\
		$034$ & $6.28$ & $114$ & $55^{62}_{50}$ \\
		$036$ & $5.93$ & $126$ & $59^{69}_{52}$ \\
		$038$ & $5.61$ & $161$ & $73^{82}_{64}$ \\
		$040$ & $5.31$ & $199$ & $85^{90}_{81}$ \\
		$042$ & $5.02$ & $229$ & $94^{107}_{85}$ \\
		$044$ & $4.76$ & $277$ & $113^{123}_{102}$ \\
		$046$ & $4.52$ & $314$ & $123^{135}_{115}$ \\
		$048$ & $4.28$ & $375$ & $142^{152}_{131}$ \\
		$050$ & $4.07$ & $432$ & $157^{174}_{145}$ \\
		$052$ & $3.86$ & $470$ & $167^{182}_{151}$ \\
		$054$ & $3.67$ & $561$ & $201^{223}_{183}$ \\
		$056$ & $3.49$ & $632$ & $214^{232}_{197}$ \\
		$058$ & $3.32$ & $701$ & $240^{264}_{212}$ \\
		$060$ & $3.16$ & $797$ & $249^{286}_{220}$ \\
		$062$ & $3.00$ & $837$ & $261^{296}_{227}$ \\
		$064$ & $2.86$ & $939$ & $284^{315}_{254}$ \\
		$066$ & $2.72$ & $1023$ & $287^{336}_{250}$ \\
		$068$ & $2.59$ & $1139$ & $310^{349}_{273}$ \\
		$070$ & $2.47$ & $1232$ & $316^{354}_{268}$ \\
		$072$ & $2.35$ & $1388$ & $323^{376}_{297}$ \\
		$074$ & $2.23$ & $1495$ & $337^{365}_{291}$ \\
		$076$ & $2.13$ & $1500$ & $325^{382}_{277}$ \\
		$078$ & $2.02$ & $1585$ & $315^{384}_{272}$ \\
		$080$ & $1.93$ & $1637$ & $329^{380}_{279}$ \\
		$082$ & $1.83$ & $1743$ & $327^{384}_{295}$ \\

		\hline
	\end{tabular}
	\begin{tabular}[t]{cccc} 
		\hline
		Snapshot & $z$ & $N_{\mathrm{galaxy,\,comoving}}$ & $N_{\mathrm{galaxy,\,lightcone}}$ \\
		 & & $\mathrm{(100 \; cMpc)^{-3}}$ & $\mathrm{(0.707 \; deg)^{-2}}$ \\
		\hline
		$084$ & $1.74$ & $1703$ & $289^{339}_{246}$ \\
		$086$ & $1.66$ & $1664$ & $281^{320}_{245}$ \\
		$088$ & $1.58$ & $1699$ & $260^{307}_{206}$ \\
		$090$ & $1.50$ & $1752$ & $254^{288}_{210}$ \\
		$092$ & $1.42$ & $1711$ & $236^{261}_{206}$ \\
		$094$ & $1.35$ & $1705$ & $220^{239}_{181}$ \\
		$096$ & $1.28$ & $1612$ & $196^{237}_{166}$ \\
		$098$ & $1.21$ & $1606$ & $180^{215}_{153}$ \\
		$100$ & $1.15$ & $1472$ & $141^{175}_{114}$ \\
		$102$ & $1.08$ & $1353$ & $119^{147}_{99}$ \\
		$104$ & $1.02$ & $1238$ & $114^{137}_{88}$ \\
		$106$ & $0.96$ & $1140$ & $85^{106}_{64}$ \\
		$108$ & $0.91$ & $967$ & $66^{86}_{48}$ \\
		$110$ & $0.85$ & $858$ & $56^{71}_{41}$ \\
		$112$ & $0.80$ & $763$ & $47^{57}_{35}$ \\
		$114$ & $0.75$ & $663$ & $33^{44}_{28}$ \\
		$116$ & $0.70$ & $535$ & $26^{34}_{20}$ \\
		$118$ & $0.65$ & $454$ & $19^{23}_{13}$ \\
		$120$ & $0.60$ & $350$ & $13^{16}_{10}$ \\
		$122$ & $0.56$ & $304$ & $8^{12}_{6}$ \\
		$124$ & $0.51$ & $253$ & $7^{10}_{3}$ \\
		$126$ & $0.47$ & $210$ & $5^{7}_{3}$ \\
		$128$ & $0.43$ & $160$ & $3^{5}_{1}$ \\
		$130$ & $0.39$ & $143$ & $2^{4}_{1}$ \\
		$132$ & $0.34$ & $109$ & $1^{2}_{0}$ \\
		$134$ & $0.31$ & $88$ & $1^{2}_{0}$ \\
		$136$ & $0.27$ & $67$ & $0^{1}_{0}$ \\
		$138$ & $0.23$ & $47$ & $0^{1}_{0}$ \\
		$140$ & $0.19$ & $28$ & $0^{0}_{0}$ \\
		$142$ & $0.16$ & $24$ & $0^{0}_{0}$ \\
		$144$ & $0.12$ & $23$ & $0^{0}_{0}$ \\
		\hline
	\end{tabular}
\end{table*}

\section{Simulation convergence test}
\label{sec:convergence}

\begin{figure}
\includegraphics[width=\columnwidth]{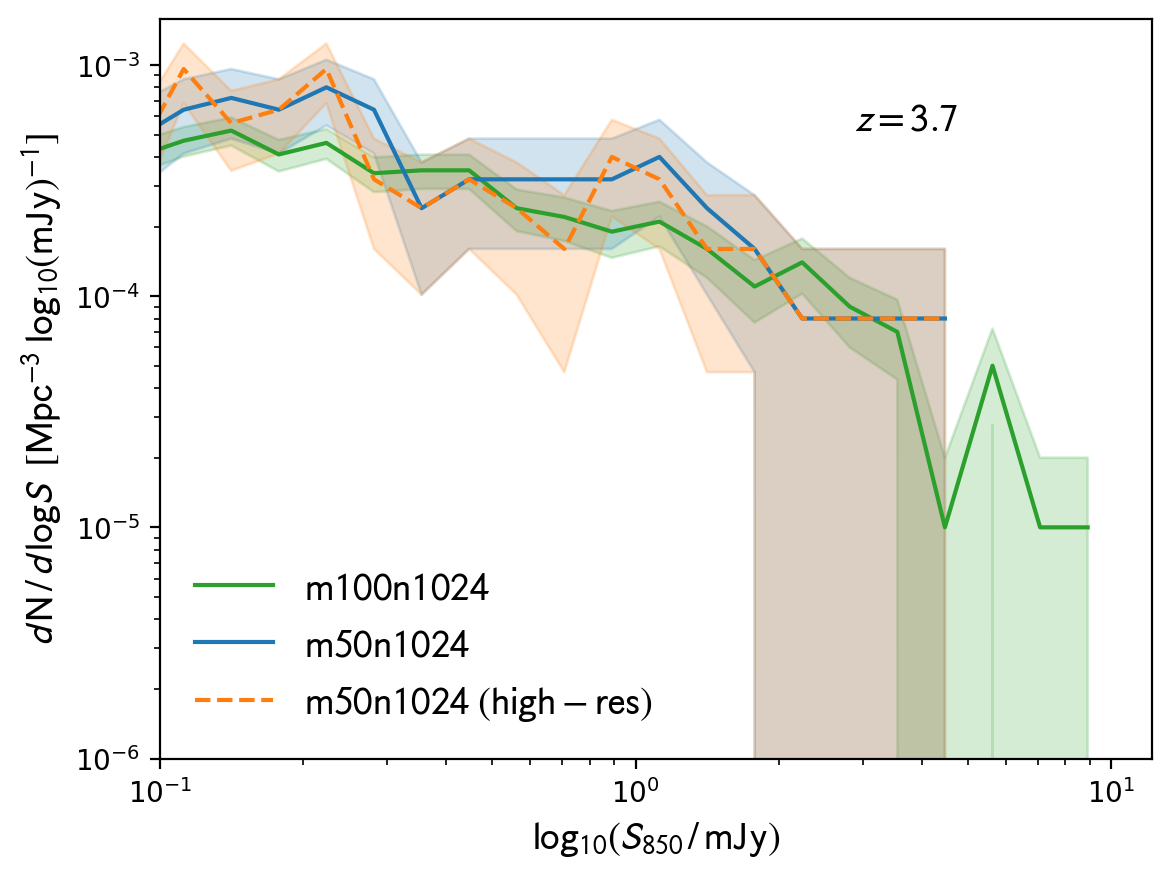}
    \caption{The $S_{850}$ luminosity function at $z = 3.7$ for the m100n1024 and m50n1024 simulations, using both the fiducial \powderday\ parameters and updated, higher resolution parameters ('high-res`).
		}
    \label{fig:sim_convergence}
\end{figure}

In \sec{method_rt} we studied the convergence of our results for increased photon number and grid resolution.
We have also tested the dependence of our results on the simulation resolution, using a 50 Mpc volume with the same number of particles as the 100 Mpc volume used throughout the rest of the analysis.
This provides eight times the mass resolution.
We label this simulation m50m1024, and the original volume m100n1024.
We do not alter the parameters of the RT, which presents a test for \textit{strong convergence}.

\fig{sim_convergence} shows the $S_{850}$ luminosity function at $z = 3.7$ for both simulations.
Both agree within $1\sigma$ poisson uncertainties at $< 1 \; \mathrm{mJy}$, though there is a slight positive offset ($\sim 0.2 \; \mathrm{dex}$) in the median around $1\, \mathrm{mJy}$.
Above this flux density there are fewer bright sources in the 50 Mpc volume, as expected.

We also show how increased photon count and grid resolution in the higher resolution volume affects our results, a test for \textit{weak convergence}.
We set $n_{\mathrm{photon}} = 5 \times 10^{6}$ and $n_{\mathrm{ref}} = 12$, and run the radiative transfer.
The resulting $S_{850}$ luminosity function, shown in \fig{sim_convergence}, is almost identical to the version using the fiducial \powderday\ parameters.
We conclude that structures below the resolution scale can have a small effect on the number counts, but this effect is mitigated by increasing the resolution of the RT (grid and photon count).

\begin{figure}
\includegraphics[width=\columnwidth]{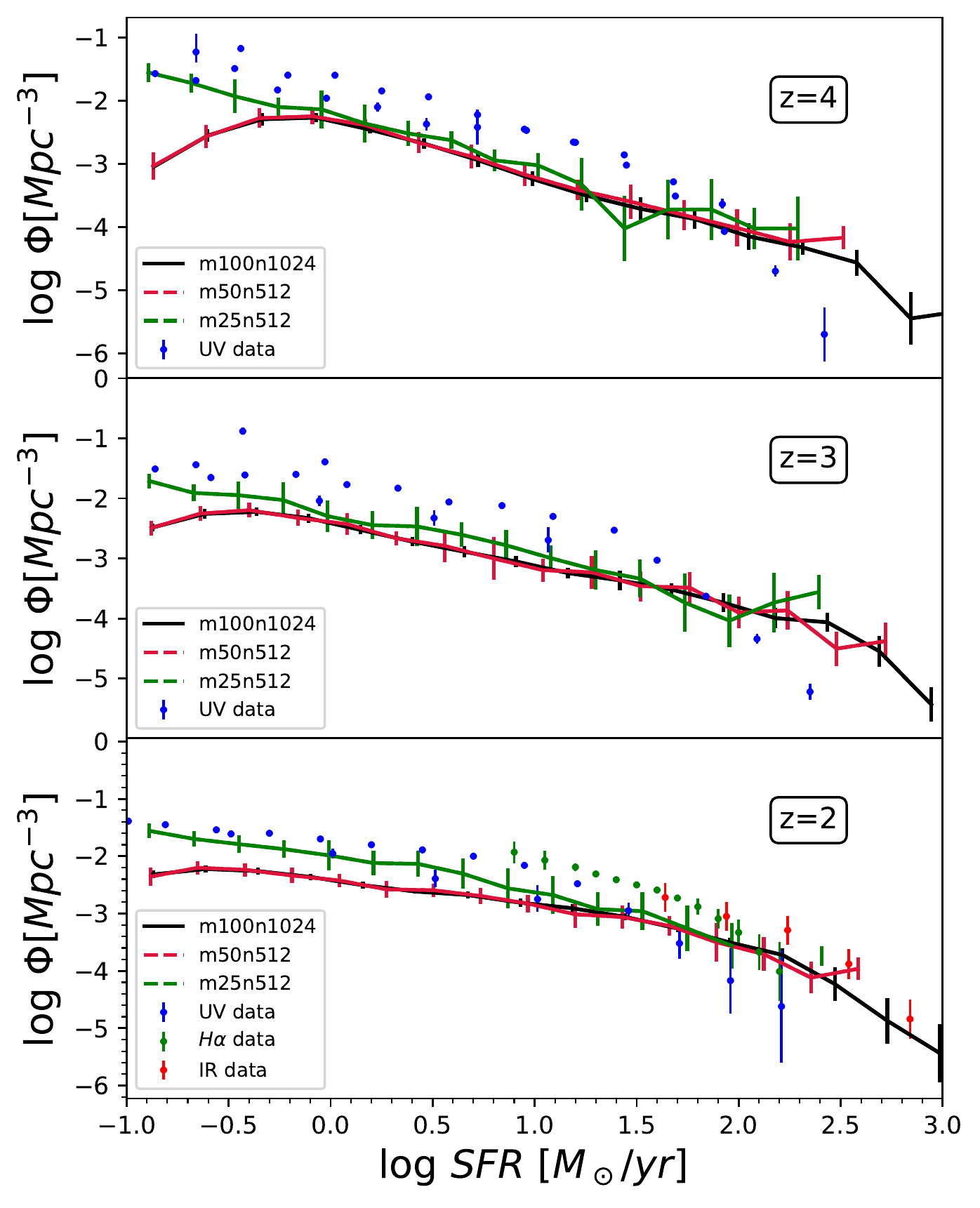}
    \caption{The SFR function at $z=4,3,2$ (top to bottom) in three \simba\ runs. Comparing the fiducial $100 \, h^{-1} \, \mathrm{Mpc}$, $2\times 1024^3$ run with the mini-me $50 \, h^{-1} \, \mathrm{Mpc}$, $2\times 512^3$ run shows excellent volume convergence, while comparison to a higher-resolution $25 \, h^{-1} \, \mathrm{Mpc}$, $2\times 512^3$ shows good resolution convergence down to our SMG limit of SFR$\geq 20 \, \mathrm{M_\odot \,yr^{-1}}$.
    }
    \label{fig:sfrfcn_res}
\end{figure}

Given that \simba's SFR function is critical for reproducing the SMG population, it is worth examining how well this is converged in terms of both box size and resolution.
For volume convergence, we compare the fiducial $100 \, h^{-1} \, \mathrm{Mpc}$, $2\times 1024^3$ particles box with ``mini-me'' \simba\ which is identical except one-eighth the volume ({\tt m50n512:} $50 \, h^{-1}\, \mathrm{Mpc}$, $2\times 512^3$).
For resolution, we further compare this to one with the same number of particles but one-eighth the volume ({\tt m25n512:} $25 \, h^{-1}\, \mathrm{Mpc}$, $2\times 512^3$).

Figure~\ref{fig:sfrfcn_res} shows this comparison.
Error bars are computed over 8 simulation sub-octants.
There is excellent agreement between {\tt m100n1024} (black line) vs. {\tt m50n512} (red) up to the highest SFR's, showing that the results are very well converged with respect to volume, even down to (at least) a $50 \, h^{-1}\, \mathrm{Mpc}$ box.

At high SFRs, the resolution convergence between {\tt m25n512} (green) vs. {\tt m50n512} (or {\tt m100n1024}) is quite good, but it begins to deviate at low SFRs.
This occurs at a higher SFR at lower redshifts: $\la 1 \; \mathrm{M_\odot \,yr^{-1}}$ at $z=4$, but $\la 10 \; \mathrm{M_\odot \, yr^{-1}}$ at $z=2$.
However, the results remain well converged for $\geq 20 \; \mathrm{M_\odot \, yr^{-1}}$, which is our (conservative) limit for studying SMGs.
We have also performed a test to see how the SFR changes with resolution at fixed halo mass. At $M_{\mathrm{halo}} \,/\, M_{\odot} \sim 10^{12}$ there is a $\sim +0.3 \; \mathrm{dex}$ offset in the SFR in the higher resolution simulation, which translates, given the sub-linear dependence on SFR, into a flux density $\sim1.4$ times higher. This cannot fully explain the offset in \fig{sim_convergence}.
Hence we do not expect resolution convergence to be an issue for the SMG population.

\section{Dependence on Stellar Population Synthesis Model}
\label{sec:bpass}

\begin{figure}
\includegraphics[width=\columnwidth]{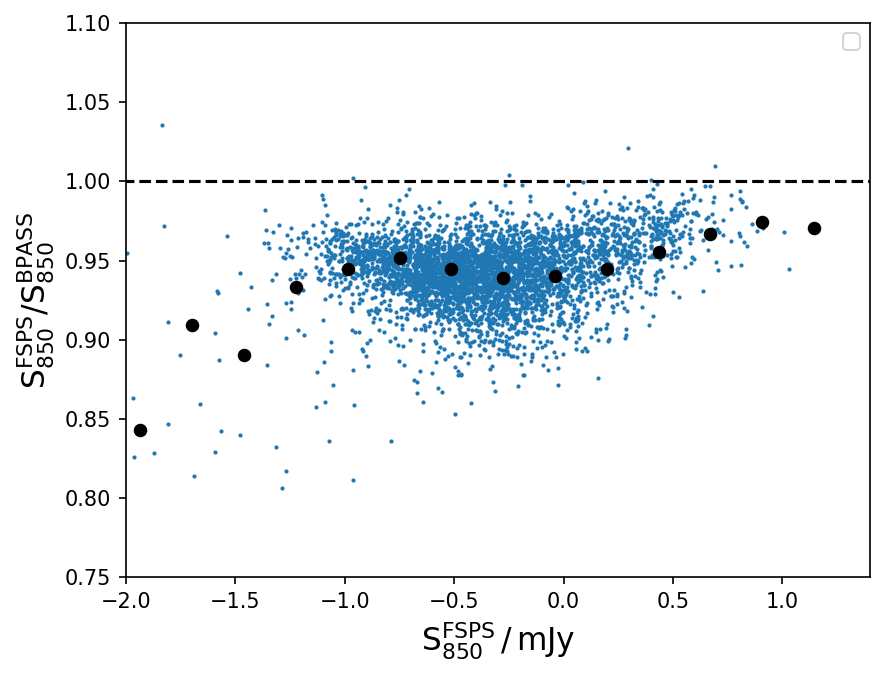}
    \caption{Ratio of the \eightfifty\ flux produced using the FSPS and BPASS models, for halos in the lightcone selection.}
    \label{fig:bpass_comparison}
\end{figure}

There are a number of different Stellar Population Synthesis (SPS) models that make different predictions for the emission from coeval populations with the same metallicity \citep{conroy_modeling_2013,wilkins_photometric_2016,lovell_sengi_2019}.
To assess the impact of SPS model choice on our measured \eightfifty\ fluxes we compare the default FSPS isochrones to those from BPASS \citep{eldridge_binary_2017,stanway_re-evaluating_2018} as a qualitative test.
A more comprehensive test, using a suite of popular SPS models, is beyond the scope of this paper, but this test provides an order of magnitude estimate of the impact of SPS model choice.

\fig{bpass_comparison} shows the ratio of \eightfifty\ fluxes obtained with the FSPS and BPASS isochrones for a selection of galaxies at $z = 2$.
The BPASS binary population fluxes are around $\sim 5\%$ higher in the mJy range.
This is even smaller than the minor offset seen between the \simba\ and observed \citep{geach_scuba-2_2017} number counts, hence our results are not sensitive to our choice of using the BPASS models.

\section{Size evolution and beam-matching}
\label{sec:size_evolution}

\begin{figure}
\includegraphics[width=\columnwidth]{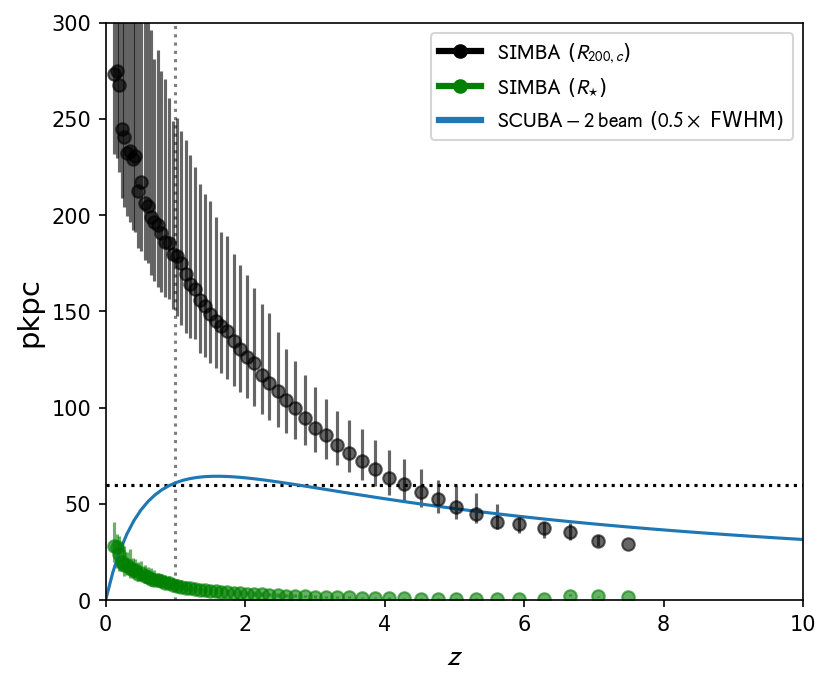}
    \caption{Redshift evolution of the SCUBA-2 beam ($\frac{1}{2} \times \mathrm{FWHM}$, blue) compared to our aperture choice ($60 \; \mathrm{pkpc}$, dotted horizontal).
		We also show the median total stellar radius (green) and the median $R_{200,c}$ of the host halo (black) for all galaxies with stellar masses $> 10^{10} \; \mathrm{M_{\odot}}$.
    }
    \label{fig:beam_size_radius_evolution}
\end{figure}

In order to provide as close to a like-for-like comparison with the S2CLS counts \citep{geach_scuba-2_2017} we employ a $D = 120\,\mathrm{pkpc}$ diameter aperture within which we measure the flux.
This broadly mimics that of the SCUBA-2 beam at $z > 1$.
We choose a fixed aperture size, rather than exactly matching the SCUBA-2 beam, so that we may compare emission properties of galaxies at different redshifts.
To show the effect such a selection would have, \fig{beam_size_radius_evolution} shows the redshift evolution of the physical size of the beam ($\frac{1}{2} \times \mathrm{FWHM}$) alongside the redshift evolution of galaxy and host halo sizes.
We also show our chosen aperture size by the horizontal line at 60 pkpc.
At all redshifts galaxies tend to be much smaller than the aperture, but at $z < 4$ their host halos extend beyond the aperture.
Other galaxies within the aperture can therefore contribute significantly to the flux density.

%% file: main.bbl
\begin{thebibliography}{}
\makeatletter
\relax
\def\mn@urlcharsother{\let\do\@makeother \do\$\do\&\do\#\do\^\do\_\do\%\do\~}
\def\mn@doi{\begingroup\mn@urlcharsother \@ifnextchar [ {\mn@doi@}
  {\mn@doi@[]}}
\def\mn@doi@[#1]#2{\def\@tempa{#1}\ifx\@tempa\@empty \href
  {http://dx.doi.org/#2} {doi:#2}\else \href {http://dx.doi.org/#2} {#1}\fi
  \endgroup}
\def\mn@eprint#1#2{\mn@eprint@#1:#2::\@nil}
\def\mn@eprint@arXiv#1{\href {http://arxiv.org/abs/#1} {{\tt arXiv:#1}}}
\def\mn@eprint@dblp#1{\href {http://dblp.uni-trier.de/rec/bibtex/#1.xml}
  {dblp:#1}}
\def\mn@eprint@#1:#2:#3:#4\@nil{\def\@tempa {#1}\def\@tempb {#2}\def\@tempc
  {#3}\ifx \@tempc \@empty \let \@tempc \@tempb \let \@tempb \@tempa \fi \ifx
  \@tempb \@empty \def\@tempb {arXiv}\fi \@ifundefined
  {mn@eprint@\@tempb}{\@tempb:\@tempc}{\expandafter \expandafter \csname
  mn@eprint@\@tempb\endcsname \expandafter{\@tempc}}}

\bibitem[\protect\citeauthoryear{Alavi et~al.,}{Alavi
  et~al.}{2014}]{alavi_ultra-faint_2014}
Alavi A.,  et~al., 2014, \mn@doi [ApJ] {10.1088/0004-637X/780/2/143}, 780, 143

\bibitem[\protect\citeauthoryear{Alexander, Bauer, Chapman, Smail, Blain,
  Brandt  \& Ivison}{Alexander et~al.}{2005}]{alexander_x-ray_2005}
Alexander D.~M.,  Bauer F.~E.,  Chapman S.~C.,  Smail I.,  Blain A.~W.,  Brandt
  W.~N.,   Ivison R.~J.,  2005, \mn@doi [ApJ] {10.1086/444342}, 632, 736

\bibitem[\protect\citeauthoryear{Alexander et~al.,}{Alexander
  et~al.}{2008}]{alexander_weighing_2008}
Alexander D.~M.,  et~al., 2008, \mn@doi [AJ] {10.1088/0004-6256/135/5/1968},
  135, 1968

\bibitem[\protect\citeauthoryear{An et~al.,}{An
  et~al.}{2019}]{an_multi-wavelength_2019}
An F.~X.,  et~al., 2019, \mn@doi [ApJ] {10.3847/1538-4357/ab4d53}, 886, 48

\bibitem[\protect\citeauthoryear{Angl{\'e}s-Alc{\'a}zar, Dav{\'e},
  Faucher-Gigu{\`e}re, {\"O}zel  \& Hopkins}{Angl{\'e}s-Alc{\'a}zar
  et~al.}{2017a}]{angles-alcazar_gravitational_2017}
Angl{\'e}s-Alc{\'a}zar D.,  Dav{\'e} R.,  Faucher-Gigu{\`e}re C.-A.,  {\"O}zel
  F.,   Hopkins P.~F.,  2017a, \mn@doi [MNRAS] {10.1093/mnras/stw2565}, 464,
  2840

\bibitem[\protect\citeauthoryear{Angl{\'e}s-Alc{\'a}zar, Faucher-Gigu{\`e}re,
  Kere{\v s}, Hopkins, Quataert  \& Murray}{Angl{\'e}s-Alc{\'a}zar
  et~al.}{2017b}]{angles-alcazar_cosmic_2017}
Angl{\'e}s-Alc{\'a}zar D.,  Faucher-Gigu{\`e}re C.-A.,  Kere{\v s} D.,  Hopkins
  P.~F.,  Quataert E.,   Murray N.,  2017b, \mn@doi [MNRAS]
  {10.1093/mnras/stx1517}, 470, 4698

\bibitem[\protect\citeauthoryear{Appleby, Dav{\'e}, Kraljic,
  Angl{\'e}s-Alc{\'a}zar  \& Narayanan}{Appleby
  et~al.}{2020}]{appleby_impact_2020}
Appleby S.,  Dav{\'e} R.,  Kraljic K.,  Angl{\'e}s-Alc{\'a}zar D.,   Narayanan
  D.,  2020, \mn@doi [MNRAS] {10.1093/mnras/staa1169}, 494, 6053

\bibitem[\protect\citeauthoryear{Asano, Takeuchi, Hirashita  \& Inoue}{Asano
  et~al.}{2013}]{asano_dust_2013}
Asano R.~S.,  Takeuchi T.~T.,  Hirashita H.,   Inoue A.~K.,  2013, \mn@doi
  [Earth, Planets, and Space] {10.5047/eps.2012.04.014}, 65, 213

\bibitem[\protect\citeauthoryear{Austermann et~al.,}{Austermann
  et~al.}{2010}]{austermann_aztec_2010}
Austermann J.~E.,  et~al., 2010, \mn@doi [MNRAS]
  {10.1111/j.1365-2966.2009.15620.x}, 401, 160

\bibitem[\protect\citeauthoryear{Baes, Tr{\v c}ka, Camps, Nersesian, Trayford,
  Theuns  \& Dobbels}{Baes et~al.}{2019}]{baes_cosmic_2019}
Baes M.,  Tr{\v c}ka A.,  Camps P.,  Nersesian A.,  Trayford J.,  Theuns T.,
  Dobbels W.,  2019, \mn@doi [MNRAS] {10.1093/mnras/stz302}, 484, 4069

\bibitem[\protect\citeauthoryear{Baes et~al.,}{Baes
  et~al.}{2020}]{baes_infrared_2020}
Baes M.,  et~al., 2020, \mn@doi [MNRAS] {10.1093/mnras/staa990}, 494, 2912

\bibitem[\protect\citeauthoryear{Bastian, Covey  \& Meyer}{Bastian
  et~al.}{2010}]{bastian_universal_2010}
Bastian N.,  Covey K.~R.,   Meyer M.~R.,  2010, \mn@doi [ARAA]
  {10.1146/annurev-astro-082708-101642}, 48, 339

\bibitem[\protect\citeauthoryear{Baugh, Lacey, Frenk, Granato, Silva, Bressan,
  Benson  \& Cole}{Baugh et~al.}{2005a}]{baugh_can_2005}
Baugh C.~M.,  Lacey C.~G.,  Frenk C.~S.,  Granato G.~L.,  Silva L.,  Bressan
  A.,  Benson A.~J.,   Cole S.,  2005a, \mn@doi [MNRAS]
  {10.1111/j.1365-2966.2004.08553.x}, 356, 1191

\bibitem[\protect\citeauthoryear{{Baugh} et~al.}{{Baugh}
  et~al.}{2005b}]{baugh05a}
{Baugh} C.~M.,  et~al., 2005b, \mn@doi [\mnras]
  {10.1111/j.1365-2966.2004.08553.x}, \href
  {http://adsabs.harvard.edu/abs/2005MNRAS.356.1191B} {356, 1191}

\bibitem[\protect\citeauthoryear{Behroozi, Wechsler  \& Conroy}{Behroozi
  et~al.}{2013}]{behroozi_average_2013}
Behroozi P.~S.,  Wechsler R.~H.,   Conroy C.,  2013, \mn@doi [ApJ]
  {10.1088/0004-637X/770/1/57}, 770, 57

\bibitem[\protect\citeauthoryear{B{\'e}thermin et~al.,}{B{\'e}thermin
  et~al.}{2015}]{bethermin_evolution_2015}
B{\'e}thermin M.,  et~al., 2015, \mn@doi [A\&A] {10.1051/0004-6361/201425031},
  573, A113

\bibitem[\protect\citeauthoryear{Bethermin et~al.,}{Bethermin
  et~al.}{2020}]{bethermin_alpine-alma_2020}
Bethermin M.,  et~al., 2020, arXiv e-prints, 2002, arXiv:2002.00962

\bibitem[\protect\citeauthoryear{Bianchi \& Schneider}{Bianchi \&
  Schneider}{2007}]{bianchi_dust_2007}
Bianchi S.,  Schneider R.,  2007, \mn@doi [MNRAS]
  {10.1111/j.1365-2966.2007.11829.x}, 378, 973

\bibitem[\protect\citeauthoryear{Blain, Smail, Ivison, Kneib  \& Frayer}{Blain
  et~al.}{2002}]{blain_submillimeter_2002}
Blain A.~W.,  Smail I.,  Ivison R.~J.,  Kneib J.-P.,   Frayer D.~T.,  2002,
  \mn@doi [Physics Reports] {10.1016/S0370-1573(02)00134-5}, 369, 111

\bibitem[\protect\citeauthoryear{Bondi \& Hoyle}{Bondi \&
  Hoyle}{1944}]{bondi_mechanism_1944}
Bondi H.,  Hoyle F.,  1944, \mn@doi [MNRAS] {10.1093/mnras/104.5.273}, 104, 273

\bibitem[\protect\citeauthoryear{Bothwell et~al.,}{Bothwell
  et~al.}{2013}]{bothwell_survey_2013}
Bothwell M.~S.,  et~al., 2013, \mn@doi [MNRAS] {10.1093/mnras/sts562}, 429,
  3047

\bibitem[\protect\citeauthoryear{Bussmann et~al.,}{Bussmann
  et~al.}{2015}]{bussmann_hermes_2015}
Bussmann R.~S.,  et~al., 2015, \mn@doi [ApJ] {10.1088/0004-637X/812/1/43}, 812,
  43

\bibitem[\protect\citeauthoryear{Camps, Trayford, Baes, Theuns, Schaller  \&
  Schaye}{Camps et~al.}{2016}]{camps_far-infrared_2016}
Camps P.,  Trayford J.~W.,  Baes M.,  Theuns T.,  Schaller M.,   Schaye J.,
  2016, \mn@doi [MNRAS] {10.1093/mnras/stw1735}, 462, 1057

\bibitem[\protect\citeauthoryear{Camps et~al.,}{Camps
  et~al.}{2018}]{camps_data_2018}
Camps P.,  et~al., 2018, \mn@doi [ApJS] {10.3847/1538-4365/aaa24c}, 234, 20

\bibitem[\protect\citeauthoryear{Carilli et~al.,}{Carilli
  et~al.}{2010}]{carilli_imaging_2010}
Carilli C.~L.,  et~al., 2010, \mn@doi [ApJ] {10.1088/0004-637X/714/2/1407},
  714, 1407

\bibitem[\protect\citeauthoryear{Casey et~al.,}{Casey
  et~al.}{2013}]{casey_characterization_2013}
Casey C.~M.,  et~al., 2013, \mn@doi [MNRAS] {10.1093/mnras/stt1673}, 436, 1919

\bibitem[\protect\citeauthoryear{Casey, Narayanan  \& Cooray}{Casey
  et~al.}{2014}]{casey_dusty_2014}
Casey C.~M.,  Narayanan D.,   Cooray A.,  2014, \mn@doi [Phys. Rep.]
  {10.1016/j.physrep.2014.02.009}, 541, 45

\bibitem[\protect\citeauthoryear{Casey et~al.,}{Casey
  et~al.}{2018}]{casey_brightest_2018}
Casey C.~M.,  et~al., 2018, \mn@doi [ApJ] {10.3847/1538-4357/aac82d}, 862, 77

\bibitem[\protect\citeauthoryear{Chabrier}{Chabrier}{2003}]{chabrier_galactic_2003}
Chabrier G.,  2003, \mn@doi [PASP] {10.1086/376392}, 115, 763

\bibitem[\protect\citeauthoryear{Chapman, Blain, Smail  \& Ivison}{Chapman
  et~al.}{2005}]{chapman_redshift_2005}
Chapman S.~C.,  Blain A.~W.,  Smail I.,   Ivison R.~J.,  2005, \mn@doi [ApJ]
  {10.1086/428082}, 622, 772

\bibitem[\protect\citeauthoryear{Chen, Cowie, Barger, Casey, Lee, Sanders, Wang
   \& Williams}{Chen et~al.}{2013}]{chen_resolving_2013}
Chen C.-C.,  Cowie L.~L.,  Barger A.~J.,  Casey C.~M.,  Lee N.,  Sanders D.~B.,
   Wang W.-H.,   Williams J.~P.,  2013, \mn@doi [ApJ]
  {10.1088/0004-637X/776/2/131}, 776, 131

\bibitem[\protect\citeauthoryear{Chen et~al.,}{Chen
  et~al.}{2015}]{chen_alma_2015}
Chen C.-C.,  et~al., 2015, \mn@doi [ApJ] {10.1088/0004-637X/799/2/194}, 799,
  194

\bibitem[\protect\citeauthoryear{Chen et~al.,}{Chen
  et~al.}{2016}]{chen_faint_2016}
Chen C.-C.,  et~al., 2016, \mn@doi [ApJ] {10.3847/0004-637X/831/1/91}, 831, 91

\bibitem[\protect\citeauthoryear{Choi, Ostriker, Naab  \& Johansson}{Choi
  et~al.}{2012}]{choi_radiative_2012}
Choi E.,  Ostriker J.~P.,  Naab T.,   Johansson P.~H.,  2012, \mn@doi [ApJ]
  {10.1088/0004-637X/754/2/125}, 754, 125

\bibitem[\protect\citeauthoryear{Christiansen, Dav{\'e}, Sorini  \&
  Angl{\'e}s-Alc{\'a}zar}{Christiansen et~al.}{2019}]{christiansen_jet_2019}
Christiansen J.~F.,  Dav{\'e} R.,  Sorini D.,   Angl{\'e}s-Alc{\'a}zar D.,
  2019, arXiv e-prints, 1911, arXiv:1911.01343

\bibitem[\protect\citeauthoryear{Conroy}{Conroy}{2013}]{conroy_modeling_2013}
Conroy C.,  2013, \mn@doi [ARAA] {10.1146/annurev-astro-082812-141017}, 51, 393

\bibitem[\protect\citeauthoryear{Conroy \& Gunn}{Conroy \&
  Gunn}{2010}]{conroy_propagation_2010}
Conroy C.,  Gunn J.~E.,  2010, \mn@doi [ApJ] {10.1088/0004-637X/712/2/833},
  712, 833

\bibitem[\protect\citeauthoryear{Conroy, Gunn  \& White}{Conroy
  et~al.}{2009}]{conroy_propagation_2009}
Conroy C.,  Gunn J.~E.,   White M.,  2009, \mn@doi [ApJ]
  {10.1088/0004-637X/699/1/486}, 699, 486

\bibitem[\protect\citeauthoryear{Coppin et~al.,}{Coppin
  et~al.}{2006}]{coppin_scuba_2006}
Coppin K.,  et~al., 2006, \mn@doi [MNRAS] {10.1111/j.1365-2966.2006.10961.x},
  372, 1621

\bibitem[\protect\citeauthoryear{Coppin et~al.,}{Coppin
  et~al.}{2010}]{coppin_mid-infrared_2010}
Coppin K.,  et~al., 2010, \mn@doi [ApJ] {10.1088/0004-637X/713/1/503}, 713, 503

\bibitem[\protect\citeauthoryear{Cowley, Lacey, Baugh  \& Cole}{Cowley
  et~al.}{2015}]{cowley_simulated_2015}
Cowley W.~I.,  Lacey C.~G.,  Baugh C.~M.,   Cole S.,  2015, \mn@doi [MNRAS]
  {10.1093/mnras/stu2179}, 446, 1784

\bibitem[\protect\citeauthoryear{Cowley, Lacey, Baugh, Cole, Frenk  \&
  Lagos}{Cowley et~al.}{2019}]{cowley_evolution_2019}
Cowley W.~I.,  Lacey C.~G.,  Baugh C.~M.,  Cole S.,  Frenk C.~S.,   Lagos C.
  d.~P.,  2019, \mn@doi [MNRAS] {10.1093/mnras/stz1398}, 487, 3082

\bibitem[\protect\citeauthoryear{Crain et~al.,}{Crain
  et~al.}{2015}]{crain_eagle_2015}
Crain R.~A.,  et~al., 2015, \mn@doi [MNRAS] {10.1093/mnras/stv725}, 450, 1937

\bibitem[\protect\citeauthoryear{Danielson et~al.,}{Danielson
  et~al.}{2017}]{danielson_alma_2017}
Danielson A. L.~R.,  et~al., 2017, \mn@doi [ApJ] {10.3847/1538-4357/aa6caf},
  840, 78

\bibitem[\protect\citeauthoryear{Dav{\'e}, Finlator, Oppenheimer, Fardal, Katz,
  Kere{\v s}  \& Weinberg}{Dav{\'e} et~al.}{2010}]{dave_nature_2010}
Dav{\'e} R.,  Finlator K.,  Oppenheimer B.~D.,  Fardal M.,  Katz N.,  Kere{\v
  s} D.,   Weinberg D.~H.,  2010, \mn@doi [MNRAS]
  {10.1111/j.1365-2966.2010.16395.x}

\bibitem[\protect\citeauthoryear{Dav{\'e}, Thompson  \& Hopkins}{Dav{\'e}
  et~al.}{2016}]{dave_mufasa:_2016}
Dav{\'e} R.,  Thompson R.~J.,   Hopkins P.~F.,  2016, \mn@doi [MNRAS]
  {10.1093/mnras/stw1862}, 462, 3265

\bibitem[\protect\citeauthoryear{Dav{\'e}, Rafieferantsoa, Thompson  \&
  Hopkins}{Dav{\'e} et~al.}{2017}]{dave_mufasa:_2017}
Dav{\'e} R.,  Rafieferantsoa M.~H.,  Thompson R.~J.,   Hopkins P.~F.,  2017,
  \mn@doi [MNRAS] {10.1093/mnras/stx108}, 467, 115

\bibitem[\protect\citeauthoryear{Dav{\'e}, Angl{\'e}s-Alc{\'a}zar, Narayanan,
  Li, Rafieferantsoa  \& Appleby}{Dav{\'e} et~al.}{2019}]{dave_simba:_2019}
Dav{\'e} R.,  Angl{\'e}s-Alc{\'a}zar D.,  Narayanan D.,  Li Q.,  Rafieferantsoa
  M.~H.,   Appleby S.,  2019, \mn@doi [MNRAS] {10.1093/mnras/stz937}, 486, 2827

\bibitem[\protect\citeauthoryear{Dav{\'e}, Crain, Stevens, Narayanan,
  Saintonge, Catinella  \& Cortese}{Dav{\'e} et~al.}{2020}]{dave_galaxy_2020}
Dav{\'e} R.,  Crain R.~A.,  Stevens A. R.~H.,  Narayanan D.,  Saintonge A.,
  Catinella B.,   Cortese L.,  2020, \mn@doi [MNRAS] {10.1093/mnras/staa1894},
  497, 146

\bibitem[\protect\citeauthoryear{Decarli et~al.,}{Decarli
  et~al.}{2019}]{decarli_alma_2019}
Decarli R.,  et~al., 2019, \mn@doi [ApJ] {10.3847/1538-4357/ab30fe}, 882, 138

\bibitem[\protect\citeauthoryear{Dekel et~al.,}{Dekel
  et~al.}{2009a}]{dekel_cold_2009}
Dekel A.,  et~al., 2009a, \mn@doi [Nature] {10.1038/nature07648}, 457, 451

\bibitem[\protect\citeauthoryear{{Dekel} et~al.,}{{Dekel}
  et~al.}{2009b}]{dekel09a}
{Dekel} A.,  et~al., 2009b, \mn@doi [\nat] {10.1038/nature07648}, \href
  {http://adsabs.harvard.edu/abs/2009Natur.457..451D} {457, 451}

\bibitem[\protect\citeauthoryear{Dempsey et~al.,}{Dempsey
  et~al.}{2013}]{dempsey_scuba-2_2013}
Dempsey J.~T.,  et~al., 2013, \mn@doi [MNRAS] {10.1093/mnras/stt090}, 430, 2534

\bibitem[\protect\citeauthoryear{Draine}{Draine}{2003}]{draine_interstellar_2003}
Draine B.~T.,  2003, \mn@doi [ARAA] {10.1146/annurev.astro.41.011802.094840},
  41, 241

\bibitem[\protect\citeauthoryear{Dudzevi{\v c}i{\=u}t{\.e} et~al.,}{Dudzevi{\v
  c}i{\=u}t{\.e} et~al.}{2020}]{dudzeviciute_alma_2020}
Dudzevi{\v c}i{\=u}t{\.e} U.,  et~al., 2020, \mn@doi [MNRAS]
  {10.1093/mnras/staa769}, 494, 3828

\bibitem[\protect\citeauthoryear{Dwek}{Dwek}{1998}]{dwek_evolution_1998}
Dwek E.,  1998, \mn@doi [ApJ] {10.1086/305829}, 501, 643

\bibitem[\protect\citeauthoryear{Eldridge, Stanway, Xiao, McClelland, Taylor,
  Ng, Greis  \& Bray}{Eldridge et~al.}{2017}]{eldridge_binary_2017}
Eldridge J.~J.,  Stanway E.~R.,  Xiao L.,  McClelland L. A.~S.,  Taylor G.,  Ng
  M.,  Greis S. M.~L.,   Bray J.~C.,  2017, \mn@doi [PASA]
  {10.1017/pasa.2017.51}, 34, e058

\bibitem[\protect\citeauthoryear{Engel et~al.,}{Engel
  et~al.}{2010}]{engel_most_2010}
Engel H.,  et~al., 2010, \mn@doi [ApJ] {10.1088/0004-637X/724/1/233}, 724, 233

\bibitem[\protect\citeauthoryear{{Fardal}, {Katz}, {Weinberg}, {Dav{\'e}}  \&
  {Hernquist}}{{Fardal} et~al.}{2001}]{fardal01a}
{Fardal} M.~A.,  {Katz} N.,  {Weinberg} D.~H.,  {Dav{\'e}} R.,   {Hernquist}
  L.,  2001, ApJ Submitted: arXiv/0107290, \href
  {http://adsabs.harvard.edu/abs/2001astro.ph..7290F} {}

\bibitem[\protect\citeauthoryear{Ferrarotti \& Gail}{Ferrarotti \&
  Gail}{2006}]{ferrarotti_composition_2006}
Ferrarotti A.~S.,  Gail H.-P.,  2006, \mn@doi [A\&A]
  {10.1051/0004-6361:20041198}, 447, 553

\bibitem[\protect\citeauthoryear{{Finlator}, {Dav{\'e}}, {Papovich}  \&
  {Hernquist}}{{Finlator} et~al.}{2006}]{finlator06a}
{Finlator} K.,  {Dav{\'e}} R.,  {Papovich} C.,   {Hernquist} L.,  2006, \mn@doi
  [\apj] {10.1086/499349}, \href
  {http://adsabs.harvard.edu/abs/2006ApJ...639..672F} {639, 672}

\bibitem[\protect\citeauthoryear{Fontanot, Monaco, Silva  \& Grazian}{Fontanot
  et~al.}{2007}]{fontanot_reproducing_2007}
Fontanot F.,  Monaco P.,  Silva L.,   Grazian A.,  2007, \mn@doi [MNRAS]
  {10.1111/j.1365-2966.2007.12449.x}, 382, 903

\bibitem[\protect\citeauthoryear{Foreman-Mackey, Sick  \&
  Johnson}{Foreman-Mackey et~al.}{2014}]{dan_foreman-mackey_python-fsps:_2014}
Foreman-Mackey D.,  Sick J.,   Johnson B.,  2014, python-fsps: {Python}
  bindings to {FSPS} (v0.1.1), \mn@doi{10.5281/zenodo.12157}, \url
  {https://zenodo.org/record/12157}

\bibitem[\protect\citeauthoryear{Furlong et~al.,}{Furlong
  et~al.}{2015}]{furlong_evolution_2015}
Furlong M.,  et~al., 2015, \mn@doi [MNRAS] {10.1093/mnras/stv852}, 450, 4486

\bibitem[\protect\citeauthoryear{Geach et~al.,}{Geach
  et~al.}{2017}]{geach_scuba-2_2017}
Geach J.~E.,  et~al., 2017, \mn@doi [MNRAS] {10.1093/mnras/stw2721}, 465, 1789

\bibitem[\protect\citeauthoryear{Granato, Lacey, Silva, Bressan, Baugh, Cole
  \& Frenk}{Granato et~al.}{2000}]{granato_infrared_2000}
Granato G.~L.,  Lacey C.~G.,  Silva L.,  Bressan A.,  Baugh C.~M.,  Cole S.,
  Frenk C.~S.,  2000, \mn@doi [ApJ] {10.1086/317032}, 542, 710

\bibitem[\protect\citeauthoryear{{Granato}, {De Zotti}, {Silva}, {Bressan}  \&
  {Danese}}{{Granato} et~al.}{2004}]{granato04a}
{Granato} G.~L.,  {De Zotti} G.,  {Silva} L.,  {Bressan} A.,   {Danese} L.,
  2004, \mn@doi [\apj] {10.1086/379875}, \href
  {http://adsabs.harvard.edu/abs/2004ApJ...600..580G} {600, 580}

\bibitem[\protect\citeauthoryear{Gruppioni et~al.,}{Gruppioni
  et~al.}{2013}]{gruppioni_herschel_2013}
Gruppioni C.,  et~al., 2013, \mn@doi [MNRAS] {10.1093/mnras/stt308}, 432, 23

\bibitem[\protect\citeauthoryear{Haardt \& Madau}{Haardt \&
  Madau}{2012}]{haardt_radiative_2012}
Haardt F.,  Madau P.,  2012, \mn@doi [ApJ] {10.1088/0004-637X/746/2/125}, 746,
  125

\bibitem[\protect\citeauthoryear{Hao, Kennicutt, Johnson, Calzetti, Dale  \&
  Moustakas}{Hao et~al.}{2011}]{hao_dust-corrected_2011}
Hao C.-N.,  Kennicutt R.~C.,  Johnson B.~D.,  Calzetti D.,  Dale D.~A.,
  Moustakas J.,  2011, \mn@doi [ApJ] {10.1088/0004-637X/741/2/124}, 741, 124

\bibitem[\protect\citeauthoryear{Hassan, Finlator, Dav{\'e}, Churchill  \&
  Prochaska}{Hassan et~al.}{2020}]{hassan_testing_2020}
Hassan S.,  Finlator K.,  Dav{\'e} R.,  Churchill C.~W.,   Prochaska J.~X.,
  2020, \mn@doi [MNRAS] {10.1093/mnras/staa056}, 492, 2835

\bibitem[\protect\citeauthoryear{Hayward, Kere{\v s}, Jonsson, Narayanan, Cox
  \& Hernquist}{Hayward et~al.}{2011}]{hayward_what_2011}
Hayward C.~C.,  Kere{\v s} D.,  Jonsson P.,  Narayanan D.,  Cox T.~J.,
  Hernquist L.,  2011, \mn@doi [ApJ] {10.1088/0004-637X/743/2/159}, 743, 159

\bibitem[\protect\citeauthoryear{Hayward, Narayanan, Kere{\v s}, Jonsson,
  Hopkins, Cox  \& Hernquist}{Hayward
  et~al.}{2013a}]{hayward_submillimetre_2013}
Hayward C.~C.,  Narayanan D.,  Kere{\v s} D.,  Jonsson P.,  Hopkins P.~F.,  Cox
  T.~J.,   Hernquist L.,  2013a, \mn@doi [MNRAS] {10.1093/mnras/sts222}, 428,
  2529

\bibitem[\protect\citeauthoryear{Hayward, Behroozi, Somerville, Primack, Moreno
   \& Wechsler}{Hayward et~al.}{2013b}]{hayward_spatially_2013}
Hayward C.~C.,  Behroozi P.~S.,  Somerville R.~S.,  Primack J.~R.,  Moreno J.,
   Wechsler R.~H.,  2013b, \mn@doi [MNRAS] {10.1093/mnras/stt1202}, 434, 2572

\bibitem[\protect\citeauthoryear{Hayward et~al.,}{Hayward
  et~al.}{2018}]{hayward_observational_2018}
Hayward C.~C.,  et~al., 2018, \mn@doi [MNRAS] {10.1093/mnras/sty304}, 476, 2278

\bibitem[\protect\citeauthoryear{Hickox et~al.,}{Hickox
  et~al.}{2012}]{hickox_laboca_2012}
Hickox R.~C.,  et~al., 2012, \mn@doi [MNRAS]
  {10.1111/j.1365-2966.2011.20303.x}, 421, 284

\bibitem[\protect\citeauthoryear{Hildebrand}{Hildebrand}{1983}]{hildebrand_determination_1983}
Hildebrand R.~H.,  1983, QJRAS, 24, 267

\bibitem[\protect\citeauthoryear{Hirashita}{Hirashita}{2000}]{hirashita_dust_2000}
Hirashita H.,  2000, \mn@doi [PASJ] {10.1093/pasj/52.4.585}, 52, 585

\bibitem[\protect\citeauthoryear{Hodge \& da Cunha}{Hodge \&
  da~Cunha}{2020}]{hodge_high-redshift_2020}
Hodge J.~A.,  da Cunha E.,  2020, arXiv e-prints, 2004, arXiv:2004.00934

\bibitem[\protect\citeauthoryear{Hodge et~al.,}{Hodge
  et~al.}{2013}]{hodge_alma_2013}
Hodge J.~A.,  et~al., 2013, \mn@doi [ApJ] {10.1088/0004-637X/768/1/91}, 768, 91

\bibitem[\protect\citeauthoryear{Hogg}{Hogg}{2000}]{hogg_distance_2000}
Hogg D.~W.,  2000, arXiv:astro-ph/9905116

\bibitem[\protect\citeauthoryear{Holland et~al.,}{Holland
  et~al.}{2013}]{holland_scuba-2_2013}
Holland W.~S.,  et~al., 2013, \mn@doi [MNRAS] {10.1093/mnras/sts612}, 430, 2513

\bibitem[\protect\citeauthoryear{Hopkins}{Hopkins}{2013}]{hopkins_variations_2013}
Hopkins P.~F.,  2013, \mn@doi [MNRAS] {10.1093/mnras/stt713}, 433, 170

\bibitem[\protect\citeauthoryear{Hopkins}{Hopkins}{2015}]{hopkins_new_2015}
Hopkins P.~F.,  2015, \mn@doi [MNRAS] {10.1093/mnras/stv195}, 450, 53

\bibitem[\protect\citeauthoryear{Hughes et~al.,}{Hughes
  et~al.}{1998}]{hughes_high-redshift_1998}
Hughes D.~H.,  et~al., 1998, \mn@doi [Nature] {10.1038/28328}, 394, 241

\bibitem[\protect\citeauthoryear{Hunter}{Hunter}{2007}]{Hunter:2007}
Hunter J.~D.,  2007, \mn@doi [Computing in Science \& Engineering]
  {10.1109/MCSE.2007.55}, 9, 90

\bibitem[\protect\citeauthoryear{Karim et~al.,}{Karim
  et~al.}{2013}]{karim_alma_2013}
Karim A.,  et~al., 2013, \mn@doi [MNRAS] {10.1093/mnras/stt196}, 432, 2

\bibitem[\protect\citeauthoryear{Katsianis, Tescari, Blanc  \&
  Sargent}{Katsianis et~al.}{2017a}]{katsianis_evolution_2017-1}
Katsianis A.,  Tescari E.,  Blanc G.,   Sargent M.,  2017a, \mn@doi [MNRAS]
  {10.1093/mnras/stw2680}, 464, 4977

\bibitem[\protect\citeauthoryear{Katsianis et~al.,}{Katsianis
  et~al.}{2017b}]{katsianis_evolution_2017}
Katsianis A.,  et~al., 2017b, \mn@doi [MNRAS] {10.1093/mnras/stx2020}, 472, 919

\bibitem[\protect\citeauthoryear{Kennicutt}{Kennicutt}{1998a}]{kennicutt_star_1998}
Kennicutt R.~C.,  1998a, \mn@doi [ARAA] {10.1146/annurev.astro.36.1.189}, 36,
  189

\bibitem[\protect\citeauthoryear{Kennicutt}{Kennicutt}{1998b}]{kennicutt_global_1998}
Kennicutt J.,  1998b, \mn@doi [ApJ] {10.1086/305588}, 498, 541

\bibitem[\protect\citeauthoryear{Kennicutt~Jr \& Evans~II}{Kennicutt~Jr \&
  Evans~II}{2012}]{kennicutt_jr_star_2012}
Kennicutt~Jr R.~C.,  Evans~II N.~J.,  2012, \mn@doi [ARAA]
  {10.1146/annurev-astro-081811-125610}, 50, 531

\bibitem[\protect\citeauthoryear{Krumholz}{Krumholz}{2014}]{krumholz_big_2014}
Krumholz M.~R.,  2014, \mn@doi [Physics Reports]
  {10.1016/j.physrep.2014.02.001}, 539, 49

\bibitem[\protect\citeauthoryear{Krumholz \& Gnedin}{Krumholz \&
  Gnedin}{2011}]{krumholz_comparison_2011}
Krumholz M.~R.,  Gnedin N.~Y.,  2011, \mn@doi [ApJ]
  {10.1088/0004-637X/729/1/36}, 729, 36

\bibitem[\protect\citeauthoryear{Lacey et~al.,}{Lacey
  et~al.}{2016}]{lacey_unified_2016}
Lacey C.~G.,  et~al., 2016, \mn@doi [MNRAS] {10.1093/mnras/stw1888}, 462, 3854

\bibitem[\protect\citeauthoryear{Lagos, Tobar, Robotham, Obreschkow, Mitchell,
  Power  \& Elahi}{Lagos et~al.}{2018}]{lagos_shark:_2018}
Lagos C. d.~P.,  Tobar R.~J.,  Robotham A. S.~G.,  Obreschkow D.,  Mitchell
  P.~D.,  Power C.,   Elahi P.~J.,  2018, \mn@doi [MNRAS]
  {10.1093/mnras/sty2440}, 481, 3573

\bibitem[\protect\citeauthoryear{Lagos et~al.,}{Lagos
  et~al.}{2019}]{lagos_far-ultraviolet_2019}
Lagos C. d.~P.,  et~al., 2019, \mn@doi [MNRAS] {10.1093/mnras/stz2427}, 489,
  4196

\bibitem[\protect\citeauthoryear{{Leja}, {Carnall}, {Johnson}, {Conroy}  \&
  {Speagle}}{{Leja} et~al.}{2019}]{leja19a}
{Leja} J.,  {Carnall} A.~C.,  {Johnson} B.~D.,  {Conroy} C.,   {Speagle} J.~S.,
   2019, \mn@doi [\apj] {10.3847/1538-4357/ab133c}, \href
  {https://ui.adsabs.harvard.edu/abs/2019ApJ...876....3L} {876, 3}

\bibitem[\protect\citeauthoryear{Li, Narayanan  \& Dav{\'e}}{Li
  et~al.}{2019}]{li_dust--gas_2019}
Li Q.,  Narayanan D.,   Dav{\'e} R.,  2019, \mn@doi [MNRAS]
  {10.1093/mnras/stz2684}, 490, 1425

\bibitem[\protect\citeauthoryear{Lim et~al.,}{Lim
  et~al.}{2020}]{lim_scuba-2_2020}
Lim C.-F.,  et~al., 2020, \mn@doi [ApJ] {10.3847/1538-4357/ab8eaf}, 895, 104

\bibitem[\protect\citeauthoryear{Lovell}{Lovell}{2019}]{lovell_sengi_2019}
Lovell C.~C.,  2019, arXiv e-prints, 1911, arXiv:1911.12713

\bibitem[\protect\citeauthoryear{Madau \& Dickinson}{Madau \&
  Dickinson}{2014}]{madau_cosmic_2014}
Madau P.,  Dickinson M.,  2014, \mn@doi [ARAA]
  {10.1146/annurev-astro-081811-125615}, 52, 415

\bibitem[\protect\citeauthoryear{Magnelli, Elbaz, Chary, Dickinson, Le~Borgne,
  Frayer  \& Willmer}{Magnelli et~al.}{2011}]{magnelli_evolution_2011}
Magnelli B.,  Elbaz D.,  Chary R.~R.,  Dickinson M.,  Le~Borgne D.,  Frayer
  D.~T.,   Willmer C. N.~A.,  2011, \mn@doi [A\&A]
  {10.1051/0004-6361/200913941}, 528, A35

\bibitem[\protect\citeauthoryear{Magnelli et~al.,}{Magnelli
  et~al.}{2020}]{magnelli_alma_2020}
Magnelli B.,  et~al., 2020, \mn@doi [ApJ] {10.3847/1538-4357/ab7897}, 892, 66

\bibitem[\protect\citeauthoryear{Mamon, Trevisan, Thuan, Gallazzi  \&
  Dav{\'e}}{Mamon et~al.}{2020}]{mamon_frequency_2020}
Mamon G.~A.,  Trevisan M.,  Thuan T.~X.,  Gallazzi A.,   Dav{\'e} R.,  2020,
  \mn@doi [MNRAS] {10.1093/mnras/stz3556}, 492, 1791

\bibitem[\protect\citeauthoryear{McAlpine et~al.,}{McAlpine
  et~al.}{2019}]{mcalpine_nature_2019}
McAlpine S.,  et~al., 2019, \mn@doi [MNRAS] {10.1093/mnras/stz1692}, 488, 2440

\bibitem[\protect\citeauthoryear{McKinnon, Torrey  \& Vogelsberger}{McKinnon
  et~al.}{2016}]{mckinnon_dust_2016}
McKinnon R.,  Torrey P.,   Vogelsberger M.,  2016, \mn@doi [MNRAS]
  {10.1093/mnras/stw253}, 457, 3775

\bibitem[\protect\citeauthoryear{{McKinnon}, {Torrey}, {Vogelsberger},
  {Hayward}  \& {Marinacci}}{{McKinnon} et~al.}{2017}]{mckinnon17a}
{McKinnon} R.,  {Torrey} P.,  {Vogelsberger} M.,  {Hayward} C.~C.,
  {Marinacci} F.,  2017, \mn@doi [\mnras] {10.1093/mnras/stx467}, \href
  {https://ui.adsabs.harvard.edu/abs/2017MNRAS.468.1505M} {468, 1505}

\bibitem[\protect\citeauthoryear{Micha{\l }owski, Dunlop, Cirasuolo, Hjorth,
  Hayward  \& Watson}{Micha{\l }owski et~al.}{2012}]{michalowski_stellar_2012}
Micha{\l }owski M.~J.,  Dunlop J.~S.,  Cirasuolo M.,  Hjorth J.,  Hayward
  C.~C.,   Watson D.,  2012, \mn@doi [A\&A] {10.1051/0004-6361/201016308}, 541,
  A85

\bibitem[\protect\citeauthoryear{Moster, Naab  \& White}{Moster
  et~al.}{2018}]{moster_emerge_2018}
Moster B.~P.,  Naab T.,   White S. D.~M.,  2018, \mn@doi [MNRAS]
  {10.1093/mnras/sty655}, 477, 1822

\bibitem[\protect\citeauthoryear{Motte et~al.,}{Motte
  et~al.}{2018}]{motte_unexpectedly_2018}
Motte F.,  et~al., 2018, \mn@doi [Nature Astronomy]
  {10.1038/s41550-018-0452-x}, 2, 478

\bibitem[\protect\citeauthoryear{{Narayanan}, {Cox}, {Hayward}, {Younger}  \&
  {Hernquist}}{{Narayanan} et~al.}{2009}]{narayanan09a}
{Narayanan} D.,  {Cox} T.~J.,  {Hayward} C.~C.,  {Younger} J.~D.,   {Hernquist}
  L.,  2009, \mn@doi [\mnras] {10.1111/j.1365-2966.2009.15581.x}, \href
  {http://adsabs.harvard.edu/abs/2009MNRAS.400.1919N} {400, 1919}

\bibitem[\protect\citeauthoryear{Narayanan, Hayward, Cox, Hernquist, Jonsson,
  Younger  \& Groves}{Narayanan et~al.}{2010a}]{narayanan_formation_2010}
Narayanan D.,  Hayward C.~C.,  Cox T.~J.,  Hernquist L.,  Jonsson P.,  Younger
  J.~D.,   Groves B.,  2010a, \mn@doi [MNRAS]
  {10.1111/j.1365-2966.2009.15790.x}, 401, 1613

\bibitem[\protect\citeauthoryear{{Narayanan} et~al.,}{{Narayanan}
  et~al.}{2010b}]{narayanan10b}
{Narayanan} D.,  et~al., 2010b, \mn@doi [\mnras]
  {10.1111/j.1365-2966.2010.16997.x}, \href
  {http://adsabs.harvard.edu/abs/2010MNRAS.407.1701N} {407, 1701}

\bibitem[\protect\citeauthoryear{{Narayanan} et~al.,}{{Narayanan}
  et~al.}{2015a}]{narayanan15a}
{Narayanan} D.,  et~al., 2015a, \mn@doi [\nat] {10.1038/nature15383}, \href
  {http://adsabs.harvard.edu/abs/2015Natur.525..496N} {525, 496}

\bibitem[\protect\citeauthoryear{Narayanan et~al.,}{Narayanan
  et~al.}{2015b}]{narayanan_formation_2015}
Narayanan D.,  et~al., 2015b, \mn@doi [Nature] {10.1038/nature15383}, 525, 496

\bibitem[\protect\citeauthoryear{{Narayanan}, {Dav{\'e}}, {Johnson},
  {Thompson}, {Conroy}  \& {Geach}}{{Narayanan} et~al.}{2018}]{narayanan18a}
{Narayanan} D.,  {Dav{\'e}} R.,  {Johnson} B.~D.,  {Thompson} R.,  {Conroy} C.,
    {Geach} J.,  2018, \mn@doi [\mnras] {10.1093/mnras/stx2860}, \href
  {http://adsabs.harvard.edu/abs/2018MNRAS.474.1718N} {474, 1718}

\bibitem[\protect\citeauthoryear{Narayanan et~al.,}{Narayanan
  et~al.}{2020}]{narayanan_powderday_2020}
Narayanan D.,  et~al., 2020, arXiv e-prints, 2006, arXiv:2006.10757

\bibitem[\protect\citeauthoryear{Park, Kim, Wyithe, Lacey, Baugh,
  Barone-Nugent, Trenti  \& Bouwens}{Park et~al.}{2016}]{park_clustering_2016}
Park J.,  Kim H.-S.,  Wyithe J. S.~B.,  Lacey C.~G.,  Baugh C.~M.,
  Barone-Nugent R.~L.,  Trenti M.,   Bouwens R.~J.,  2016, \mn@doi [MNRAS]
  {10.1093/mnras/stw1316}, 461, 176

\bibitem[\protect\citeauthoryear{Parsa, Dunlop, McLure  \& Mortlock}{Parsa
  et~al.}{2016}]{parsa_galaxy_2016}
Parsa S.,  Dunlop J.~S.,  McLure R.~J.,   Mortlock A.,  2016, \mn@doi [MNRAS]
  {10.1093/mnras/stv2857}, 456, 3194

\bibitem[\protect\citeauthoryear{{Planck Collaboration} et~al.,}{{Planck
  Collaboration} et~al.}{2016}]{planck_collaboration_planck_2016}
{Planck Collaboration} et~al., 2016, \mn@doi [A\&A]
  {10.1051/0004-6361/201525830}, 594, A13

\bibitem[\protect\citeauthoryear{Popping et~al.,}{Popping
  et~al.}{2019}]{popping_alma_2019}
Popping G.,  et~al., 2019, \mn@doi [ApJ] {10.3847/1538-4357/ab30f2}, 882, 137

\bibitem[\protect\citeauthoryear{Popping et~al.,}{Popping
  et~al.}{2020}]{popping_alma_2020}
Popping G.,  et~al., 2020, \mn@doi [ApJ] {10.3847/1538-4357/ab76c0}, 891, 135

\bibitem[\protect\citeauthoryear{Privon, Narayanan  \& Dav{\'e}}{Privon
  et~al.}{2018}]{privon_interpretation_2018}
Privon G.~C.,  Narayanan D.,   Dav{\'e} R.,  2018, \mn@doi [ApJ]
  {10.3847/1538-4357/aae485}, 867, 102

\bibitem[\protect\citeauthoryear{Rahmati, Pawlik, Rai{\v c}evic?  \&
  Schaye}{Rahmati et~al.}{2013}]{rahmati_evolution_2013}
Rahmati A.,  Pawlik A.~H.,  Rai{\v c}evic? M.,   Schaye J.,  2013, \mn@doi
  [MNRAS] {10.1093/mnras/stt066}, 430, 2427

\bibitem[\protect\citeauthoryear{Reddy, Steidel, Pettini, Adelberger, Shapley,
  Erb  \& Dickinson}{Reddy et~al.}{2008}]{reddy_multiwavelength_2008}
Reddy N.~A.,  Steidel C.~C.,  Pettini M.,  Adelberger K.~L.,  Shapley A.~E.,
  Erb D.~K.,   Dickinson M.,  2008, \mn@doi [ApJS] {10.1086/521105}, 175, 48

\bibitem[\protect\citeauthoryear{R{\'e}my-Ruyer et~al.,}{R{\'e}my-Ruyer
  et~al.}{2014}]{remy-ruyer_gas--dust_2014}
R{\'e}my-Ruyer A.,  et~al., 2014, \mn@doi [A\&A] {10.1051/0004-6361/201322803},
  563, A31

\bibitem[\protect\citeauthoryear{Riechers et~al.,}{Riechers
  et~al.}{2010}]{riechers_massive_2010}
Riechers D.~A.,  et~al., 2010, \mn@doi [ApJL] {10.1088/2041-8205/720/2/L131},
  720, L131

\bibitem[\protect\citeauthoryear{Riechers et~al.,}{Riechers
  et~al.}{2019}]{riechers_coldz_2019}
Riechers D.~A.,  et~al., 2019, \mn@doi [ApJ] {10.3847/1538-4357/aafc27}, 872, 7

\bibitem[\protect\citeauthoryear{Robitaille}{Robitaille}{2011}]{robitaille_hyperion_2011}
Robitaille T.~P.,  2011, \mn@doi [A\&A] {10.1051/0004-6361/201117150}, 536, A79

\bibitem[\protect\citeauthoryear{Robitaille et~al.,}{Robitaille
  et~al.}{2013}]{robitaille_astropy:_2013}
Robitaille T.~P.,  et~al., 2013, \mn@doi [A\&A] {10.1051/0004-6361/201322068},
  558, A33

\bibitem[\protect\citeauthoryear{Rodr{\'i}guez~Montero, Dav{\'e}, Wild,
  Angl{\'e}s-Alc{\'a}zar  \& Narayanan}{Rodr{\'i}guez~Montero
  et~al.}{2019}]{rodriguez_montero_mergers_2019}
Rodr{\'i}guez~Montero F.,  Dav{\'e} R.,  Wild V.,  Angl{\'e}s-Alc{\'a}zar D.,
  Narayanan D.,  2019, \mn@doi [MNRAS] {10.1093/mnras/stz2580}, 490, 2139

\bibitem[\protect\citeauthoryear{Rowan-Robinson et~al.,}{Rowan-Robinson
  et~al.}{2018}]{rowan-robinson_extreme_2018}
Rowan-Robinson M.,  et~al., 2018, \mn@doi [A\&A] {10.1051/0004-6361/201832671},
  619, A169

\bibitem[\protect\citeauthoryear{Safarzadeh, Lu  \& Hayward}{Safarzadeh
  et~al.}{2017}]{safarzadeh_is_2017}
Safarzadeh M.,  Lu Y.,   Hayward C.~C.,  2017, \mn@doi [MNRAS]
  {10.1093/mnras/stx2172}, 472, 2462

\bibitem[\protect\citeauthoryear{Salim \& Narayanan}{Salim \&
  Narayanan}{2020}]{salim_dust_2020}
Salim S.,  Narayanan D.,  2020, arXiv e-prints, 2001, arXiv:2001.03181

\bibitem[\protect\citeauthoryear{Salpeter}{Salpeter}{1955}]{salpeter_luminosity_1955}
Salpeter E.~E.,  1955, \mn@doi [ApJ] {10.1086/145971}, 121, 161

\bibitem[\protect\citeauthoryear{S{\'a}nchez-Bl{\'a}zquez
  et~al.,}{S{\'a}nchez-Bl{\'a}zquez
  et~al.}{2006}]{sanchez-blazquez_medium-resolution_2006}
S{\'a}nchez-Bl{\'a}zquez P.,  et~al., 2006, \mn@doi [MNRAS]
  {10.1111/j.1365-2966.2006.10699.x}, 371, 703

\bibitem[\protect\citeauthoryear{Sanders \& Mirabel}{Sanders \&
  Mirabel}{1996}]{sanders_luminous_1996}
Sanders D.~B.,  Mirabel I.~F.,  1996, \mn@doi [ARAA]
  {10.1146/annurev.astro.34.1.749}, 34, 749

\bibitem[\protect\citeauthoryear{Schaye et~al.,}{Schaye
  et~al.}{2015}]{schaye_eagle_2015}
Schaye J.,  et~al., 2015, \mn@doi [MNRAS] {10.1093/mnras/stu2058}, 446, 521

\bibitem[\protect\citeauthoryear{Schneider et~al.,}{Schneider
  et~al.}{2018}]{schneider_excess_2018}
Schneider F. R.~N.,  et~al., 2018, \mn@doi [Science] {10.1126/science.aan0106},
  359, 69

\bibitem[\protect\citeauthoryear{Scott, Dunlop  \& Serjeant}{Scott
  et~al.}{2006}]{scott_combined_2006}
Scott S.~E.,  Dunlop J.~S.,   Serjeant S.,  2006, \mn@doi [MNRAS]
  {10.1111/j.1365-2966.2006.10478.x}, 370, 1057

\bibitem[\protect\citeauthoryear{Scott et~al.,}{Scott
  et~al.}{2012}]{scott_source_2012}
Scott K.~S.,  et~al., 2012, \mn@doi [MNRAS] {10.1111/j.1365-2966.2012.20905.x},
  423, 575

\bibitem[\protect\citeauthoryear{Shimizu, Yoshida  \& Okamoto}{Shimizu
  et~al.}{2012}]{shimizu_submillimetre_2012}
Shimizu I.,  Yoshida N.,   Okamoto T.,  2012, \mn@doi [MNRAS]
  {10.1111/j.1365-2966.2012.22107.x}, 427, 2866

\bibitem[\protect\citeauthoryear{{Silva}, {Granato}, {Bressan}  \&
  {Danese}}{{Silva} et~al.}{1998}]{silva98a}
{Silva} L.,  {Granato} G.~L.,  {Bressan} A.,   {Danese} L.,  1998, \mn@doi
  [\apj] {10.1086/306476}, \href
  {http://adsabs.harvard.edu/abs/1998ApJ...509..103S} {509, 103}

\bibitem[\protect\citeauthoryear{Simpson et~al.,}{Simpson
  et~al.}{2014}]{simpson_alma_2014}
Simpson J.~M.,  et~al., 2014, \mn@doi [ApJ] {10.1088/0004-637X/788/2/125}, 788,
  125

\bibitem[\protect\citeauthoryear{Simpson et~al.,}{Simpson
  et~al.}{2015}]{simpson_scuba-2_2015}
Simpson J.~M.,  et~al., 2015, \mn@doi [ApJ] {10.1088/0004-637X/807/2/128}, 807,
  128

\bibitem[\protect\citeauthoryear{Simpson et~al.,}{Simpson
  et~al.}{2017}]{simpson_scuba-2_2017}
Simpson J.~M.,  et~al., 2017, \mn@doi [ApJ] {10.3847/1538-4357/aa65d0}, 839, 58

\bibitem[\protect\citeauthoryear{Simpson et~al.,}{Simpson
  et~al.}{2019}]{simpson_east_2019}
Simpson J.~M.,  et~al., 2019, \mn@doi [ApJ] {10.3847/1538-4357/ab23ff}, 880, 43

\bibitem[\protect\citeauthoryear{Siringo et~al.,}{Siringo
  et~al.}{2009}]{siringo_large_2009}
Siringo G.,  et~al., 2009, \mn@doi [A\&A] {10.1051/0004-6361/200811454}, 497,
  945

\bibitem[\protect\citeauthoryear{Smail, Ivison  \& Blain}{Smail
  et~al.}{1997}]{smail_deep_1997}
Smail I.,  Ivison R.~J.,   Blain A.~W.,  1997, \mn@doi [ApJL] {10.1086/311017},
  490, L5

\bibitem[\protect\citeauthoryear{Smit, Bouwens, Franx, Illingworth, Labb{\'e},
  Oesch  \& Dokkum}{Smit et~al.}{2012}]{smit_star_2012}
Smit R.,  Bouwens R.~J.,  Franx M.,  Illingworth G.~D.,  Labb{\'e} I.,  Oesch
  P.~A.,   Dokkum P. G.~v.,  2012, \mn@doi [ApJ] {10.1088/0004-637X/756/1/14},
  756, 14

\bibitem[\protect\citeauthoryear{Smith et~al.,}{Smith
  et~al.}{2017}]{smith_grackle_2017}
Smith B.~D.,  et~al., 2017, \mn@doi [MNRAS] {10.1093/mnras/stw3291}, 466, 2217

\bibitem[\protect\citeauthoryear{Smol{\v c}i{\'c} et~al.,}{Smol{\v c}i{\'c}
  et~al.}{2012}]{smolcic_millimeter_2012}
Smol{\v c}i{\'c} V.,  et~al., 2012, \mn@doi [A\&A]
  {10.1051/0004-6361/201219368}, 548, A4

\bibitem[\protect\citeauthoryear{Sobral, Smail, Best, Geach, Matsuda, Stott,
  Cirasuolo  \& Kurk}{Sobral et~al.}{2013}]{sobral_large_2013}
Sobral D.,  Smail I.,  Best P.~N.,  Geach J.~E.,  Matsuda Y.,  Stott J.~P.,
  Cirasuolo M.,   Kurk J.,  2013, \mn@doi [MNRAS] {10.1093/mnras/sts096}, 428,
  1128

\bibitem[\protect\citeauthoryear{Somerville \& Dav{\'e}}{Somerville \&
  Dav{\'e}}{2015}]{somerville_physical_2015}
Somerville R.~S.,  Dav{\'e} R.,  2015, \mn@doi [ARAA]
  {10.1146/annurev-astro-082812-140951}, 53, 51

\bibitem[\protect\citeauthoryear{Somerville, Gilmore, Primack  \&
  Dom{\'i}nguez}{Somerville et~al.}{2012}]{somerville_galaxy_2012}
Somerville R.~S.,  Gilmore R.~C.,  Primack J.~R.,   Dom{\'i}nguez A.,  2012,
  \mn@doi [MNRAS] {10.1111/j.1365-2966.2012.20490.x}, 423, 1992

\bibitem[\protect\citeauthoryear{Springel et~al.,}{Springel
  et~al.}{2005}]{springel_simulations_2005}
Springel V.,  et~al., 2005, \mn@doi [Nature] {10.1038/nature03597}, 435, 629

\bibitem[\protect\citeauthoryear{Stach et~al.,}{Stach
  et~al.}{2018}]{stach_alma_2018}
Stach S.~M.,  et~al., 2018, \mn@doi [ApJ] {10.3847/1538-4357/aac5e5}, 860, 161

\bibitem[\protect\citeauthoryear{Stach et~al.,}{Stach
  et~al.}{2019}]{stach_alma_2019}
Stach S.~M.,  et~al., 2019, \mn@doi [MNRAS] {10.1093/mnras/stz1536}, 487, 4648

\bibitem[\protect\citeauthoryear{Stanway \& Eldridge}{Stanway \&
  Eldridge}{2018}]{stanway_re-evaluating_2018}
Stanway E.~R.,  Eldridge J.~J.,  2018, \mn@doi [MNRAS] {10.1093/mnras/sty1353},
  479, 75

\bibitem[\protect\citeauthoryear{Swinbank, Smail, Chapman, Blain, Ivison  \&
  Keel}{Swinbank et~al.}{2004}]{swinbank_rest-frame_2004}
Swinbank A.~M.,  Smail I.,  Chapman S.~C.,  Blain A.~W.,  Ivison R.~J.,   Keel
  W.~C.,  2004, \mn@doi [ApJ] {10.1086/425171}, 617, 64

\bibitem[\protect\citeauthoryear{Swinbank et~al.,}{Swinbank
  et~al.}{2008}]{swinbank_properties_2008}
Swinbank A.~M.,  et~al., 2008, \mn@doi [MNRAS]
  {10.1111/j.1365-2966.2008.13911.x}, 391, 420

\bibitem[\protect\citeauthoryear{Tacconi et~al.,}{Tacconi
  et~al.}{2008}]{tacconi_submillimeter_2008}
Tacconi L.~J.,  et~al., 2008, \mn@doi [ApJ] {10.1086/587168}, 680, 246

\bibitem[\protect\citeauthoryear{Thomas, Dav{\'e}, Angl{\'e}s-Alc{\'a}zar  \&
  Jarvis}{Thomas et~al.}{2019}]{thomas_black_2019}
Thomas N.,  Dav{\'e} R.,  Angl{\'e}s-Alc{\'a}zar D.,   Jarvis M.,  2019,
  \mn@doi [MNRAS] {10.1093/mnras/stz1703}, 487, 5764

\bibitem[\protect\citeauthoryear{Trayford et~al.,}{Trayford
  et~al.}{2017}]{trayford_optical_2017}
Trayford J.~W.,  et~al., 2017, \mn@doi [MNRAS] {10.1093/mnras/stx1051}, 470,
  771

\bibitem[\protect\citeauthoryear{Trayford, Lagos, Robotham  \&
  Obreschkow}{Trayford et~al.}{2020}]{trayford_fade_2020}
Trayford J.~W.,  Lagos C. d.~P.,  Robotham A. S.~G.,   Obreschkow D.,  2020,
  \mn@doi [MNRAS] {10.1093/mnras/stz3234}, 491, 3937

\bibitem[\protect\citeauthoryear{Tr{\v c}ka et~al.,}{Tr{\v c}ka
  et~al.}{2020}]{trcka_reproducing_2020}
Tr{\v c}ka A.,  et~al., 2020, \mn@doi [MNRAS] {10.1093/mnras/staa857}, 494,
  2823

\bibitem[\protect\citeauthoryear{Turk, Smith, Oishi, Skory, Skillman, Abel  \&
  Norman}{Turk et~al.}{2010}]{turk_yt_2010}
Turk M.~J.,  Smith B.~D.,  Oishi J.~S.,  Skory S.,  Skillman S.~W.,  Abel T.,
  Norman M.~L.,  2010, \mn@doi [ApJS] {10.1088/0067-0049/192/1/9}, 192, 9

\bibitem[\protect\citeauthoryear{{Virtanen} et~al.,}{{Virtanen}
  et~al.}{2020}]{2020SciPy-NMeth}
{Virtanen} P.,  et~al., 2020, \mn@doi [Nature Methods]
  {https://doi.org/10.1038/s41592-019-0686-2}, \href {https://rdcu.be/b08Wh}
  {17, 261}

\bibitem[\protect\citeauthoryear{Wang, Cowie, Barger  \& Williams}{Wang
  et~al.}{2011}]{wang_sma_2011}
Wang W.-H.,  Cowie L.~L.,  Barger A.~J.,   Williams J.~P.,  2011, \mn@doi
  [ApJL] {10.1088/2041-8205/726/2/L18}, 726, L18

\bibitem[\protect\citeauthoryear{Wang et~al.,}{Wang
  et~al.}{2013}]{wang_alma_2013}
Wang S.~X.,  et~al., 2013, \mn@doi [ApJ] {10.1088/0004-637X/778/2/179}, 778,
  179

\bibitem[\protect\citeauthoryear{Wang, Pearson, Cowley, Trayford,
  B{\'e}thermin, Gruppioni, Hurley  \& Micha{\l }owski}{Wang
  et~al.}{2019}]{wang_multi-wavelength_2019}
Wang L.,  Pearson W.~J.,  Cowley W.,  Trayford J.~W.,  B{\'e}thermin M.,
  Gruppioni C.,  Hurley P.,   Micha{\l }owski M.~J.,  2019, \mn@doi [A\&A]
  {10.1051/0004-6361/201834093}, 624, A98

\bibitem[\protect\citeauthoryear{Wardlow et~al.,}{Wardlow
  et~al.}{2011}]{wardlow_laboca_2011}
Wardlow J.~L.,  et~al., 2011, \mn@doi [MNRAS]
  {10.1111/j.1365-2966.2011.18795.x}, 415, 1479

\bibitem[\protect\citeauthoryear{Wardlow et~al.,}{Wardlow
  et~al.}{2018}]{wardlow_alma_2018}
Wardlow J.~L.,  et~al., 2018, \mn@doi [MNRAS] {10.1093/mnras/sty1526}, 479,
  3879

\bibitem[\protect\citeauthoryear{Wei{\ss} et~al.,}{Wei{\ss}
  et~al.}{2009}]{weis_large_2009}
Wei{\ss} A.,  et~al., 2009, \mn@doi [ApJ] {10.1088/0004-637X/707/2/1201}, 707,
  1201

\bibitem[\protect\citeauthoryear{Wilkins, Feng, Di-Matteo, Croft, Stanway,
  Bunker, Waters  \& Lovell}{Wilkins et~al.}{2016}]{wilkins_photometric_2016}
Wilkins S.~M.,  Feng Y.,  Di-Matteo T.,  Croft R.,  Stanway E.~R.,  Bunker A.,
  Waters D.,   Lovell C.,  2016, \mn@doi [MNRAS] {10.1093/mnras/stw1154}, 460,
  3170

\bibitem[\protect\citeauthoryear{Wilkins, Lovell  \& Stanway}{Wilkins
  et~al.}{2019}]{wilkins_recalibrating_2019}
Wilkins S.~M.,  Lovell C.~C.,   Stanway E.~R.,  2019, \mn@doi [MNRAS]
  {10.1093/mnras/stz2894}, 490, 5359

\bibitem[\protect\citeauthoryear{Wu, Dav{\'e}, Tacchella  \& Lotz}{Wu
  et~al.}{2020}]{wu_photometric_2020}
Wu X.,  Dav{\'e} R.,  Tacchella S.,   Lotz J.,  2020, \mn@doi [MNRAS]
  {10.1093/mnras/staa1044}, 494, 5636

\bibitem[\protect\citeauthoryear{Zhang, Romano, Ivison, Papadopoulos  \&
  Matteucci}{Zhang et~al.}{2018}]{zhang_stellar_2018}
Zhang Z.-Y.,  Romano D.,  Ivison R.~J.,  Papadopoulos P.~P.,   Matteucci F.,
  2018, \mn@doi [Nature] {10.1038/s41586-018-0196-x}, 558, 260

\bibitem[\protect\citeauthoryear{da Cunha et~al.,}{da~Cunha
  et~al.}{2015}]{da_cunha_alma_2015}
da Cunha E.,  et~al., 2015, \mn@doi [ApJ] {10.1088/0004-637X/806/1/110}, 806,
  110

\bibitem[\protect\citeauthoryear{van~der Burg, Hildebrandt  \& Erben}{van~der
  Burg et~al.}{2010}]{van_der_burg_uv_2010}
van~der Burg R. F.~J.,  Hildebrandt H.,   Erben T.,  2010, \mn@doi [A\&A]
  {10.1051/0004-6361/200913812}, 523, A74

\makeatother
\end{thebibliography}
